\documentclass[twocolumn,preprintnumbers,amsmath,amssymb]{revtex4}

\usepackage{graphicx}
\usepackage{dcolumn}
\usepackage{bm}
\usepackage[latin1]{inputenc}

\newcommand{\ud}[1]{{#1^{\dagger}}}
\newcommand{\bra}[1]{\left\langle #1\right|}
\newcommand{\ket}[1]{\left| #1\right\rangle}

\newcommand\Tr{\mathrm{Tr}}
\newcommand{\mean}[1]{\langle#1\rangle}

\begin{document}


\title{Regimes of strong light-matter coupling under incoherent
  excitation}

\author{E. del Valle} 

\affiliation{Physikdepartment, TU M\"unchen, James-Franck-Str. 1,
  85748 Garching, Germany}%

\author{F. P. Laussy}

\affiliation{Walter Schottky Institut, Technische Universit\"at
  M\"unchen, Am Coulombwall 3, 85748 Garching, Germany}

\date{\today}

\begin{abstract}
  We study a two-level system (atom, superconducting qubit or quantum
  dot) strongly coupled to the single photonic mode of a cavity, in
  the presence of incoherent pumping and including detuning and
  dephasing. This system displays a striking quantum to classical
  transition. On the grounds of several approximations that reproduce
  to various degrees exact results obtained numerically, we separate
  five regimes of operations, that we term ``linear'', ``quantum'',
  ``lasing'', ``quenching'' and ``thermal''.  In the fully quantized
  picture, the lasing regime arises as a condensation of dressed
  states and manifests itself as a Mollow triplet structure in the
  direct emitter photoluminescence spectrum, which embeds fundamental
  features of the full-field quantization description of light-matter
  interactions.
\end{abstract}


\maketitle


\section{Introduction}

The strong coupling regime is the ultimate limit of light-matter
interaction, at the level of a single quantum or a few quanta of
excitations. This gave rise to the field of \emph{cavity Quantum
  Electrodynamics} (cavity QED)~\cite{haroche_book06a}, which in the
recent years has blossomed in a large variety of physical systems,
from atoms~\cite{raimond01a} to semiconductors~\cite{khitrova06a}
passing by superconducting circuits~\cite{makhlin01a} and
nanomechanical oscillators~\cite{kippenberg08a}. A fascinating aspect
of this fundamental problem is how it bridges the gap between quantum
and classical coherence. In the former case, one has quantum
superpositions of light and matter, entangling photons with the ground
and excited states of the emitter. In the latter case, one has a
classical photon field, a continuous function of a continuous
variable, fully specified in all its attributes.  The passage from one
to the other can be tracked in the one-atom lasing transition. The
\emph{one-atom laser} is a concept first proposed and theoretically
studied by Mu and Savage~\cite{mu92a}, with the aim of achieving lower
thresholds for lasing. They encouraged experimentalists to bring the
number of atoms $N$ in a conventional laser to unity. In a very high
quality factor cavity, the emitter reaches the strong coupling regime
at the single excitation level. They showed that in this regime, a
single incoherently excited emitter (a two level system in the
simplest case) can constitute the whole gain medium and populate
singlehandedly the cavity with a very large number of photons. If the
spontaneous emission rate of the atom into other modes than the cavity
is small, the growth in the population of photons exhibits no
threshold as a function of the pumping rate~\cite{rice94a} and
develops a classical (Poissonian) field statistics, thanks to the
efficient periodic exchange of excitations with the atom.


On the theoretical side, the single-atom laser has been extensively
studied for its qualities as a
laser~\cite{mu92a,ginzel93a,loffler97a,jones99a,benson99a,karlovich01a,kilin02a,clemens04a,karlovich08a,delvalle09a,poddubny10a,gonzaleztudela10b,ritter10a,auffeves10a,delvalle10d,lsc_gartner11a},
mostly solving its steady state numerically but also through
analytical techniques, such as the continued fraction
expansion~\cite{lsc_gartner11a} or phase-space
representation~\cite{karlovich01a,kilin02a,lsc_gartner11a}, and under
different approximations such as the
few-photon~\cite{loffler97a,karlovich08a,auffeves10a,lsc_gartner11a}
or
semiclassical~\cite{mu92a,benson99a,poddubny10a,delvalle10d,lsc_gartner11a}
dynamics. On the other hand, the spectral properties, that is, atom
and cavity photoluminescence emission spectra, have been studied only
numerically~\cite{mu92a,ginzel93a,loffler97a,benson99a,clemens04a,delvalle09a,poddubny10a},
although some approximations for the linewidth of the cavity spectrum,
that converges to a single peak and exhibits the standard
line-narrowing, provided useful analytical
expressions~\cite{benson99a,poddubny10a}. Given that the cavity field
is coherent in the lasing regime, the back-action of the field on the
emitter leads to the formation of a \emph{Mollow triplet} in its
spontaneous emission
spectrum~\cite{loffler97a,clemens04a,delvalle09a}, with similar
properties to that theoretically predicted by Mollow for an atom under
resonant laser excitation~\cite{mollow69a}. This structure is of a
great fundamental interest as it represents a pinnacle of nonlinear
quantum optics.

On the experimental side, the strong coupling regime is now firmly
established at the single and few photon level with
atoms~\cite{boca04a} as well as with \emph{artificial atoms},
superconducting qubits~\cite{wallraff04a,fink08a} or semiconductor
quantum dots~\cite{reithmaier04a,yoshie04a,peter05a}, and the one atom
laser as described above has been realised in all these
systems~\cite{mckeever03a,astafiev07a,nomura10a}.  The Mollow triplet
under incoherent pumping has not yet been reported in any experiment,
one reason being that they typically focus on the cavity field lasing
properties.  Only the original configuration proposed by Mollow, under
resonant coherent excitation, has been reported, also in all the
systems above, namely with a single atom~\cite{wu75a}, a
molecule~\cite{wrigge08a}, a superconducting qubit~\cite{astafiev10a}
and a quantum dot~\cite{muller07a,vamivakas09a,flagg09a,ates09a}.

In this text, we consider in detail the quantum to classical
transition that leads to lasing in strong-coupling, starting from the
fully quantized description of Jaynes and Cummings~\cite{jaynes63a}.
We show that quantization breaks down as coherence is formed and a
classical description becomes more appropriate. We provide several
limiting cases that describe the system in its various regimes of
operation. Varying degrees of agreement are afforded depending on the
complexity of the approximation. We provide compact analytical
approximations in the simplest cases and a straightforward numerical
procedure that leads to an excellent quantitative agreement. We
include pure dephasing, important in solid state
systems~\cite{lsc_ulrich11a,lsc_roy11a}, and arbitrary detuning
between the modes. We focus more particularly on the lasing regime at
resonance, where we show that the emergence of a Mollow triplet
manifests classical nonlinearities of strong light-matter coupling in
cavity-QED. As a whole, we show that the lasing transition is a rich
and complex one and we hope to give a rather comprehensive view of its
various limits.

The remaining of this text is organized as follows. As the most
striking and characteristic manifestation of lasing in strong coupling
is the Mollow triplet formed under incoherent pumping, we first
revisit its coherent excitation counterpart, this is done in
Sec.~\ref{sec:1}, extending it to include pure dephasing and
detuning. Starting from Sec.~\ref{sec:2}, we turn to the case of
incoherent excitation exclusively and obtain analytically the steady
state, mode populations and photon counting statistics (\ref{sec:21}),
the system full density matrix (\ref{sec:22}) and the two-time
correlators needed to compute the power (or luminescence) spectra
(\ref{sec:23}). We derive the expressions for the lasing properties by
applying the \emph{semiclassical approximation} that we compare to
other approximations that describe the transition into and out-of
lasing. In Sec.~\ref{sec:3}, we put all these elements together and
derive the spectra of emission for both the cavity and the emitter.
We analyse the resonances of the system (\ref{sec:31}) as well as the
elastic scattering component (\ref{sec:33}). We apply again the
semiclassical approximation to simplify the emitter spectra into a
compact closed-form expression for the Mollow triplet under incoherent
pumping (\ref{sec:34}). This expression is used to explore the
parameters where the Mollow can be observed experimentally. In
Sec.~\ref{sec:4}, we summarize our main findings.

\section{Mollow triplet under coherent excitation}
\label{sec:1}

The analysis of light-scattering by a two-level system (representing
an atomic transition) was first given by Mollow~\cite{mollow69a}, who
reported the antibunching of the scattered light, as well as the
spectral structure now known as the \emph{Mollow
  triplet}~\cite{loudon_book00a}. It results in the case where a
strong-beam of light, with frequency~$\omega_\mathrm{L}$, impinges on the
emitter which has a natural frequency~$\omega_\sigma$ ($\hbar=1$). The
Hamiltonian reads:
\begin{equation}
  \label{eq:WedFeb24185855GMT2010}
  H_\mathrm{L}(t)=\omega_\sigma\ud{\sigma}\sigma+\Omega_\mathrm{L}(e^{i\omega_\mathrm{L} t}\sigma+e^{-i\omega_\mathrm{L} t}\ud{\sigma})
\end{equation}
where~$\sigma$ is the pseudo spin operator for the two-level system
and $\Omega_\mathrm{L}$ is its coupling strength with the optical
laser field. Note that the later, described by a complex ($c$-number)
wave $E(t)\approx \Omega_\mathrm{L}(e^{i\omega_\mathrm{L}
  t}+e^{-i\omega_\mathrm{L} t})$, is thus entirely classical.  The
explicit (and fast) time dependence in $H_\mathrm{L}$ can be removed
by going into a frame rotating with the laser
($\Delta=\omega_\mathrm{L}-\omega_\sigma$):
\begin{equation}
  \label{eq:WedFeb24185716GMT2010}
  H_\mathrm{L}=-\Delta\ud{\sigma}\sigma+\Omega_\mathrm{L}(\sigma+\ud{\sigma})\,.
\end{equation}
The spontaneous decay and the pure dephasing suffered by the emitter
can be described by two Lindblad terms in the master equation:
\begin{equation}
  \label{eq:TueNov30150109CET2010}
  \partial_t\rho=i[\rho,H_\mathrm{L}]+\Big[\frac{\gamma_\sigma}{2}\mathcal{L}_\sigma
  +\frac{\gamma_\phi}{2} \mathcal{L}_{\ud{\sigma}\sigma}\Big]\rho\,,
\end{equation}
where $\mathcal{L}_\sigma (\rho)=(2\sigma\rho
\ud{\sigma}-\ud{\sigma}\sigma\rho-\rho\ud{\sigma}\sigma)$. The steady
state of this simple system ($\partial_t\rho=0$) can be solved
analytically (see Appendix~\ref{ap:qrf}) in terms of the emitter
population and coherence~\cite{mollow69a,loudon_book00a}:
\begin{subequations}
  \label{eq:SunJun27175004MSD2010}
  \begin{align}
    &n_\sigma \equiv
    \mean{\ud{\sigma}\sigma}=\frac{(\Omega_\mathrm{L}^\mathrm{eff})^2}{2(\Omega_\mathrm{L}^\mathrm{eff})^2+\frac{\gamma_\sigma}{2}\frac{\gamma_\sigma+\gamma_\phi}{2}}\,, \\
    &\mean{\ud{\sigma}}=i\frac{\gamma_\sigma/2}{\Omega_\mathrm{L}}n_\sigma(1-i\frac{2\Delta}{\gamma_\sigma+\gamma_\phi})\,.
  \end{align}
\end{subequations}
The effective coupling to the laser (intensity that effectively
excites the emitter) is:
\begin{equation}
  \label{eq:TueNov30150310CET2010}
  \Omega_\mathrm{L}^\mathrm{eff} \equiv\frac{\Omega_\mathrm{L}}{\sqrt{1+\big(\frac{2\Delta}{\gamma_\sigma+\gamma_\phi}\big)^2}}\,.
\end{equation}
At resonance, $\Omega_\mathrm{L}^\mathrm{eff}\rightarrow
\Omega_\mathrm{L}$ and $\mean{\ud{\sigma}}$ is pure imaginary.  The
effective laser intensity $\Omega_\mathrm{L}^\mathrm{eff}$ is reduced
with detuning by an amount that depends on the overlap in frequency
between the laser and the emitter lineshapes. As the laser has no
linewidth, the total emitter width $\gamma_\sigma+\gamma_\phi$
determines the overlap. Therefore pure dephasing compensates for
detuning by increasing this overlap.

\begin{figure}[t] 
  \centering 
  \includegraphics[width=\linewidth]{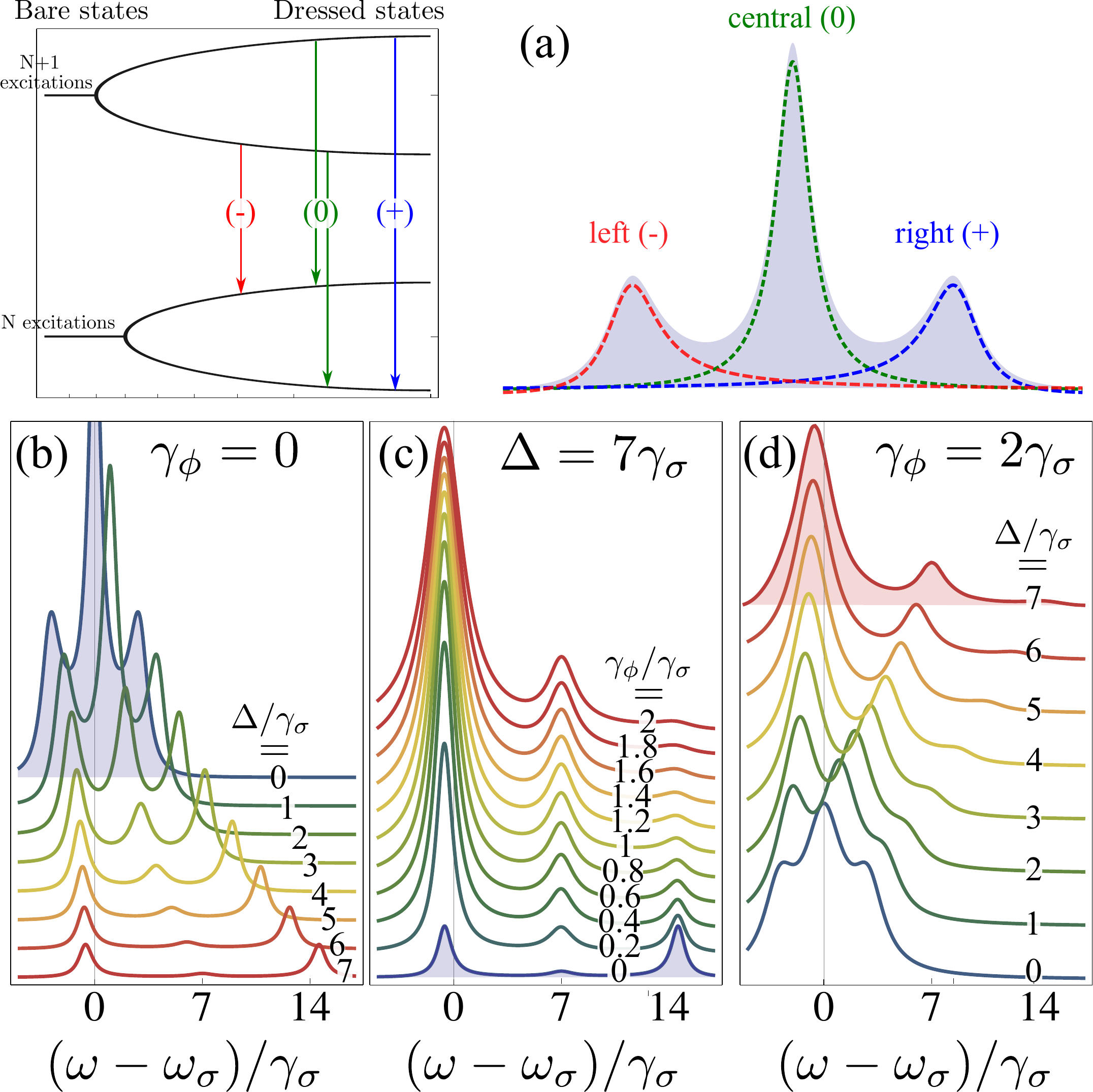}
  \caption{(Color online) (a) Origin of the peaks in the Mollow
    triplet: the three different frequencies ($p=0,\pm$) are found in
    the four possible transitions between two Jaynes-Cummings rungs at
    high intensities. Below, different Mollow triplets when varying
    detuning (b, d) or dephasing (c), with
    $\Omega_\mathrm{L}=1.5\gamma_\sigma$. The symmetry of the triplet
    is broken only under the combined action of detuning and
    dephasing.}
  \label{fig:WedMay4170658CEST2011}
\end{figure} 

The normalized spectra of emission reads in the steady state (that we
set as $t=0$):
\begin{equation} 
\label{eq:MonJul21132824CEST2008}
S_\sigma(\omega)=\frac{1}{\pi n_\sigma}\Re\int_{0}^{\infty}\mean{\ud{\sigma}(0)\sigma(\tau)}e^{i\omega\tau}d\tau\,.
\end{equation} 

Any two-time correlators can always be decomposed as a sum of complex
damped exponentials~\cite{delvalle_book09a}:
\begin{equation}
  \label{eq:SunNov16135056GMT2008}
  \mean{\ud{\sigma}(0)\sigma(\tau)}=n_\sigma \sum_p (L_p +i K_p) e^{-i\omega_p\tau}e^{-\frac{\gamma_p}{2}\tau}\,,
\end{equation}
where all the parameters, weights $L_p$, $K_p$, frequencies $\omega_p$
and effective decay rates $\gamma_p$, are real. They can be obtained
by means of the quantum regression formula (see
Appendix~\ref{ap:qrf}).  Eq.~(\ref{eq:MonJul21132824CEST2008}) leads
to the spectrum,
\begin{equation}
  \label{eq:SunNov16140314GMT2008}
  S_\sigma(\omega)=\frac{1}{\pi}
  \sum_p
  \frac{L_p\frac{\gamma_p}{2}-K_p(\omega-\omega_p)}{\big(\frac{\gamma_p}{2}\big)^2+(\omega-\omega_p)^2}\,.
\end{equation}
In the case of
Eqs.~(\ref{eq:WedFeb24185855GMT2010}--\ref{eq:TueNov30150109CET2010}),
the spectrum has four components, that we label $p=\mathrm{coh}$, $0$,
$+$, $-$. The elastic scattering component ($p=\mathrm{coh}$) is a
delta peak $\delta(\omega)$ at the laser frequency
($\omega_\mathrm{coh}=\gamma_\mathrm{coh}=K_\mathrm{coh}=0$) with
weight:
\begin{equation}
  \label{eq:FriMay6214110CEST2011}
    L_\mathrm{coh}=\frac{|\mean{\ud{\sigma}}|^2}{n_\sigma}=\frac{\gamma_\sigma^2}{8 (\Omega_\mathrm{L}^\mathrm{eff})^2+\gamma_\sigma(\gamma_\sigma+\gamma_\phi)}\,,
\end{equation}
The inelastic scattering part is a triplet with a central peak ($p=0$)
and two sidebands $p=\pm$ that carry the information about the
light-matter interaction. The physical origin of these two peaks is in
the transitions between the dressed states of the Jaynes-Cummings
Hamiltonian at high number of excitations~\cite{cohentannoudji77a}, as
shown in Fig.~\ref{fig:WedMay4170658CEST2011}(a). The two transitions
between different types of dressed states become degenerate and form
the central peak~$(0)$, while transitions between the same type of
dressed states give rise to the side bands. The full expression of the
spectrum out of resonance is too lengthy to be given here.  The
resonant case formula is shorter. It reads:
\begin{subequations}
  \label{eq:SunJun27200837MSD2010}
  \begin{align}
    &L_0+iK_0=\frac{1}{2}\,,\\
    &\omega_0=0\,,\quad\gamma_0=\gamma_\sigma+\gamma_\phi \,,\label{eq:TueMay17163344CEST2011}\\
    &L_{\pm}+iK_\pm=\\
    &\frac{\frac{8\Omega_\mathrm{L}^2}{\gamma_\sigma(\gamma_\sigma+\gamma_\phi)}\big[1 \pm \frac{5\gamma_\sigma-\gamma_\phi}{4 R_\mathrm{L}}\big]-\frac{\gamma_\sigma-\gamma_\phi}{\gamma_\sigma+\gamma_\phi}\big[1\pm  i\frac{\gamma_\sigma-\gamma_\phi}{4R_\mathrm{L}}\big]}{4\big(1+\frac{8 \Omega_\mathrm{L}^2}{\gamma_\sigma(\gamma_\sigma+\gamma_\phi)}\big)}\,,\nonumber\\
    &\omega_\pm=\pm\Re (R_\mathrm{L})\,,\quad\gamma_\pm=\frac{3\gamma_\sigma+\gamma_\phi}{2}\pm 2\Im(R_\mathrm{L})\label{eq:TueMay17163448CEST2011} \,,
  \end{align}
\end{subequations}
where we have defined the (half) \emph{Mollow splitting}:
\begin{equation} 
  \label{eq:FriApr1183145CEST2011}
  R_\mathrm{L}=\sqrt{(2\Omega_\mathrm{L})^2-\big(\frac{\gamma_\sigma-\gamma_\phi}{4}\big)^2}\,.
\end{equation}

Strong coupling, where the character of the dynamics of the
two-time correlator is oscillating rather than damped, is defined by
the appearance of this splitting ($\Re(R_\mathrm{L})\neq 0$), that is:
\begin{equation}
  \label{eq:ThuFeb25182119GMT2010}
  2\Omega_\mathrm{L}>|\gamma_\sigma-\gamma_\phi|/4\,.
\end{equation}
%

We see from Eqs.~(\ref{eq:TueMay17163344CEST2011}) and
(\ref{eq:TueMay17163448CEST2011}) that, beyond the expected broadening
of the lines, dephasing also shifts the two satellite peaks.  In
general, this shift brings the side peaks closer to each other,
inducing the transition into weak coupling when $\gamma_\phi >
\gamma_\sigma+8\Omega_\mathrm{L}$.  However, surprisingly, the maximum
splitting, for a fixed $\gamma_\sigma$ and $\Omega_\mathrm{L}$,
corresponds to a nonzero dephasing, $\gamma_\phi=\gamma_\sigma$. In
fact, the splitting remains different from zero, in the presence of
dephasing, as long as $\gamma_\sigma-8\Omega_\mathrm{L} < \gamma_\phi
< \gamma_\sigma+8\Omega_\mathrm{L}$. If the driving field is too weak
to bring by itself the system to strong coupling
($\Omega_\mathrm{L}<\gamma_\sigma/8$), it can be aided by increasing
dephasing. However, higher dephasing also blurs the spectral
features. For the regimes of excitation where dephasing induces
strong-coupling ($0<\gamma_\sigma-8\Omega_\mathrm{L} < \gamma_\phi\leq
\gamma_\sigma$), the observed lineshape always remains single peaked.

The final expression for the Mollow triplet spectrum at resonance in
presence of dephasing reads:
\begin{multline}
  \label{eq:SunJun27235639MSD2010}
  S_\sigma(\omega)=L_\mathrm{coh}\delta(\omega)+\frac{1}{2\pi}\frac{\frac{\gamma_\sigma+\gamma_\phi}{2}}{\big(\frac{\gamma_\sigma+\gamma_\phi}{2}\big)^2+\omega^2}+\\
  \frac{1}{\pi}\Bigg(\gamma_\sigma  \Omega_\mathrm{L}^2-\frac{\gamma_\sigma -\gamma_\phi }{16}(\gamma_\sigma^2+\omega^2)\Bigg)\Bigg/\\
\Bigg(\frac{\gamma_\sigma^2+\omega^2}{16} \big[(\gamma_\sigma +\gamma_\phi)^2+4 \omega^2\big]\\+\big[\gamma_\sigma  (\gamma_\sigma +\gamma_\phi )-2 \omega^2\big] \Omega_\mathrm{L}^2+4 \Omega_\mathrm{L}^4\Bigg)\,.
\end{multline}
The $\delta$ scattering peak and the central peak in the first line
are neatly set apart from the two side bands in the rest of the
expression.  This decomposition is shown in
Fig.~\ref{fig:WedMay4170658CEST2011}(a).

\begin{figure}[t] 
  \centering 
  \includegraphics[width=.8\linewidth]{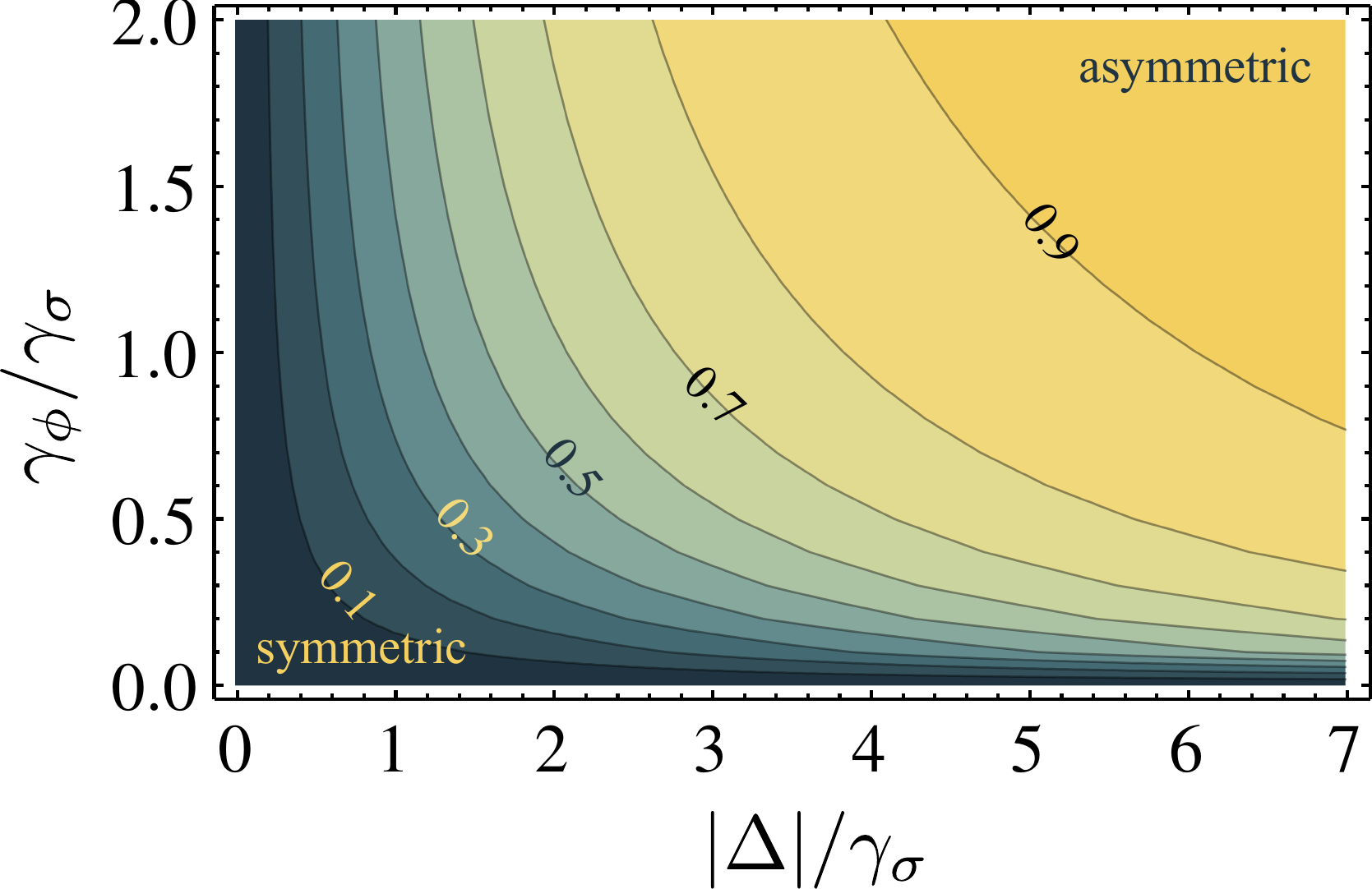}
  \caption{(Color online) Degree of asymmetry~$\mathcal{V}$ of the
    Mollow triplet under coherent excitation, as a function of
    detuning~$\Delta$ and pure dephasing~$\gamma_\phi$ for
    $\Omega_\mathrm{L}=1.5\gamma_\sigma$. It is 0 (dark blue) when the
    side peaks have the same intensity and 1 (bright yellow) when one
    of them disappears completely.}
  \label{fig:SunMay8004044CEST2011}
\end{figure} 

The effect on the Mollow triplet of detuning the laser from the
emitter is shown in Fig.~\ref{fig:WedMay4170658CEST2011}(b). It
spreads the side bands apart, with asymptotes $\omega_\sigma$ and
$\omega_\sigma+2\Delta$, while the central peak, pinned at the driving
laser frequency $\omega_\mathrm{L}$, gets suppressed. The scattering
peak, not shown, ultimately dominates the spectrum over its incoherent
part which fades away.  In any case, the lineshape remains always
symmetric with respect to the laser frequency (central peak), a
characteristic proper to coherent excitation as we shall see later.

Out of resonance, pure dephasing has a strong qualitative effect: it
breaks the above symmetry, bringing the spectrum towards the uncoupled
case with only one Lorentzian peak at $\omega_\sigma$ with FWHM
$\gamma_\sigma +\gamma_\phi$. Even a small dephasing enhances
considerably the emitter peak relatively to the others, as shown in
Fig.~\ref{fig:WedMay4170658CEST2011}(c). This asymmetry becomes larger
the weaker the effective laser drive, that is, for lower
$\Omega_\mathrm{L}$ and larger $\Delta$ and dephasing. One can
quantify the visibility of this asymmetry as the difference between
the intensities of the two side peaks:
\begin{equation}
  \label{eq:TueMay17172235CEST2011}
  \mathcal{V}=\frac{|L_+-L_-|}{|L_+|+|L_-|}\,.
\end{equation}
This is plotted in Fig.~\ref{fig:SunMay8004044CEST2011}, where lighter
(yellow) colors refer to smaller degrees of symmetry (minimum when only
one peak of the two side bands survives). As a practical application,
one can measure the magnitude of pure dephasing as a function of
detuning from the degree of asymmetry.

\section{One-atom laser}
\label{sec:2}

In the cavity QED version of this physics, the system is described by
the Jaynes-Cummings Hamiltonian~\cite{shore93a}:
\begin{equation}
  \label{eq:FriOct2103450BST2009}
  H=\omega_a\ud{a}a+\omega_\sigma\ud{\sigma}\sigma+g(\ud{a}\sigma+a\ud{\sigma})\,,
\end{equation}
where also the light field is quantized, through the annihilation
operator~$a$. The detuning is now $\Delta=\omega_a-\omega_\sigma$ and
we consider $\omega_a=0$ as the reference energy. The Liouvillian,
$\partial_t\rho=\mathcal{L}\rho$, to describe this system in a
dissipative context with decay ($\gamma_c$), incoherent pumping
($P_c$) and pure dephasing ($\gamma_\phi$) has the
form~\cite{delvalle09a}:
\begin{subequations}
  \label{eq:ThuOct18162449UTC2007}
  \begin{align}
    \mathcal{L} O=i[O,H]&+\sum_{c=a,\sigma}\frac{\gamma_c}2(2cO\ud{c}-\ud{c}cO-O\ud{c}c)\\
    &+\sum_{c=a,\sigma}\frac{P_c}2(2\ud{c}Oc-c\ud{c}O-Oc\ud{c})\\
    &+\frac{\gamma_\phi}2(2\ud{\sigma}\sigma O\ud{\sigma}\sigma-\ud{\sigma}\sigma O-O\ud{\sigma}\sigma)\,,
  \end{align}
\end{subequations}
where~$\rho$ is the density matrix for the combined emitter/cavity
system.  The effective broadenings of the uncoupled modes are defined
by $\Gamma_a=\gamma_a-P_a$ and
$\Gamma_\sigma=\gamma_\sigma+P_\sigma$.

\subsection{One-time correlators: populations and statistics}
\label{sec:21}

\begin{figure*}[t] 
  \centering 
  \includegraphics[width=\linewidth]{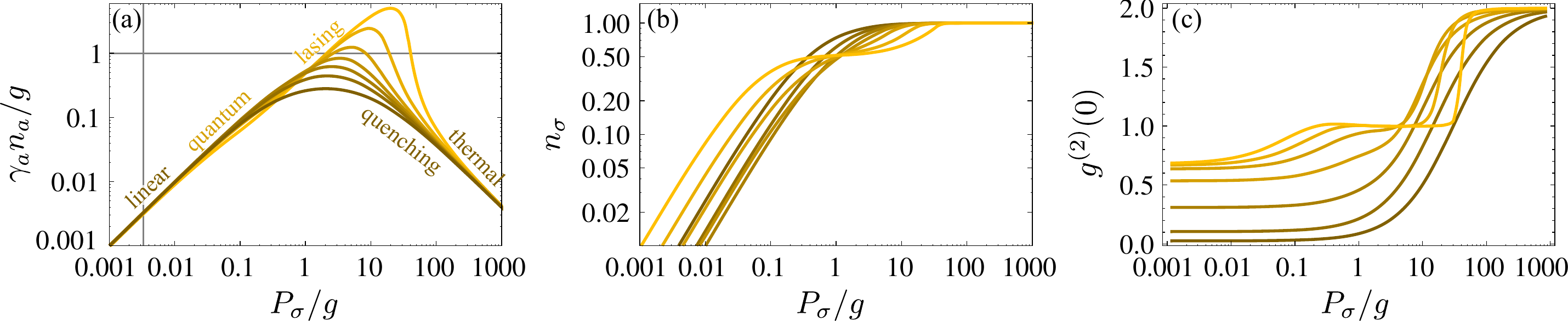}  
  \caption{(Color online) Exact results computed numerically for (a)
    $\gamma_a n_a/g$, (b) $n_\sigma$ and (c) $g^{(2)}$, as a function
    of pumping for various systems, with $\gamma_a/g\in\{0.1, 0.22,
    0.46, 1, 2.15, 4.64, 10\}$ from lighter to darker shades (top to
    bottom in (a)).  Other parameters are $\gamma_\sigma=0.00334g$,
    from Ref.~\cite{laucht09b}, and $P_a=\gamma_\phi=\Delta=0$. The
    different regimes of operation are designated in (a).}
  \label{fig:MoMar22195241WET2010}
\end{figure*} 

The light field that was previously a classical laser field was fully
characterised by its intensity ($|\Omega_\mathrm{L}|^2$) and its
frequency ($\omega_\mathrm{L}$). In the fully quantized description,
correlations between the fields should be taken into account, namely,
in the steady state:
\begin{subequations}
  \label{eq:MoMar22172729WET2010}
  \begin{align}
    & N_a[n]=\mean{\ud{a}^na^n}\,,\quad
    N_\sigma[n]=\mean{\ud{a}^{n-1}a^{n-1}\ud{\sigma}\sigma}\,,\\
    &\tilde N_{a\sigma}[n]=\mean{\ud{a}^na^{n-1}\sigma}=N^r_{a\sigma}[n]+iN^i_{a\sigma}[n]\,,
\end{align}
\end{subequations}
with $N_a[n]$ and $N_\sigma[n]$ real and $i\tilde N_{a\sigma}[n]$
complex in general but real at resonance, all others being zero. The
main observables that characterize the system are:
\begin{equation}
  \label{eq:SunJul20214055CEST2008}
  n_a=N_a[1]\,,\quad n_\sigma=N_\sigma[1]\quad\text{and}\quad g^{(2)}=N_a[2]/n_a^2\,.
\end{equation}
In the following, we provide exact implicit expressions for the
correlators (\ref{eq:MoMar22172729WET2010}), that allow an efficient
numerical solution, and derive approximate analytical expressions for
different regimes of excitation.

In the case without any direct cavity pumping, $P_a=0$, the field
correlators admit a simple expression in terms of $N_a[n]$ (the
general equations are given in
Appendix~\ref{sec:ThuJul1223020MSD2010}):
\begin{subequations}
  \label{eq:MoMar22180002WET2010}
  \begin{align}
    N_\sigma[n]=&\frac{P_\sigma N_{a}[n-1]-\gamma_a
      N_{a}[n]}{\Gamma_\sigma+\gamma_a(n-1)}    \,,\\
    N^i_{a\sigma}[n]=&\frac{\gamma_a}{2g}N_{a}[n]\,,\\
    N^r_{a\sigma}[n]=&\frac{-\Delta \gamma_a
      N_{a}[n]/g}{\Gamma_\sigma+\gamma_\phi+\gamma_a(2n-1)}\,.
  \end{align}
\end{subequations}
This allows to obtain a single equation for $N_a[n]$:
\begin{multline}
  \label{eq:MoMar22173241WET2010}
  0=-\Big[\frac{1}{C_\mathrm{eff}[n]}+\frac{n\gamma_a}{\Gamma_\sigma+(n-1)\gamma_a}-\frac{2P_\sigma}{\Gamma_\sigma+n\gamma_a}+1\Big]N_{a}[n]\\
  +\frac{nP_\sigma}{\Gamma_\sigma+(n-1)\gamma_a}N_a[n-1]-\frac{2\gamma_a}{\Gamma_\sigma+n\gamma_a}N_a[n+1]\,.
\end{multline}
where we have introduced, respectively, the effective cooperativity,
effective coupling and total decoherence (in the presence of detuning
and pure dephasing):
\begin{subequations}
  \label{eq:MonMar28154711CEST2011}
  \begin{align}
    C_\mathrm{eff}[n]=&\frac{4 (g_\mathrm{eff}[n])^2}{\gamma_a\Gamma_\mathrm{T}[n]} \,,\\
    g_\mathrm{eff}[n]=&\frac{g}{\sqrt{1+(2\Delta/\Gamma_\mathrm{T}[n])^2}} \,,\\
    \Gamma_\mathrm{T}[n]=&\Gamma_\sigma+\gamma_\phi+(2n-1)\gamma_a\,.
  \end{align}
\end{subequations}
Let us note that $g_\mathrm{eff}[n]=g$ for all~$n$ at resonance or
when decoherence is large as compared to the detuning. As in the case
of laser excitation, the effect of detuning is to effectively diminish
the coherent coupling, which magnitude is linked to the spectral
overlap between modes (represented by $\Gamma_\mathrm{T}$). Here too,
the decoupling caused by detuning can be compensated by increasing
decoherence (decay or pure dephasing) since in this case the spectral
overlap between the cavity and the emitter increases, bringing them
effectively back to resonance.  Detuning and pure dephasing only
appear in the cooperativity parameter $C_\mathrm{eff}[n]$, as noted
also by Auffèves \emph{et al.}~\cite{auffeves10a}. This situation is
similar to the case of two coupled harmonic modes~\cite{laussy09a}.

We reduced the whole steady-state problem of the Jaynes--Cummings with
emitter pumping and decay to a single equation,
Eq.~(\ref{eq:MoMar22173241WET2010}), which is however a nonlinear
recurrence equation with non-constant coefficients, for which there is
no general method leading to an exact solution. The problem put in
this form is nevertheless quite tractable numerically and we shall in
the following present various limiting cases which will spell out the
physics of this problem.

Since $N_a[0]=1$ by definition, Eq.~(\ref{eq:MoMar22173241WET2010})
can be easily iterated numerically to provide $N_a[n]$ for all $n$ as
a function only of the mean number of photons in the cavity,
$n_a$. The small $n$ equations are the most important since they
capture the dominant few-photons correlations. The exact expressions
for $n_\sigma$ and $g^{2}$ in terms of $n_a$ are, in any case, simple
enough:
\begin{subequations}
  \label{eq:TueMar29094201CEST2011}
  \begin{align}
    n_\sigma=&\frac{P_\sigma -\gamma_a n_a}{\Gamma_\sigma}  \,,\\
    g^{(2)}=&\frac{\Gamma_\sigma+\gamma_a}{2\gamma_an_a}\label{eq:ThuMay12200410CEST2011}\\
    &\times\Big(\frac{P_\sigma}{n_a\Gamma_\sigma}+\frac{2P_\sigma}{\Gamma_\sigma+\gamma_a}
    -\frac{\Gamma_\sigma+\gamma_\phi+\gamma_a}{\kappa_\sigma}-\frac{\gamma_a+\Gamma_\sigma}{\Gamma_\sigma}\Big)\,.\nonumber
  \end{align}
\end{subequations}
They are given in terms of a key parameter of the system, the Purcell
rate of transfer of population from emitter to the cavity mode:
\begin{equation}
  \label{eq:MonMar28155703CEST2011}
  \kappa_\sigma=\frac{4(g_\mathrm{eff}[1])^2}{\gamma_a}\,.
\end{equation}
This parameter is large for good cavities, when cavity QED is realized
at its fullest: $\kappa_\sigma \gg g$.

One can obtain $n_a$ self-consistently by truncating $N_a[n]$ at a
sufficiently high number of photons, $n_\mathrm{max}$, and solving
numerically the resulting finite set of equations. This gives the
results plotted in Fig.~\ref{fig:MoMar22195241WET2010}, for (a)
$\gamma_a n_a/g$, (b) $n_\sigma$ and (c) $g^{(2)}$. In the best
systems ($\gamma_a,\gamma_\sigma\ll g$), we distinguish five regions
in these plots, which we shall investigate in more details in the
remaining of this text:
\begin{enumerate}
\item \emph{Linear quantum regime} (or simply ``\emph{linear}''),
  where $P_\sigma\ll \gamma_\sigma$, keeping the emitter essentially
  in its ground state, very rarely excited.
\item \emph{Nonlinear quantum regime} (or simply ``\emph{quantum}''),
  where $P_\sigma\sim \gamma_\sigma$ is enough to probe higher ($n>1$)
  rungs of the Jaynes-Cummings ladder, without climbing it too highly
  so that few photons effects remain the dominant ones.
\item \emph{Lasing} or \emph{nonlinear classical regime}, when
  $P_\sigma\gg \gamma_\sigma$ and the emitter population $\approx
  0.5$, the cavity can accumulate a great number of photons and the
  field becomes Poissonian.
\item \emph{Self-quenching regime}, when $P_\sigma\gtrapprox
  \kappa_\sigma/2$ starts to drive the emitter to saturation,
  $n_\sigma>1/2$, reducing the number of photons.
\item \emph{Thermal regime} or \emph{linear classical regime}, when
  $P_\sigma>\kappa_\sigma$, the emitter is always in its excited
  state, $n_\sigma\rightarrow1$, and the dephasing induced by the pump
  disrupts the coherent coupling, so that the number of photons is
  very low again and the field becomes thermal.
\end{enumerate}

Similar classifications have been proposed, for instance by Poddubny
\emph{et al.}~\cite{poddubny10a}. In the following, we address several
types of approximations, that perform to varying degrees of accuracy
depending on the level of complexity involved and the regime under
consideration. Some of our approximations recover known
results~\cite{mu92a,benson99a,poddubny10a,delvalle10d,lsc_gartner11a},
however, this comparative analysis will give us, beyond good
approximated formulas, valuable insights into the underlying
physics. It will also allow us to determine the pumping ranges that
determine each regime. As the expression for $n_\sigma$ and~$g^{(2)}$
follow straightforwardly from that of $n_a$, we will not provide their
explicit form in most of the cases analysed below.

\begin{figure}[hbpt] 
  \centering 
  \includegraphics[width=.95\linewidth]{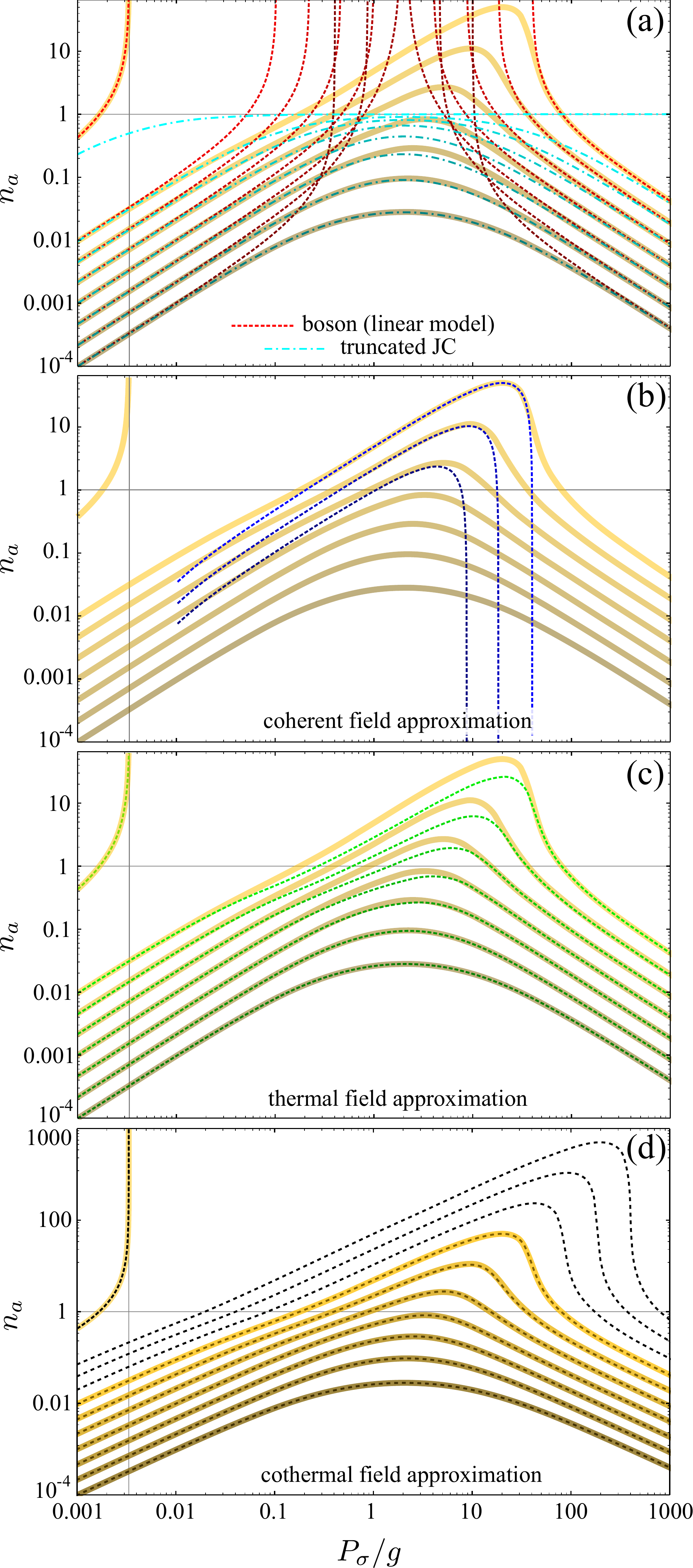}  
  \caption{(Color online) Comparison between the numerical results for
    $n_a$ (from Fig.~\ref{fig:MoMar22195241WET2010}(a)), in thick
    yellowish lines, with different approximated solutions. In each
    panel, $\gamma_a$ decreases exponentially from bottom to top
    curves, ranking from $10g$ (darkest curve) to $0.1g$ (lightest
    curve). The limiting case $\gamma_a=0$ is also shown on the left
    upper corner, where the divergence is a feature of the
    Jaynes--Cummings model. In (a), we superimpose the linear model
    (dotted red) and Jaynes-Cummings truncated at one excitation
    (dashed blue), given by Eq.~(\ref{eq:ThuMay12181806CEST2011})(a).
    In (b), the approximated semiclassical solution,
    Eq.~(\ref{eq:SunJun27143144MSD2010}), which provides an accurate
    description in the lasing regime given by
    Eq.~(\ref{eq:TueJul6130009CEST2010}).  In (c), the thermal
    approximation, given by Eq.~(\ref{eq:SatMay14022519CEST2011}) that
    converges to the models in (a) at low and high pumps.  In (d), the
    cothermal approximation, given by the numerical solution of
    Eqs.~(\ref{eq:MonApr11102948CEST2011}). In this case, we extend
    solutions to values of $\gamma_a/g$ till $0.01g$, out of reach
    numerically. A vertical guideline marks the value of
    $\gamma_\sigma=0.00334g$. Other parameters are
    $P_a=\Delta=\gamma_\phi=0$.}
  \label{fig:ThuMay12185542CEST2011}
\end{figure} 

\subsubsection{Linear model approximations}
\label{sec:210}

In the linear regime, where the emitter is excited with very low
probability, it can be well approximated by another harmonic
oscillator~\cite{laussy09a}. This allows to find a closed-form
analytical solution:
\begin{subequations}
  \label{eq:ThuMay12181806CEST2011}
  \begin{align}
    &n_a\approx\frac{\kappa_\sigma}{\kappa_\sigma(\Gamma_\sigma+\gamma_a)+\Gamma_\sigma(\Gamma_\sigma+\gamma_a+\gamma_\phi)}P_\sigma\,,\\
    &n_\sigma\approx\frac{\kappa_\sigma+\gamma_a+\gamma_\sigma+\gamma_\phi}{\kappa_\sigma(\Gamma_\sigma+\gamma_a)+\Gamma_\sigma(\Gamma_\sigma+\gamma_a+\gamma_\phi)}P_\sigma\,.
  \end{align}
\end{subequations}

Equivalent expressions for~$n_a$ and~$n_\sigma$ are obtained in the
first order truncation of a continuous fraction expansion of these
quantities, as recently shown by Gartner~\cite{lsc_gartner11a}. They
are also formally identical to those obtained by truncating the
Jaynes-Cummings model at the first rung of
excitation~\cite{laucht09b,auffeves09a,auffeves10a}\footnote{The
  truncation cannot be done at the level of
  Eq.~(\ref{eq:MoMar22173241WET2010}), since this one related to the
  number of photons $n$, but rather with
  Eqs.~(\ref{eq:MoMar22174241WET2010}). At high numbers, truncation in
  manifolds of excitations or of photons gives the same result but not
  at the single particle limit.}.  The only difference is in the
effective broadening $\Gamma_\sigma$, that appears with a $-$ sign in
the case of coupled bosons, $\Gamma_\sigma\rightarrow
\gamma_\sigma-P_\sigma$~\cite{laussy09a}, and a $+$ sign in the
truncated Jaynes--Cummings model,
$\Gamma_\sigma\rightarrow\gamma_\sigma+P_\sigma$. At $P_\sigma \ll
\gamma_\sigma$, the sign becomes irrelevant with $\Gamma_\sigma\approx
\gamma_\sigma$ and the population grows linearly with pumping,
$n_a\approx C_1P_\sigma$, with the slope:
\begin{equation}
  \label{eq:TueMay17234800CEST2011}
  C_1=\frac{\kappa_\sigma}{\kappa_\sigma(\gamma_\sigma+\gamma_a)+\gamma_\sigma(\gamma_\sigma+\gamma_a+\gamma_\phi)}\,.
\end{equation}
This agrees with the numerical results for
$P_\sigma\lessapprox\gamma_\sigma$ as shown in
Fig.~\ref{fig:ThuMay12185542CEST2011}(a). In the case where
$\gamma_\sigma=0$, we simply have $n_a\approx
P_\sigma/\gamma_a$. Interestingly, the two models also provide the
right formula in the high pumping regime where the number of photons
is low again,
\begin{equation}
  \label{eq:SatMay14025421CEST2011}
  \lim_{P_\sigma \rightarrow \infty}n_a= \frac{\kappa_\sigma}{P_\sigma}\,,
\end{equation}
but the emitter is completely saturated.

The bosonic populations diverge at two values of pumping (where the
denominators vanish):
\begin{equation}
  \label{eq:SunMay15171526CEST2011}
  P_\pm= \gamma_\sigma+ \frac{\kappa_\sigma+\gamma_a+\gamma_\phi}{2}\left( 1\pm \sqrt{1-\frac{4\kappa_\sigma \gamma_a}{\kappa_\sigma+\gamma_a+\gamma_\phi}} \right)\,.
\end{equation}
For good systems with small cavity decay rates, we have $P_-\approx
\gamma_\sigma$ and $P_+\approx \kappa_\sigma+\gamma_\sigma
+\gamma_\phi$. At $P_-$, the two populations diverge but remain
positive, a manifestation in this model that the system enters the
lasing oscillations where both populations are ``inverted''. For
intermediate pumpings, where the system is in the lasing regime
$P_-<P_\sigma<P_+$, both populations are negative, meaning that the
physics in this regime is out of reach of this model. For large enough
pumping, $P_\sigma>P_+$, the system exits the lasing regime as $n_a$
in the bosonic model converges again to the exact numerical result
following Eq.~(\ref{eq:SatMay14025421CEST2011}). However, $n_\sigma$
remains negative for all $P_\sigma>P_-$, meaning that its population
remains thereafter inverted.  The bosonic model thus provides an
accurate description of the transition \emph{in} and \emph{out} of
lasing though the appearance of divergences and negative populations,
as we will confirm later when linking it to the full Jaynes--Cummings
system. The truncated Jaynes-Cummings formulas remain always positive
and instead of a divergence, the two-level system becomes inverted,
with $n_\sigma\rightarrow 1$ from below. Given that the number of
photons is always below one, this model is not suited to describe the
transitions in and out the lasing regime.

The bosonic and fermionic models provide opposite statistics for the
cavity field. Two coupled harmonic oscillators under incoherent
excitation are always thermal with a photon distribution
$T(n)=n_a^n/(n_a+1)^{n+1}$ and $g^{(2)}=2$. On the other hand,
truncating the Jaynes-Cummings model at one excitation means that one
excludes the possibility to have two photons at a time in the cavity,
which results in perfect photon antibunching, $g^{(2)}=0$.  Both
models provide the exact $n_a$ solution of the Jaynes--Cummings model
in two and opposite limiting cases, namely, $\gamma_a\rightarrow 0$
for the bosonic model, where
\begin{equation}
  \label{eq:SunMay15163638CEST2011}
  n_a=\frac{P_\sigma}{\gamma_\sigma-P_\sigma}\,,
\end{equation}
with $g^{(2)}=2$, and  $\gamma_a\rightarrow \infty$ for the
truncated Jaynes--Cummings model, where
\begin{equation}
  \label{eq:SunMay15234113CEST2011}
  n_a=\frac{P_\sigma}{\gamma_\sigma+P_\sigma}\frac{\kappa_\sigma}{\gamma_a}\,,
\end{equation}
with $g^{(2)}=0$. This is seen in
Fig.~\ref{fig:ThuMay12185542CEST2011}(a), where $\gamma_a/g=10$
(lowest curve) is already a good approximation for~$\infty$, while
$\gamma_a/g=0$ (uppermost curve) is exact. In the first limiting case,
recovered by the bosonic model, the system accumulates photons in the
most effective way possible, which leads to a divergence in the number
of photons, also in the exact Jaynes-Cummings model. The Rabi delivery
of photons is so efficient in the one-atom laser that, unless there is
a leakage of photons from another channel, the accumulation of photons
is unbounded. It is unlimited by the strong-coupling feed.

In the opposite limit of weak and inefficient coupling, the emitter
undergoes population inversion unaffected by the cavity while the
cavity gets an effective pumping of photons from $n_\sigma$ through a
very weak Purcell rate
$\kappa_\sigma$~\cite{auffeves10a,lsc_gartner11a}.  For $\gamma_a>4g$
the truncated Jaynes-Cummings model provides a quantitatively good
agreement for the entire range of pumping, as shown in
Fig.~\ref{fig:ThuMay12185542CEST2011}(a), if one excludes the
behaviour of $g^{(2)}$. This limit is also accounted for exactly by a
series expansion in $P_\sigma$, given in
Appendix~\ref{sec:ThuMay26100821CEST2011}, since the coupling is
perturbative. In this case, the system goes directly from the quantum
linear to the classical linear regime.

One can obtain an exact expression for $g^{(2)}$ in the linear regime
solving Eqs.~(\ref{eq:MoMar22173241WET2010}) truncated, not at one,
but at two photons, that is, for $n=1,2$ (assuming $N_a[3]= 0$). Note
that at $P_\sigma=0$, $N_a[n]=0$ for all $n$, but we are interested in
the limit:
\begin{equation}
  \label{eq:FriMay13022715CEST2011}
  g_{P_\sigma\rightarrow0}^{(2)}=\lim_{P_\sigma \rightarrow 0}\frac{N_a[2]}{n_a^2}\,,
\end{equation}
that remains different from zero, since $N_a[n]\propto
P_\sigma^n$. This method leads, to first order in $P_\sigma$, to the
populations of Eq.~(\ref{eq:ThuMay12181806CEST2011}) and to:
\begin{equation}
  \label{eq:FriMay13005728CEST2011}
  g_{P_\sigma\rightarrow0}^{(2)}=2\frac{\kappa_\sigma(\gamma_a+\gamma_\sigma)+\gamma_\sigma(\gamma_a+\gamma_\sigma+\gamma_\phi)}{\kappa_\sigma(3\gamma_a+\gamma_\sigma)+(\gamma_a+\gamma_\sigma)(3\gamma_a+\gamma_\sigma+\gamma_\phi)}\,.
\end{equation}
This result is exact and valid for any set of parameters, as one can
check by simply solving the equations to the next order of truncation
($n=1,2,3$ and $N_a[4]=0$). In general, $g^{(2)} \in [0,2]$ in the
linear regime of the Jaynes--Cummings dynamics, although in
Fig.~\ref{fig:MoMar22195241WET2010}(c) we only show $0\le
g^{(2)}\le\frac{2}{3}$ since we have chosen $\gamma_\sigma\approx0$.

\subsubsection{Semiclassical approximation}
\label{sec:211}

Given that, as shown in Fig.~\ref{fig:MoMar22195241WET2010}(c), the
cavity field becomes Poissonian in the lasing regime, we can find good
approximated solutions for the steady state under the assumption
$\mathrm{T}[n]=e^{-n_a}n_a^n/n!$, which leads to $N_a[n]=n_a^n$. To
establish the lasing in strong-coupling regime, one also needs a good
cavity, so we shall assume $\gamma_a\ll g$, and high enough pumping,
$P_\sigma\gg \gamma_\sigma,\gamma_a$. Plugging $N_a[n]=n_a^n$ into
Eq.~(\ref{eq:MoMar22173241WET2010}) and solving the resulting equation
for $n=1$, or equivalently, imposing $g^{(2)}=1$ in
Eq.~(\ref{eq:ThuMay12200410CEST2011}), gives
\begin{equation}
  \label{eq:SunJun27143144MSD2010}
  n_{a}\approx\frac{\Gamma_\sigma}{2\gamma_a}\Big(1-\frac{2\gamma_\sigma}{\Gamma_\sigma} -\frac{\Gamma_\sigma+\gamma_\phi}{\kappa_\sigma}\Big)\,.
\end{equation}
This is a very good approximation for the region where the cavity
field behaves classically~\cite{mu92a,karlovich01a}, equivalent to
solving the $n=0$ equation in
Eqs.~(\ref{eq:MoMar22173241WET2010})~\cite{delvalle10d,lsc_gartner11a}. The
probability of finding the emitter in its excited state reads:
\begin{equation}
  \label{eq:SunJun27143720MSD2010}
  n_\sigma\approx\frac{1}{2}\Big(1+\frac{\Gamma_\sigma+\gamma_\phi}{\kappa_\sigma}\Big)\,.
\end{equation}
The approximated $n_a$ is plotted with blue dashed lines in
Fig.~\ref{fig:ThuMay12185542CEST2011}(b) for comparison with the
numerical results and exhibits a remarkable agreement for a large
pumping range of practical interest (the axis is in log
scale). Similar agreement is found for $n_\sigma$~\cite{delvalle10d}.

The two expressions for the populations have a straightforward
interpretation. Two parameters determine the populations (neglecting
for the sake of simplicity the small correction brought by the emitter
decay, $2\gamma_\sigma/\Gamma_\sigma\ll 1$): the ``\emph{cavity
  feeding}'' and the ``\emph{emitter feeding}'' efficiencies, defined
as $F_a=\Gamma_\sigma/(2\gamma_a)$ and
$F_\sigma=(\Gamma_\sigma+\gamma_\phi)/\kappa_\sigma$,
respectively. They follow as:
\begin{equation}
  \label{eq:SunJun27153112MSD2010}
  n_a\approx F_a (1-F_\sigma)\quad  \mathrm{and} \quad  n_\sigma\approx(1+F_\sigma)/2\,.
\end{equation}
%

The cavity population increases linearly with pumping ($n_a\approx
F_a$) while the emitter is half occupied ($n_\sigma\approx 1/2$). This
is the range of pumping with the most effective accumulation of
photons in the cavity, as the incoherent processes are small enough
not to disrupt the coherent coupling dynamics. All the excitations
injected into the emitter are transferred into the cavity.  We already
have seen that there is a linear relationship between $n_a$ and
$P_\sigma$ in the---aptly denominated---linear regime, as given by
Eq.~(\ref{eq:TueMay17234800CEST2011}). A similar linear relationship
$n_a\approx C_2P_\sigma$ also holds in the lasing regime, when
$P_\sigma\ll \kappa_\sigma$ but beyond the quantum regime, $P_\sigma >
\gamma_a,\gamma_\sigma$, this time with a slope $C_2$ defined as:
\begin{equation}
  \label{eq:TueMay17235117CEST2011}
  C_2=\frac{1}{2\gamma_a}.
\end{equation}
The transition between the two types of linear behaviours $n_a\approx
C_iP_\sigma$, $i=1 \rightarrow 2$, and the question of the threshold
in this process is an interesting subject~\cite{clemens04a} that would
bring us too far astray and that we postpone to another
work~\cite{elena_delvalle11a}. We will only comment in the present
text that this intermediate region is the less liable to the types of
approximation that we derive here, since it lies at the frontier
between the very few and the very large number of excitations, and no
good approximation can reproduce it to a high degree of accuracy other
than by keeping track of all correlations between the particles,
which, being a $N$ body type of problem, implies a numerical
procedure. We call this intermediate region the ``quantum nonlinear''
or simply the ``quantum'' regime.

When the pumping is sizable as compared to $\kappa_\sigma$, the
emitter occupation starts to show signs of saturation, increasing
linearly with pumping, and quenching the linear increase of the cavity
population. $F_\sigma$ represents therefore the degree to which the
pumping succeeds in populating the emitter itself, against the
coherent exchange of population that feeds the cavity with efficiency
$F_a$. The maximum population of the cavity
\begin{equation}
  \label{eq:SatMay14012622CEST2011}
  \max(n_a)\approx \frac{\kappa_\sigma}{8\gamma_a}\big(1-\frac{4\gamma_\sigma+2\gamma_\phi}{\kappa_\sigma}\big)\,,
\end{equation}
is reached at the intermediate rate 
\begin{equation}
  \label{eq:SatMay14012718CEST2011}
  P_\sigma |_{\max(n_a)}\approx \frac{\kappa_\sigma}{2}\big(1- \frac{2\gamma_\sigma+\gamma_\phi}{\kappa_\sigma}\big)\,.
\end{equation}
The present approximated expressions are valid until $n_\sigma$
approaches 1, then, the self-quenching dominates the dynamics and
$n_a\rightarrow 0$, at around:
\begin{equation}
  \label{eq:SatMay14013147CEST2011}
  P_\mathrm{max}\approx \kappa_\sigma-3\gamma_\sigma-2\gamma_\phi\,,  
\end{equation}
that is, when the pump reaches the effective transfer rate of
excitation towards the cavity mode. At this point the statistics
changes to thermal due to pump-induced decoherence. Note that
$P_\mathrm{max}\approx P_+$, found in the previous subsection when
analysing the bosonic model.


\subsubsection{Thermal approximation}
\label{sec:213}

Assuming a thermal state for the cavity field, with statistics
$T(n)=n_a^n/(n_a+1)^{n+1}$ and thus with $N_a[n]=n!n_a^n$, satisfies
in good approximation Eqs.~(\ref{eq:MoMar22173241WET2010}) when the
system undergoes self-quenching and gets driven into the thermal
regime ($P_\sigma>\kappa_\sigma$). The $n_a$ obtained from the
equation ($n=1$), or equivalently, imposing $g^{(2)}=2$ in
Eq.~(\ref{eq:ThuMay12200410CEST2011}), reads:
\begin{multline}
  \label{eq:SatMay14022519CEST2011}
  n_a\approx \frac{1}{8\gamma_a}\Big\{(\Gamma_\sigma+\gamma_a)\times \\
  \sqrt{\frac{16 P_\sigma \gamma_a/\Gamma_\sigma}{\Gamma_\sigma+\gamma_a}+\big[1+ \frac{\Gamma_\sigma+\gamma_a+\gamma_\phi}{\kappa_\sigma}-\frac{2P_\sigma}{\Gamma_\sigma+\gamma_a}+\frac{\gamma_a}{\Gamma_\sigma}\big]^2}\\
  -\Gamma_\sigma\big(\frac{\Gamma_\sigma+\gamma_\phi}{\kappa_\sigma}+\frac{2\gamma_\sigma}{\Gamma_\sigma}-1\big)
  -\gamma_a\big(\frac{2\Gamma_\sigma+\gamma_\phi}{\kappa_\sigma}+2\big)\\
  -\frac{\gamma_a^2}{\Gamma_\sigma}\big(\frac{\Gamma_\sigma}{\kappa_\sigma}+1\big)\Big\}\,.
\end{multline}
%
%
This solution converges to the linear result of
Eq.~(\ref{eq:ThuMay12181806CEST2011}) to first order in $P_\sigma$
when $P_\sigma<\gamma_\sigma$. The opposite limit, well into the
thermal region, is that already found with
Eq.~(\ref{eq:SatMay14025421CEST2011}) to first order in $1/P_\sigma$,
which provides the decreasing tail at large pumpings, as shown in
Fig.~\ref{fig:ThuMay12185542CEST2011}. The lines converge to the same
universal curve when plotting $\gamma_an_a/g$ in
Fig.~\ref{fig:MoMar22195241WET2010}(a). In the intermediate region, it
provides a good approximation when the system is not good enough to
lase, at large dissipation rates, similarly to the truncated
Jaynes-Cummings formula. Of course, having $g^{(2)}=2$ throughout,
does not, in general, represent well the statistics.

Fig.~\ref{fig:ThuMay12185542CEST2011}(c) may seem to shown that
Eq.~(\ref{eq:SatMay14022519CEST2011}) captures the qualitative
behaviour of $n_a$ even in the lasing regime. It does not, however,
provide the change into the linear lasing slope from $n_a \sim
C_1P_\sigma$ to $C_2P_\sigma$, discussed in the previous subsection,
which is a serious conceptual shortcoming.

In two particular cases, the thermal solution becomes exact for
Eq.~(\ref{eq:MoMar22173241WET2010}), namely, $\gamma_c=P_c=0$ with
$c=a$ (previously discussed) on the one hand and its counterpart
$c=\sigma$ on the other hand. In both cases $N_{a\sigma}=0$ and
$N_\sigma[n]=\bar n_\sigma N_a[n-1]$ with $N_a[n]=n!\bar n_a^n$,
where:
\begin{equation}
  \label{eq:FriJul2000551MSD2010}
   \bar n_a=\frac{P_c}{\gamma_c-P_c}\quad\text{and}\quad\bar n_\sigma=\frac{P_c}{\gamma_c+P_c}\,.
\end{equation}
The steady state in this case is two uncorrelated thermal fields at
the same temperature. This is independent of the coupling
strength~$g$, which only determines the speed at which this steady
state is achieved. Although the discontinuity at $g=0$ might seem
counterintuitive, it is physically clear that the thermal equilibrium
does not otherwise depend on details of the microscopic couplings. The
case $P_\sigma=\gamma_\sigma=0$, that corresponds to thermal
excitation of the cavity mode only, presents interesting aspects which
we also postpone to a future work.

The case $P_a=0=\gamma_a=0$, appearing in
Fig.~\ref{fig:ThuMay12185542CEST2011}(c) as the uppermost exact curve,
diverges, as we noted previously, at $P_\sigma\geq \gamma_\sigma$,
meaning that the system does not have a steady state but rather its
intensity grows without bounds. This is to be interpreted as the
instability accompanying the transition into
lasing~\cite{karlovich01a}.

\subsubsection{Cothermal approximation}
\label{sec:212}

\begin{figure}[t]
  \centering 
  \includegraphics[width=\linewidth]{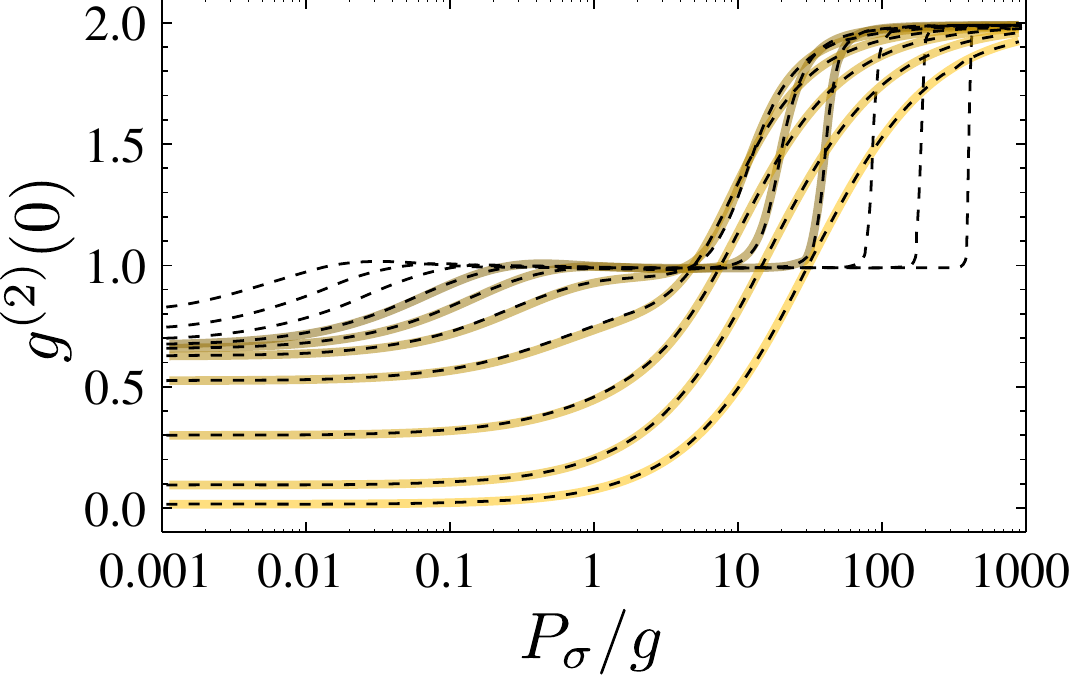}
  \caption{(Color online) Comparison between the numerical $g^{(2)}$
    (in thick lines) and the cothermal approximation (dashed lines),
    from Eq.~(\ref{eq:SatMay14032426CEST2011}) with $n_a$ of
    Fig.~\ref{fig:ThuMay12185542CEST2011}. It interpolates neatly
    between the lasing and the thermal regions ($1\rightarrow 2$) but
    also provides a very good agreement in the linear regime.}
  \label{fig:MonMay16183014CEST2011}
\end{figure} 

\begin{figure*}[t] 
  \centering
  \includegraphics[width=1\linewidth]{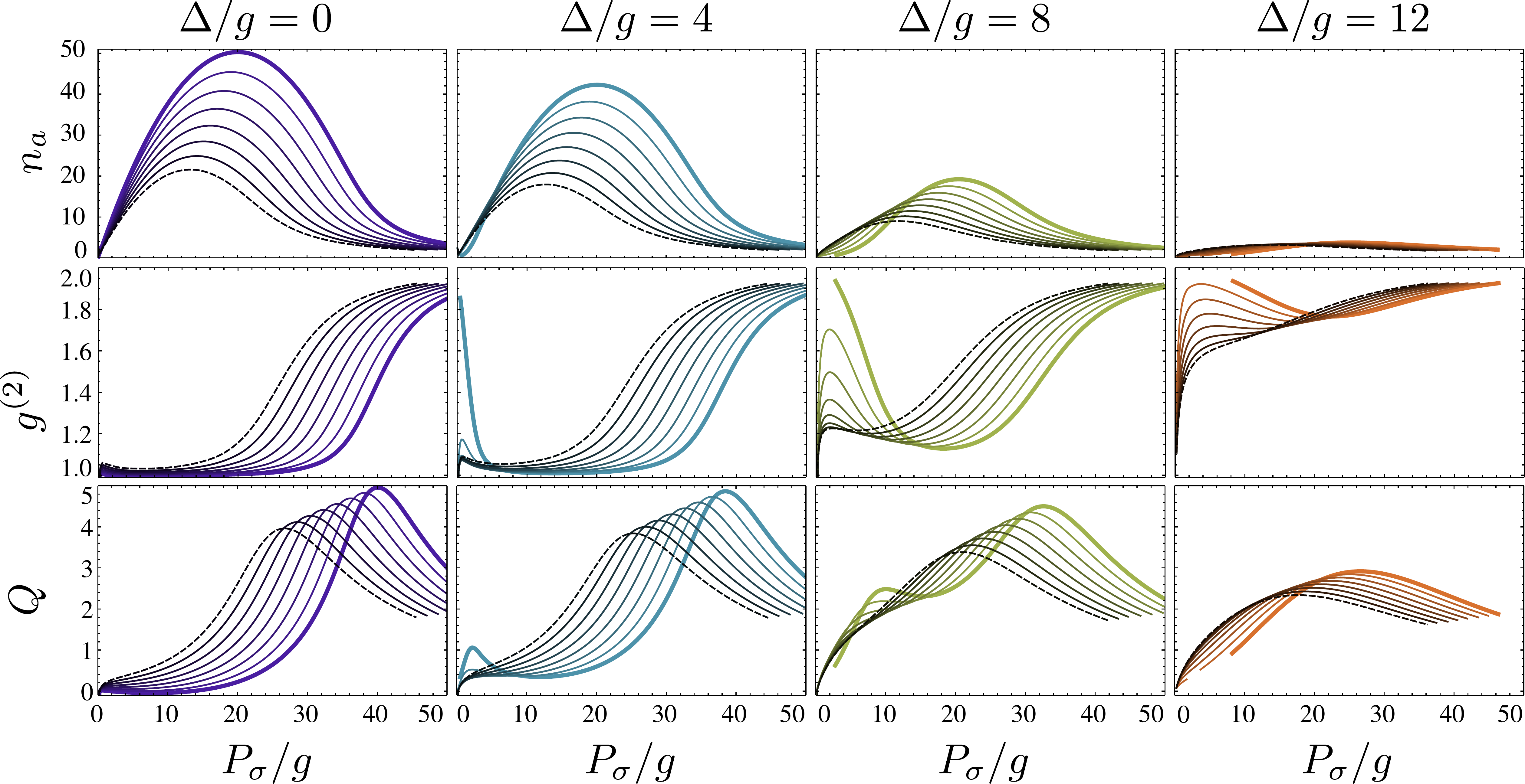}
  \caption{(Color online) (a) $n_a$, (b) $g^{(2)}$ and (c) $Q$-Mandel
    parameter, as a function of the pumping rate, extracted with the
    cothermal approximation. Detuning is fixed in each column to the
    value while pure dephasing increases from $0$ (thick curves) to
    $12g$ (dashed curves) in steps of $2g$. Detuning brings a
    threshold for lasing and while dephasing can help to reduce it, it
    also makes the system less robust to quenching. Parameters are
    $\gamma_a=0.1g$, $P_a=0=\gamma_\sigma=0$.}
  \label{fig:ThuApr28162759CEST2011}
\end{figure*}

A very good quantitative approximation for the whole pumping range is
obtained assuming that the field is in a so-called \emph{cothermal
  state}~\cite{laussy06a}, which is the field with, in average,
$n_a^\mathrm{coh}$ coherent photons and $n_a^{th}$ thermal photons (with a
total cavity population $n_a=n_a^{th}+n_a^\mathrm{coh}$). In this case, the
photonic statistics reads:
\begin{equation}
  \label{eq:MonMay2144201CEST2011}
  \mathrm{T}[n]=e^{-n_a^\mathrm{coh}/(1+n_a^{th})}\frac{(n_a^{th})^n}{(1+n_a^{th})^{n+1}} \mathrm{L}_n[\frac{-n_a^\mathrm{coh}}{n_a^{th}(1+n_a^{th})}]\,,
\end{equation}
and moments of the distribution are:
\begin{equation}
  \label{eq:MoMar22191031WET2010}
  N_{a}[n]=n! (n_a^{th})^n \mathrm{L}_n[-\frac{n_a^\mathrm{coh}}{n_a^{th}}]\,,
\end{equation}
with $\mathrm{L}_n$ the $n$-th Laguerre polynomial. The second order
coherence function is
\begin{equation}
  \label{eq:SatMay14032426CEST2011}
  g^{(2)}=2-\big(\frac{n_a^\mathrm{coh}}{n_a}\big)^2\,.
\end{equation}
One can write the set of Eqs.~(\ref{eq:MoMar22173241WET2010}) for
$n=$1,~2 using the parametrization of
Eq.~(\ref{eq:MoMar22191031WET2010}):
\begin{subequations}
  \label{eq:MonApr11102948CEST2011}
  \begin{align}
    &0=\Big[\frac{1}{C_\mathrm{eff}[1]}+\frac{\gamma_a}{\Gamma_\sigma}-\frac{2P_\sigma}{\Gamma_\sigma+\gamma_a}+1\Big]n_a\nonumber\\
    &-\frac{P_\sigma}{\Gamma_\sigma}+\frac{2\gamma_a}{\Gamma_\sigma+\gamma_a}[2n_a^2-(n_a^\mathrm{coh})^2]\,,\\
    &0=\Big[\frac{1}{C_\mathrm{eff}[2]}+\frac{2\gamma_a}{\Gamma_\sigma+2\gamma_a}-\frac{2P_\sigma}{\Gamma_\sigma+2\gamma_a}+1\Big][2n_a^2-(n_a^\mathrm{coh})^2]\nonumber \\
    &-\frac{2P_\sigma}{\Gamma_\sigma+\gamma_a}n_a+\frac{2\gamma_a}{\Gamma_\sigma+2\gamma_a}[6n_a^3-9n_a(n_a^\mathrm{coh})^2+4(n_a^\mathrm{coh})^3]\,.
  \end{align}
\end{subequations}
Equations~(\ref{eq:MonApr11102948CEST2011}) can be easily solved
numerically (two coupled nonlinear equations), extracting the physical
solution $n_a$, $n_a^\mathrm{coh}\geq 0$. They yield an excellent
agreement with the complete numerical solution as shown in
Fig.~\ref{fig:ThuMay12185542CEST2011}(d) for the cavity population and
Fig.~\ref{fig:MonMay16183014CEST2011} for~$g^{(2)}$. We took advantage
of this to extend the results to cases out of reach of our numerical
procedure, namely, $\gamma_a/g$ down to $0.01$, shown on
Figs.~\ref{fig:ThuMay12185542CEST2011} and
\ref{fig:MonMay16183014CEST2011} as the dashed lines only, without the
corresponding numerical calculation. The transitions from the linear
regime to lasing and from lasing to the thermal regime are well
accounted for within this approximation. Although they are not exact,
they provide a fairly good quantitative description of the entire
pumping range for any set of parameters, particularly of the quantum
regime which eludes most approximation schemes.

The transition into a thermal field can be induced by increasing the
incoherent pumping but also other sources of decoherence such as
dissipation, pure dephasing or detuning. In
Fig.~\ref{fig:ThuApr28162759CEST2011}, we plot the extracted $n_a$ and
$g^{(2)}$, respectively, under the cothermal approximation, as a
function of pumping, when increasing detuning and pure dephasing. The
lasing to thermal transition is even more apparent in the Mandel-$Q$
parameter, defined as $Q=n_a(g^{(2)}-1)$, featuring it with a broad
peak, as shown in the third row of
Fig.~\ref{fig:ThuApr28162759CEST2011}. Both detuning and dephasing are
clearly detrimental for the intensity and coherence in the
self-quenching regime but they may compensate each other in the
transition from the linear regime into
lasing. Fig.~\ref{fig:ThuApr28162759CEST2011} shows the appearance of
a threshold in the presence of detuning (at small pumping the thermal
fraction increases), due to the reduction of
$g_\mathrm{eff}$~\cite{rice94a}. Adding dephasing, $\gamma_\phi$, may
compensate this reduction and smoothen the threshold up to a certain
point (as it also increases to the total decoherence rate
$\Gamma_\mathrm{T}$). This observation was made by Auffèves \emph{et
  al.}~\cite{auffeves10a}. In general, the maximum intensity,
max($n_a$), and coherence achieved, always decrease with both detuning
and dephasing. However, detuning makes it occur at higher pumpings
while dephasing does at lower ones. Pure dephasing is thus indeed
making both transitions, from quantum to lasing and from lasing to
self-quenching, occur at smaller pumping.

\subsection{Density matrix}
\label{sec:22}

We have obtained in the previous section good approximations for the
statistics of the cavity field, $\mathrm{T}[n]$, as well as the main
quantities of interest such as $n_a$. To compute the photoluminescence
spectrum, what we will do in next section, we also need the full
density matrix in the steady state.  In the very strong-coupling
regime, we can relate it analytically to $\mathrm{T}[n]$, as we show
in this section.

The full statistics is most conveniently obtained from the master
equation with elements $\rho_{m,i;n,j}$ for $m$, $n$ photons and~$i$,
$j$ excitation of the emitter ($m, n\in\mathbf{N}$, $i,
j\in\{0,1\}$). Rather than to consider the equations of motion for the
matrix elements directly, it is clearer and more efficient to consider
only elements that are nonzero in the steady state. These are:
\begin{equation} 
\label{eq:SunJul20203845CEST2008}
  \mathrm{p}_0[n]=\rho_{n,0;n,0}\,,\quad  \mathrm{p}_1[n]=\rho_{n,1;n,1}\,,\quad\mathrm{q}[n]=\rho_{n,0;n-1,1}\,,
\end{equation}
and correspond to, respectively, the probability to have~$n$ photons
with ($\mathrm{p}_1$) or without ($\mathrm{p}_0$) excitation of the
emitter, and the coherence element between the states~$\ket{n,0}$
and~$\ket{n-1,1}$, linked by the Hamiltonian
Eq.~(\ref{eq:FriOct2103450BST2009}). Both~$p_0$ and~$p_1$ are
real. It is convenient to separate~$q$ into its real and imaginary
parts, $\mathrm{q}[n]=\mathrm{q}_\mathrm{r}[n]+i
\mathrm{q}_\mathrm{i}[n]$, as they play different roles in the
dynamics. The equations for these quantities are derived from the
Liouvillian equation~(\ref{eq:ThuOct18162449UTC2007}). They read
(their full form is given in Appendix~\ref{sec:ThuJul1223020MSD2010}):
\begin{subequations} 
\label{eq:MoMar22200554WET2010} 
\begin{align}
  \partial_t\mathrm{p}_0[n+1]=&\mathcal{D}_\mathrm{phot}\{\mathrm{p}_0[n+1]\}\\
  &-P_\sigma\mathrm{p}_0[n+1]+\gamma_\sigma
  \mathrm{p}_1[n+1]\nonumber\\
  &-2g\sqrt{n+1}\mathrm{q}_\mathrm{i}[n+1]\,,\nonumber\\
  \partial_t\mathrm{p}_1[n]=&\mathcal{D}_\mathrm{phot}\{\mathrm{p}_1[n]\}\\
  &-\gamma_\sigma\mathrm{p}_1[n]+P_\sigma
  \mathrm{p}_0[n]+2g\sqrt{n+1}\mathrm{q}_\mathrm{i}[n+1]\,,\label{eq:MoMar22202656WET2010}\nonumber\\
  \partial_t\mathrm{q}_\mathrm{i}[n+1]=&\mathcal{D}_\mathrm{phot}\{\mathrm{q}_\mathrm{i}[n+1]\}\nonumber\\
  &-\Big[\frac{\Gamma_\sigma+\gamma_\phi}{2}+\frac{2\Delta^2}{\Gamma_\sigma+\gamma_\phi}\Big]\mathrm{q}_\mathrm{i}[n+1]\nonumber\\
  &+g\sqrt{n+1}(\mathrm{p}_0[n+1]-\mathrm{p}_1[n])\,,
\end{align} 
\end{subequations}
where we have separated the cavity dynamics into a superoperator
$\mathcal{D}_\mathrm{phot}$, which exact expression is given in the
Appendix. If this photonic dynamics is much slower than the rest, that
is $\gamma_a,P_a\ll \Gamma_\sigma,g$, one can solve the emitter
dynamics separately, ignoring
$\mathcal{D}_\mathrm{phot}$~\cite{scully_book02a,mu92a}.  This assumes
that the photon distribution
$\mathrm{T}[n]=\mathrm{p}_0[n]+\mathrm{p}_1[n]$ does not change during
the excitation and interaction with the emitter that happens at a
timescale, of order $1/P_\sigma$ or $1/g$, much faster than the photon
one, of order $1/\gamma_a$. This approximation becomes exact for the
perfect cavity, when $\gamma_a=0$ (and $P_a=0$, which we already
assumed), which is why the system admits an exact solution
(cf.~Eq.~(\ref{eq:FriJul2000551MSD2010})). Neglecting the photon
dynamics allows one to link all the density matrix elements with the
photon statistics $\mathrm{T}[n]$:
\begin{subequations} 
\label{eq:MoMar22203849WET2010} 
\begin{align}
  &\mathrm{p}_0[n+1]\approx\\
  &\frac{\kappa_a(n+1)\Big(\frac{P_\sigma}{\Gamma_\sigma}\mathrm{T}[n]+\frac{\gamma_\sigma}{\Gamma_\sigma}\mathrm{T}[n+1]\Big)+\gamma_\sigma\mathrm{T}[n+1]}{2\kappa_a(n+1)+\Gamma_\sigma}\,,\nonumber\\
  &\mathrm{p}_1[n]\approx\\
  &\frac{\kappa_a(n+1)\Big(\frac{P_\sigma}{\Gamma_\sigma}\mathrm{T}[n]+\frac{\gamma_\sigma}{\Gamma_\sigma}\mathrm{T}[n+1]\Big)+P_\sigma\mathrm{T}[n]}{2\kappa_a(n+1)+\Gamma_\sigma}\,,\nonumber\\
  &\mathrm{q}_\mathrm{i}[n+1]\approx\frac{\kappa_a\sqrt{n+1}}{2g}(\mathrm{p}_0[n+1]-\mathrm{p}_1[n])\\
  &=-\frac{\kappa_a\sqrt{n+1}}{2g}\frac{P_\sigma\mathrm{T}[n]-\gamma_\sigma\mathrm{T}[n+1]}{2\kappa_a(n+1)+\Gamma_\sigma}\nonumber\,.
\end{align}
\end{subequations}
where 
\begin{equation}
  \label{eq:TueMar29183012CEST2011}
  \kappa_a=\frac{4(g_\mathrm{eff}[1])^2}{\Gamma_\sigma+\gamma_\phi}
\end{equation}
is the Purcell rate of transfer of population from the cavity mode to
the emitter (cf.~Eq.~(\ref{eq:MonMar28155703CEST2011})) and
$g_\mathrm{eff}$ follows the definition in
Eq.~(\ref{eq:MonMar28154711CEST2011}) for negligible $\gamma_a$. Note
that $\mathrm{p}_0[n]+\mathrm{p}_1[n]$ is not strictly equal to
$\mathrm{T}[n]$, due to our approximations, but the numerical
discrepancy is small in the regime of interest where the number of
photons is high and $n\approx n+1$.  In particular, the equality holds
exactly in the aforementioned case of $\gamma_a=P_a=0$, thanks to
some nontrivial simplifications of the expressions when
$\mathrm{T}[n]$ is thermal.

\subsection{Two-time correlators}
\label{sec:23}

We now turn to the problem of the steady state optical emission
spectrum, that consists in computing two-time correlators of the type
$\mean{\ud{c}(0)c(\tau)}$ with $c=a$, $\sigma$. We can link the two
time correlators to the quantities derived in the previous sections
following an implementation of the quantum regression theorem that
relies explicitly on the density matrix~$\rho$:
\begin{equation} 
\label{eq:SunNov16185703GMT2008}
\mean{\ud{c}(0)c(\tau)}= \sum_{k,l}\rho^c_{[k;l]} (\tau)\bra{l}c\ket{k}\,,
\end{equation} 
where $\rho^c_{[k;l]} (\tau)=\mean{\ud{c}(\ket{l}\bra{k})(\tau)}$ is
in the Schr\"odinger picture, where the states evolve and operators
have their steady state values~\cite{elena_molmer96a}. The indices
$k$, $l$ go through all the states in the system Hilbert space
($k=(k_1,k_2)$ with $k_1=0,1,\hdots, n_\mathrm{max}$ for the photons
and $k_2=0,1$ for the emitter). The elements $\rho^c_{[k;l]}$ follow
the same master equation as the density matrix elements
$\rho_{[k;l]}$. Since similar approximations can also be naturally
implemented, this will allow us to provide closed-form solutions for
the two-time correlators, as is detailed in
Appendix~\ref{sec:SuSep26172348WEST2010}. We give here the main lines
of the derivations and introduce the key quantities that lead to the
final result. We single out, again, only the nonzero elements. For
two-time correlators, they can be gathered in four functions of $n$:
\begin{subequations} 
  \label{eq:TueMar23162915WET2010}
\begin{align} 
  &S_i[n]\equiv \rho^c_{[n,i;n-1,i]}\,,\quad i=0,1\,,  \quad n\geq 0\,,\\
  &Q[n]\equiv \rho^c_{[n,1;n,0]}\,,\quad n\geq 0\,,\\
  &V[n]\equiv \rho^c_{[n,0;n-2,1]}\,,\quad n\geq 1\,.
\end{align}
\end{subequations} 

The two-times correlators follow from these quantities (with
$c=\sigma$, $a$ respectively) as:
\begin{subequations}
    \label{eq:WedMar24180922WET2010}
  \begin{align}
    &\langle\ud{\sigma}(0)\sigma(\tau)\rangle=\sum_{n=0}^{\infty}Q[n]\,,\label{eq:WedMar30182503CEST2011}\\
    &\langle\ud{a}(0)a(\tau)\rangle=\sum_{n=0}^{\infty}(\sqrt{n+1}S_0[n+1]+\sqrt{n}S_1[n])\,.\label{eq:WedMar30182553CEST2011}
  \end{align}
\end{subequations}
Each term $n$ of these sums accounts for the transitions between the
rungs $n+1$ and $n$, as in the case of spontaneous
emission~\cite{delvalle09a}, the first rung, being given by $n=0$.

The equations of motion for the quantities
in Eqs.~(\ref{eq:TueMar23162915WET2010}) are extracted from the master
equation (cf.  Eq.~\ref{eq:SunJul20204529CEST20082}), and can also be,
like for the single time dynamics, separated into a slow photonic
dynamics embedded in a superoperator $\mathcal{D}_\mathrm{phot}$ on
the one hand, and a fast emitter and coupling dynamics on the other
hand:
\begin{subequations} 
  \label{eq:TueMar23165009WET2010} 
  \begin{align}
    \partial_\tau S_0[n+1]=&\mathcal{D}_\mathrm{phot}\{S_0[n+1]\}\\
    &-P_\sigma S_0[n+1]+\gamma_\sigma S_1[n+1]\nonumber\\
    &+ig(\sqrt{n}V[n+1]-\sqrt{n+1}Q[n])\,,\nonumber\\
    \partial_\tau S_1[n]=&\mathcal{D}_\mathrm{phot}\{S_1[n]\}\\
    &+P_\sigma S_0[n]-\gamma_\sigma S_1[n]\nonumber\\
    &-ig(\sqrt{n+1}V[n+1]-\sqrt{n}Q[n])\,,\nonumber\\
    \partial_\tau Q[n]=&\mathcal{D}_\mathrm{phot}\{Q[n]\}\\
    &-\Big(\frac{\Gamma_\sigma+\gamma_\phi}{2}-i\Delta\Big)Q[n]\nonumber\\
    &+ig(\sqrt{n}S_1[n]-\sqrt{n+1}S_0[n+1])\,,\nonumber\\
    \partial_\tau V[n+1]=&\mathcal{D}_\mathrm{phot}\{V[n+1]\}\\
    &-\Big(\frac{\Gamma_\sigma+\gamma_\phi}{2}+i\Delta\Big)V[n+1]\nonumber\\
    &+ig(\sqrt{n}S_0[n+1]-\sqrt{n+1}S_1[n])\,,\nonumber
\end{align} 
\end{subequations}
for $n\geq 1$. After some long, but straightforward algebra, we can
express $S_{0,1}[n]$ and $Q[n]$ in terms of $\mathrm{p}_{0,1}[n]$ and
$\mathrm{q}_i[n]$, which, in turn, are expressed in terms of the
statistics $\mathrm{T}[n]$. This can be done for arbitrary parameters,
including detuning. However, simple expressions are possible only at
resonance, where we can write Eq.~(\ref{eq:SunNov16185703GMT2008}) as:
\begin{align}
  \label{eq:ThuMar25104239WET2010}
  \mean{\ud{c}(0)c(\tau)}&=E ^c+n_c\sum_{n=0}^{\infty} [\nonumber\\
  &C_\mathrm{I}^c[n] e^{-iR_\mathrm{I}[n]\tau}e^{-(3\Gamma_\sigma+\gamma_\phi)\tau/4}+\text{R.s.i.}\nonumber\\
  &C_\mathrm{O}^c[n] e^{-iR_\mathrm{O}[n]\tau}e^{-(3\Gamma_\sigma+\gamma_\phi)\tau/4}+\text{R.s.i.}]\,.
\end{align}
``R.s.i'' stands for ``Rabi sign inversion'' and is the operation that
consists in changing the sign of $R_{\mathrm{I},\mathrm{O}}$ keeping all other
quantities the same.  The first term $E^c$ that factors out of the
sum, is independent of $\tau$, due to the approximation of very large
photonic lifetime. It is discussed separately in
section~\ref{sec:33}. The most fundamental quantities that arise in
the above treatment are the \emph{$n$th manifold inner} and
\emph{outer (half) Rabi frequencies}:
\begin{equation}
  \label{eq:ThuMar25104706WET2010}
    R_{\mathrm{O},\mathrm{I}}[n]=\sqrt{g^2(\sqrt{n+1}\pm\sqrt{n})^2-\Big(\frac{\Gamma_\sigma-\gamma_\phi}{4}\Big)^2}\,.
\end{equation}

The (half) Rabi frequency of the linear regime, $R_0$ is recovered as
the particular case $n=0$ with
\begin{equation}
  \label{eq:SuSep26163817WEST2010}
R_0=R_\mathrm{O}[0]=R_\mathrm{I}[0]=\sqrt{g^2-\Big(\frac{\Gamma_\sigma-\gamma_\phi}{4}\Big)^2}\,.
\end{equation}
The coefficient $C_\mathrm{I}^c[n]$ (id. for $C_\mathrm{O}^c[n]$) are complex
quantities in general that we decompose into their real and imaginary
part as 
\begin{equation}
C_\mathrm{I}^c[n]=L_\mathrm{I}^c[n]+i K_\mathrm{I}^c[n]\label{eq:ThuMar31183059CEST2011}
\end{equation}
They are a function of $R_\mathrm{I}[n]$ (and $C_\mathrm{O}$ of
$R_\mathrm{O}[n]$). 

As stated previously, the coefficients
$C_{\mathrm{I},\mathrm{O}}^c[n]$ are written in terms of the system
parameters and the steady state photon distribution $\mathrm{T}[n]$
only.  Their general full expression are too long to be given.  Only
for the simplest case of $\gamma_\sigma$, $\gamma_\phi=0$ and
$c=\sigma$, the expressions simplify sufficiently to be reproduced
here:
\begin{equation}
  \label{eq:SunJul4144504CEST2010}
  C_{\mathrm{I},\mathrm{O}}^\sigma[n]=\frac{\alpha_{\mathrm{I},\mathrm{O}}^\sigma[n]}{n_\sigma} \mathrm{T}[n]+\frac{\beta_{\mathrm{I},\mathrm{O}}^\sigma[n]}{n_\sigma}\mathrm{T}[n-1]
\end{equation}
where:
\begin{subequations}
  \label{eq:SuMar28193842WEST2010}
  \begin{align}
    \alpha_{\mathrm{I},\mathrm{O}}^\sigma[n]&=\frac{\left(\frac{P_\sigma}{2}\right)^2+g^2(1+n)}{P_\sigma^2+8g^2(1+n)}+\frac{iP_\sigma}{4R_{\mathrm{I},\mathrm{O}}[n]}\times\\
    &\frac{\left(\frac{P_\sigma}{2}\right)^2-g^2(1+n\mp 2\sqrt{n(1+n)})}{P_\sigma^2+8g^2(1+n)}\,,\nonumber\\
    \beta_{\mathrm{I},\mathrm{O}}^\sigma[n]&=\pm g^2 P_\sigma^2 (4+3\frac{i P_\sigma}{4R_{\mathrm{I},\mathrm{O}}[n]})\times\\
    &\frac{2 g^2(\sqrt{n(1+n)}\pm n)+P_\sigma^2(\sqrt{n(1+n)}\mp n)}{4(8 g^2n+P_\sigma^2)(4g^4+4g^2P_\sigma^2(1+2n)+P_\sigma^4)}\,,\nonumber
   \end{align}
\end{subequations}
where the notations $\pm$ and $\mp$ associate $I$ to the upper sign
and $O$ to the lower one.

\section{Mollow triplet under incoherent excitation}
\label{sec:3}

\begin{figure}[t]
  \centering
  \includegraphics[width=\linewidth]{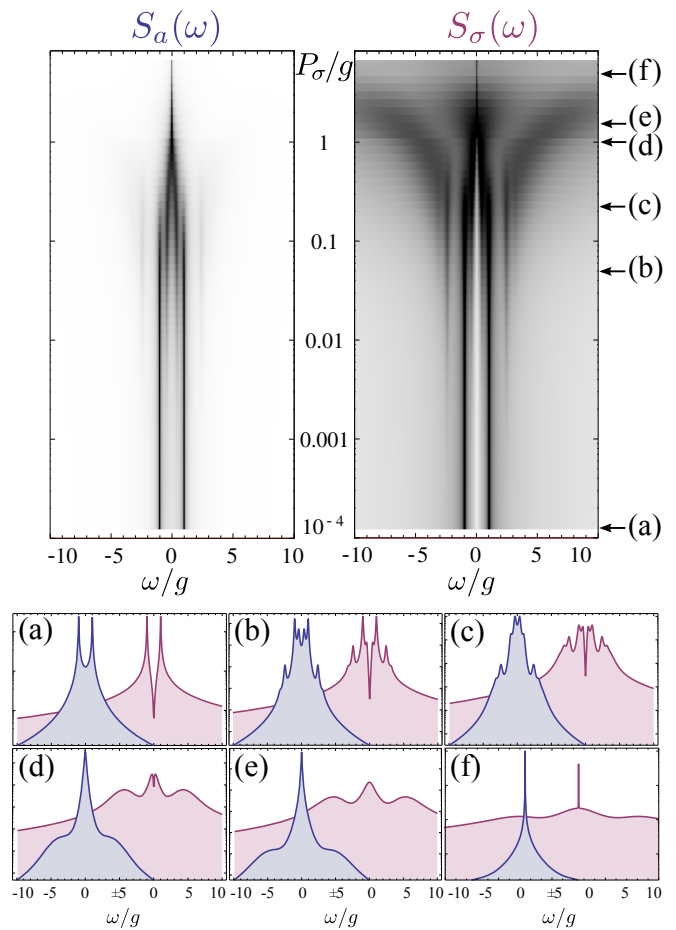}
  \caption{(Color online) Cavity (left blue) and emitter (right pink)
    spectra of emission computed numerically. As a function of
    pumping, a transition from (a) the quantum linear regime to (f)
    lasing can be followed passing through (b) the quantum nonlinear
    regime that (c) melts into (d) and (e) structures of much reduced
    complexity, namely, triplets. A Mollow triplet is neatly visible
    in the emitter spectrum, whereas the cavity gives rise to a single
    narrowing line. Note that in the cuts (a--f), spectra are
    displayed in log-scale. Parameters are $\gamma_a=0.1g$,
    $\gamma_\sigma=0.00334g$, $P_a=\gamma_\phi=\Delta=0$.}
  \label{fig:MonMay16180915CEST2011}
\end{figure}

\begin{figure*}[thbp]
  \centering
  \includegraphics[width=.9\linewidth]{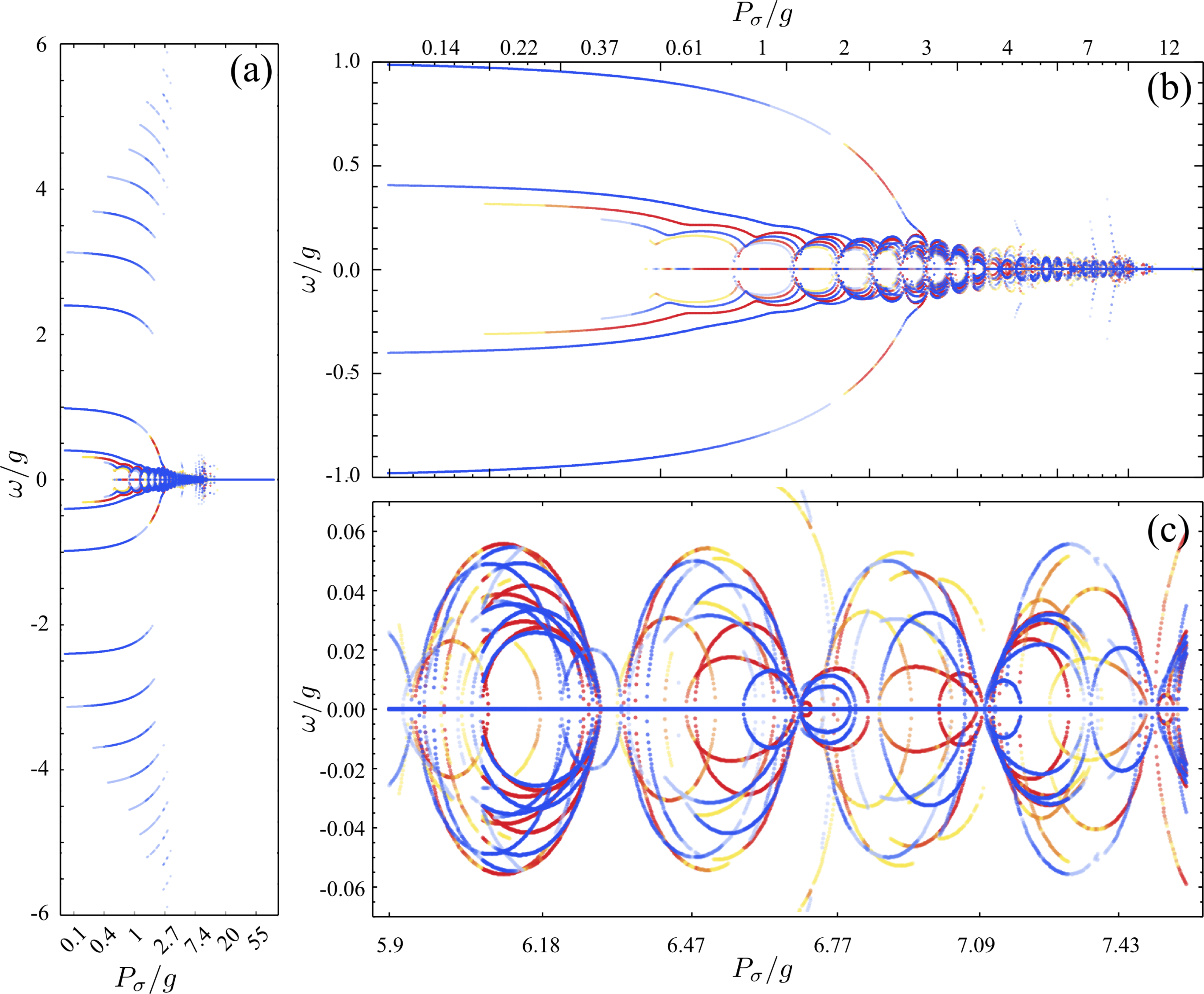}
  \caption{(Color online) Lasing as a condensation of dressed states:
    transitions energies between the dressed states when solving
    numerically the dynamics exactly are shown. The color code
    (online) has blue shades corresponding to a positive weight of the
    transition in the cavity emission and red shades corresponding to
    a negative weight. In (a) all regimes are shown over all energies,
    in (b) a close-up of~(a) is given for frequencies $|\omega_p|<g$
    lying between the vacuum Rabi splitting, in (c) a close-up of~(b)
    is given in the lasing regime, showing an extremely complex
    structure of the exact solution obtained in the full-quantization
    picture, although the final result washes out completely most of
    this underlying information, to provide only the single narrow
    line of a lasing system.}
  \label{fig:FriApr1111520CEST2011}
\end{figure*}

The luminescence spectra can, like all other quantities, be obtained
``exactly'' through numerical
computations~\cite{loffler97a,delvalle09a}. In our case, computing
Eqs.~(\ref{eq:TueMar23165009WET2010}) and applying the definition of
Eq.~(\ref{eq:SunNov16140314GMT2008}) for the power spectra, one
arrives to results such as those shown in
Fig.~\ref{fig:MonMay16180915CEST2011}, where PL spectra are computed
both for the cavity and the emitter, as a function of pumping. In
insets (a--f), we select various snapshots plotted in log-scale that
illustrate the regimes discussed previously, namely: (a) shows a case
from the linear regime, where only the first rung of the
Jaynes--Cummings ladder is occupied, the system being otherwise in
vacuum. This yields the vacuum Rabi splitting. The differences in
lineshapes in this regime have been amply discussed
elsewhere~\cite{laussy08a}. In (b), one is in the quantum nonlinear
regime, where the spectrum has a complex structure featuring the many
peaks that arise from transitions between the lower rungs of the
Jaynes--Cummings ladder.  In (c), one leaves the quantum regime, with
a collapse of the quantum nonlinear peaks that ``melt'' to form an
emerging structure of much reduced complexity, namely, a triplet, as
seen in (d) where one enters the lasing regime that develops in (e)
and is fully formed in (f). The triplet is particularly evident in the
emitter power spectrum, where it is neatly visible also in a linear
scale.  Quenching is not shown explicitly in this plot but is
elsewhere~\cite{delvalle09a}.

A remarkable feature of lasing in strong-coupling arises in the form
of a scattering peak, well known from Mollow's results
(cf. Eq.~(\ref{eq:SunJun27235639MSD2010})) where it is due to the
driving laser scattering photons off the atom. This $\delta$ peak also
forms in the cavity QED version of this problem, as seen in
Fig.~\ref{fig:MonMay16180915CEST2011}(f) on the emitter spectrum as
the very sharp peak sitting on top of the incoherent Mollow
triplet. Note that this peak arises from a numerical computation. It
is a close counterpart of the Rayleigh scattering peak of resonance
fluorescence, although this is the coherent field grown by the emitter
itself in the lasing process that is the source of the scattered
photons, on the very emitter that created them in the first
place. More interestingly, this peak which, in the lasing regime, is
maximum and narrower (with the same linewidth as the cavity), forms
smoothly from a similar depletion when approaching the lasing regime
from below, as seen in Fig.~\ref{fig:MonMay16180915CEST2011}(d), where
an equally narrow absorption peak is carved in the emerging
triplet. As opposed to scattering, in this case, the cavity is
coherently and resonantly absorbing excitations from the emitter (this
depletion is pinned at the cavity energy). As the field is building
its coherence, it sucks energy very efficiently from its source, until
a point, shown in Fig.~\ref{fig:MonMay16180915CEST2011}(e), where the
cavity does not require such a coherent absorption to keep building in
intensity. This marks the transition between the point where the
system is building its coherent field to the one where it is fully
formed and acting back on the emitter.

These results are certainly beautiful (an animation of this transition
is available in the supplementary material of Ref.~\cite{laussy10b})
but obtaining them numerically is an intensive computer task since the
Hilbert space becomes very large, whereas the final output certainly
looks amenable to a simple analytical description. Solving the system
numerically nevertheless gives access to every single transition that
occur in the many rungs of the Jaynes--Cummings ladder through the
dressed state decomposition of
Eq.~(\ref{eq:SunNov16140314GMT2008}). This yields a surprisingly
complex structure, shown in Fig.~\ref{fig:FriApr1111520CEST2011},
where we plot the positions $\omega_p$ where the system emits,
weighted by the intensity $L_p$ of the transition such that resonances
disappear with vanishing
intensities. Fig.~\ref{fig:FriApr1111520CEST2011}(a) gives an overall
picture while a zoom of the transitions lying between the Rabi doublet
is given in Fig.~\ref{fig:FriApr1111520CEST2011}(b), showing an
emerging and intricate structure, further zoomed in (c) in the lasing
region.  The inner peaks form ``bubbles'' that open and collapse
around the origin, where lies the lasing mode. Such behaviours of the
dressed states also appear in simpler systems such as two coupled
two-level emitters incoherently pumped, which can be solved fully
analytically~\cite{delvalle10b}. The bubbles formation result from a
complex interplay between pumping and decay, which open new channels
of coherence flow in the
system. Figure~\ref{fig:FriApr1111520CEST2011}(a), shows clearly the
satellite peaks of the Mollow triplet. Although the lines are neatly
splitted the one from the other, their increasing broadening allows
the formation of a smooth spectral shape in the lasing regime, that
one can follows with the naked eye from the ``melting'' of the
quantized structure, as shown in
Fig.~\ref{fig:MonMay16180915CEST2011}.
Fig.~\ref{fig:FriApr1111520CEST2011}(a) also shows how the inner peaks
ultimately all converge at the origin, thereby forming the lasing
mode. In the lasing in strong coupling scenario, lasing can thus be
seen as a Bose condensate of the dressed states~\cite{imamoglu96a,
  imamoglu96b, laussy04a}. It is fascinating to follow the formation
of a coherent and classical field from a fully quantized picture, but
this brings little insights into the actual phenomenon. Beside hinting
at its underlying richness and complexity,
Fig.~\ref{fig:FriApr1111520CEST2011} essentially shows that a complete
and fully quantized description of a system that is behaving basically
classically is hopelessly complicated, keeping track of a huge amount
of irrelevant details, while the behaviour is well accounted for by a
few macroscopic degrees of freedom, such as an intensity $n_a$ and a
off-diagonal coherence element $\langle
a\rangle$. Fig.~\ref{fig:MonMay16180915CEST2011} thus illustrates, in
one of the most fundamental system of quantum optics, the breakdown of
the quantum picture in a quantum to classical transition. Even in the
simplest and exactly solvable system, it is difficult to read much,
and we surmise that the condensation of dressed states in the lasing
process is out of reach of the present understanding of dissipative
quantum optics, calling for a framework such as that developed for
conservative systems~\cite{berry87a, gutzwiller92a,elena_haake01a}.

Since the intricate patterns of Fig.~\ref{fig:FriApr1111520CEST2011}
occur at a different energy scale than that of the observables and do
not show up in the spectra, one can hope in the wake of the excellent
approximations derived previously to get similar analytical results
also in the spectral domain. In the following, we derive such an
approximate description of the exact picture presented
above~\cite{delvalle10d}, allowing us to read the essential physics of
this transition in the lasing regime.

\subsection{Spectral decomposition}

The Fourier transform in Eq.~(\ref{eq:MonJul21132824CEST2008}) of the
two-time correlators~(\ref{eq:ThuMar25104239WET2010}) provides an
approximated expression for the spectrum of emission:
\begin{align}
  \label{eq:WedMar24190127WET2010}
  S_c(\omega)=&\frac{\Re(E^c)}{n_c}\delta(\omega)+\frac{1}{\pi}\sum_{n=0}^{\infty}[\nonumber\\
  &+\frac{L_\mathrm{I}^c[n]\frac{\gamma_\mathrm{I}[n]}{2}-K_\mathrm{I}^c[n](\omega-\omega_\mathrm{I}[n])}{\big(\frac{\gamma_\mathrm{I}[n]}{2}\big)^2+(\omega-\omega_\mathrm{I}[n])^2}+\text{R.s.i.}\nonumber\\
  &+\frac{L_\mathrm{O}^c[n]\frac{\gamma_\mathrm{O}[n]}{2}-K_\mathrm{O}^c[n](\omega-\omega_\mathrm{O}[n])}{\big(\frac{\gamma_\mathrm{O}[n]}{2}\big)^2+(\omega-\omega_\mathrm{O}[n])^2}+\text{R.s.i.}\left.\right]\,,
\end{align}
where we have introduced the positions and broadenings:
\begin{subequations}
  \begin{align}
    \label{eq:SunJul4155121CEST2010}
    \omega_{\mathrm{I},\mathrm{O}}[n]&= \Re(R_{\mathrm{I},\mathrm{O}}[n])\,,\\
    \gamma_{\mathrm{I},\mathrm{O}}[n]&=
    \frac{3\Gamma_\sigma+\gamma_\phi}2-2\Im(R_{\mathrm{I},\mathrm{O}}[n])\,,
  \end{align}
\end{subequations}
so that the optical spectrum is composed of a series of Lorentzian
lines at frequencies~$\omega_p$ with linewidths~$\gamma_p$ and weight
$L_p$ plus interference terms weighted by $K_p$. These lines arise
from transitions between rungs $n+1$ and $n$ of the Jaynes-Cummings
ladder.  The spectra are normalized to unity, therefore, the
\emph{incoherent} part of the spectra (second and third lines) is
normalized to $1-\Re(E^c)/n_c$.  Each transition can exhibit weak or
strong coupling, that is, the rungs being split or not into dressed
states, similarly to the case with no pumping~\cite{delvalle09a}. When
split, all peaks broadenings are the same, regardless of the rung:
$\gamma_\mathrm{I}[n]=\gamma_\mathrm{O}[n]=(3\Gamma_\sigma+\gamma_\phi)/2$.

\begin{figure}[t]
  \centering
  \includegraphics[width=\linewidth]{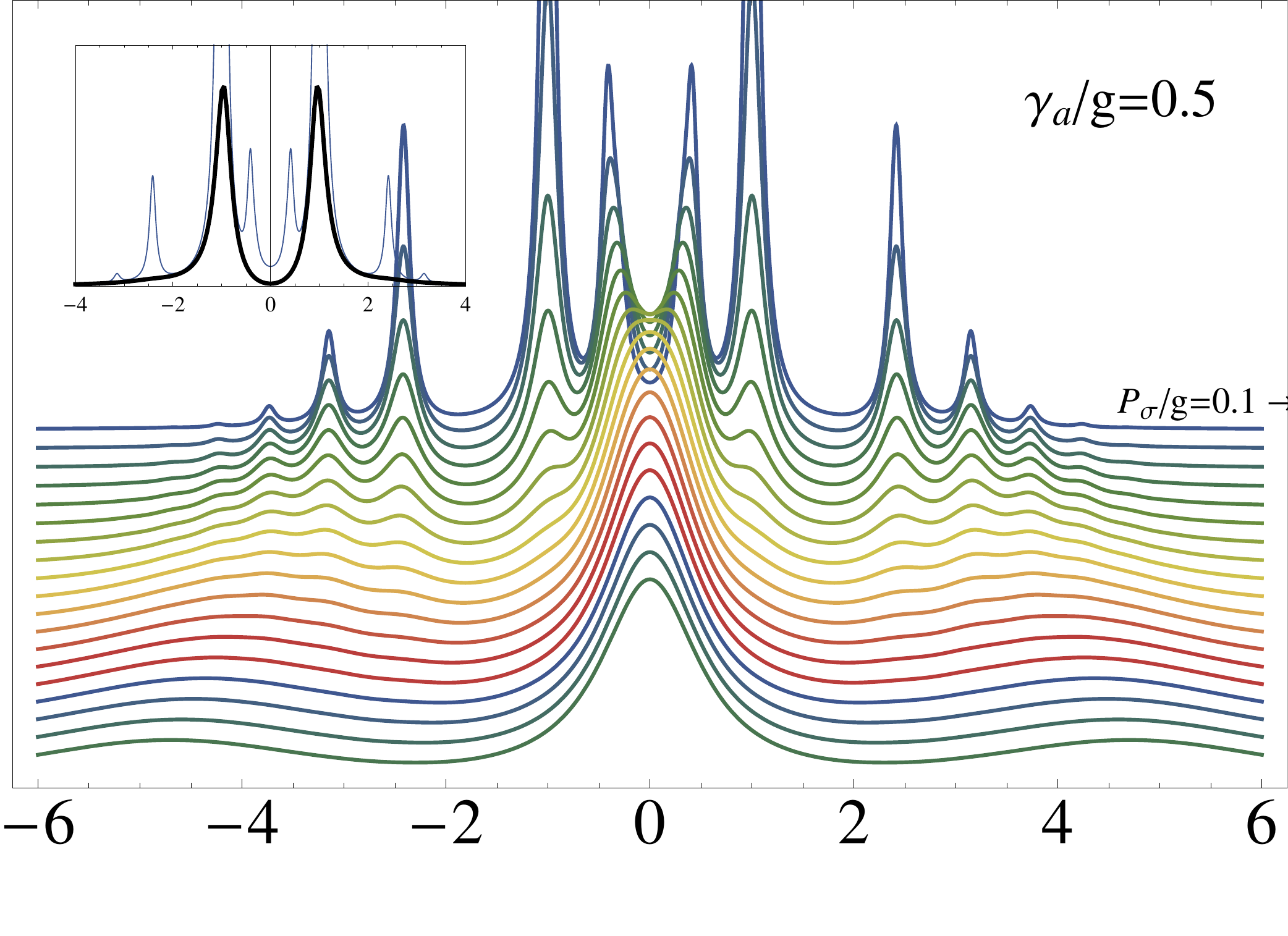}
  \includegraphics[width=\linewidth]{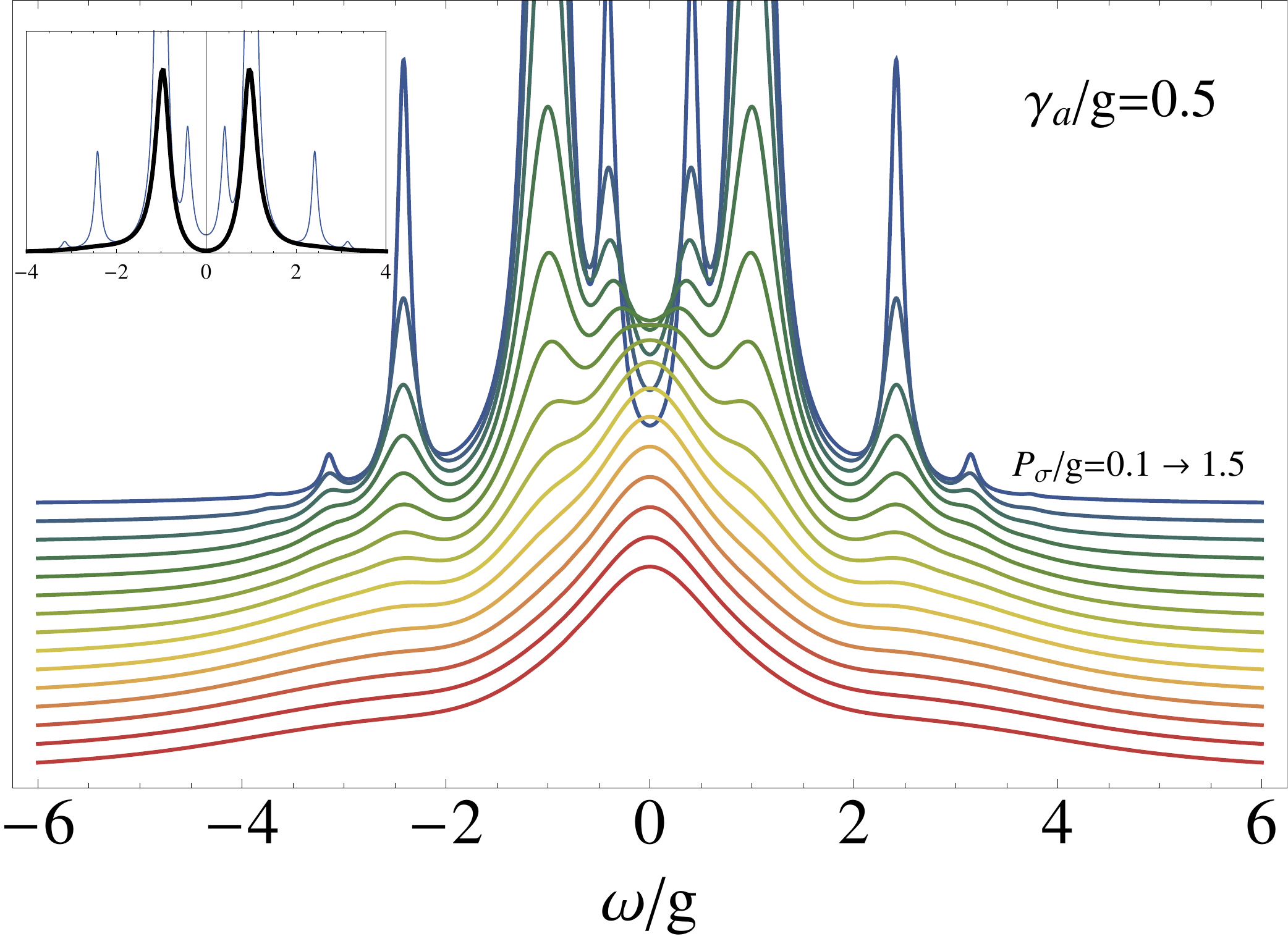}
  \caption{(Color online) Approximated spectra of emission,
    Eq.~(\ref{eq:WedMar24190127WET2010}), as pumping is increased and
    the system undergoes the quantum-to-classical transition. The
    plots correspond to two different cavity lifetimes
    ($\gamma_a=0.1g$ and $0.5g$). Parameters, not specified in the
    plots, are $P_a=\gamma_\sigma=\gamma_\phi=\Delta=0$. In inset, the
    comparison between the approximated-analytical (thin blue) and
    exact-numerical (thick black) spectra for the lowest pump
    $P_\sigma=0.1g$. Although the approximation breaks in the quantum
    regime, it shows how the Mollow triplet is formed.}
  \label{fig:FriApr1105117CEST2011}
\end{figure}

In Fig.~\ref{fig:FriApr1105117CEST2011}, we see two examples of
spectra computed as in Eq.~(\ref{eq:WedMar24190127WET2010}) with
$\mathrm{T}[n]$ taken as Poissonian. The first case, with
$\gamma_a=0.1g$, corresponds to a very good cavity well in the strong
coupling regime with the emitter, such as is realized in circuit
QED. The second case, with $\gamma_a=0.5g$, corresponds to a less
favourable situation representative of the state of the art of systems
such as quantum dots in microcavities~\cite{ohta11a}. In both plots,
the pump increases from the top to the bottom curves and features the
quantum to classical crossover. The transitions between the
Jaynes-Cummings rungs are first resolved individually, at low pumping,
and then merge into a Mollow triplet. The approximated spectrum of
emission differs from the numerical exact result at low pumpings,
since the assumption of much larger pumping than decay does not apply
here. The regime of validity for our approximated spectrum and that
for the observation of the Mollow triplet with incoherent excitation
is:
\begin{equation}
  \label{eq:TueJul6130009CEST2010}
  \gamma_a\,,\gamma_\sigma\,,\gamma_\phi\ll g<P_\sigma\ll \kappa_\sigma\,.
\end{equation}
Note that out of resonance, one must consider $g_\mathrm{eff}[1]$
instead of~$g$ in Eq.~(\ref{eq:TueJul6130009CEST2010}).  As the
position of the peaks is still well approximated, however, this
approximation provides an instructive and physically transparent
picture of the Mollow triplet formation. We analyse these peak
positions in more details in the next subsection.

\subsection{Peak positions}
\label{sec:31}

\begin{figure}[t]
  \centering
  \includegraphics[width=\linewidth]{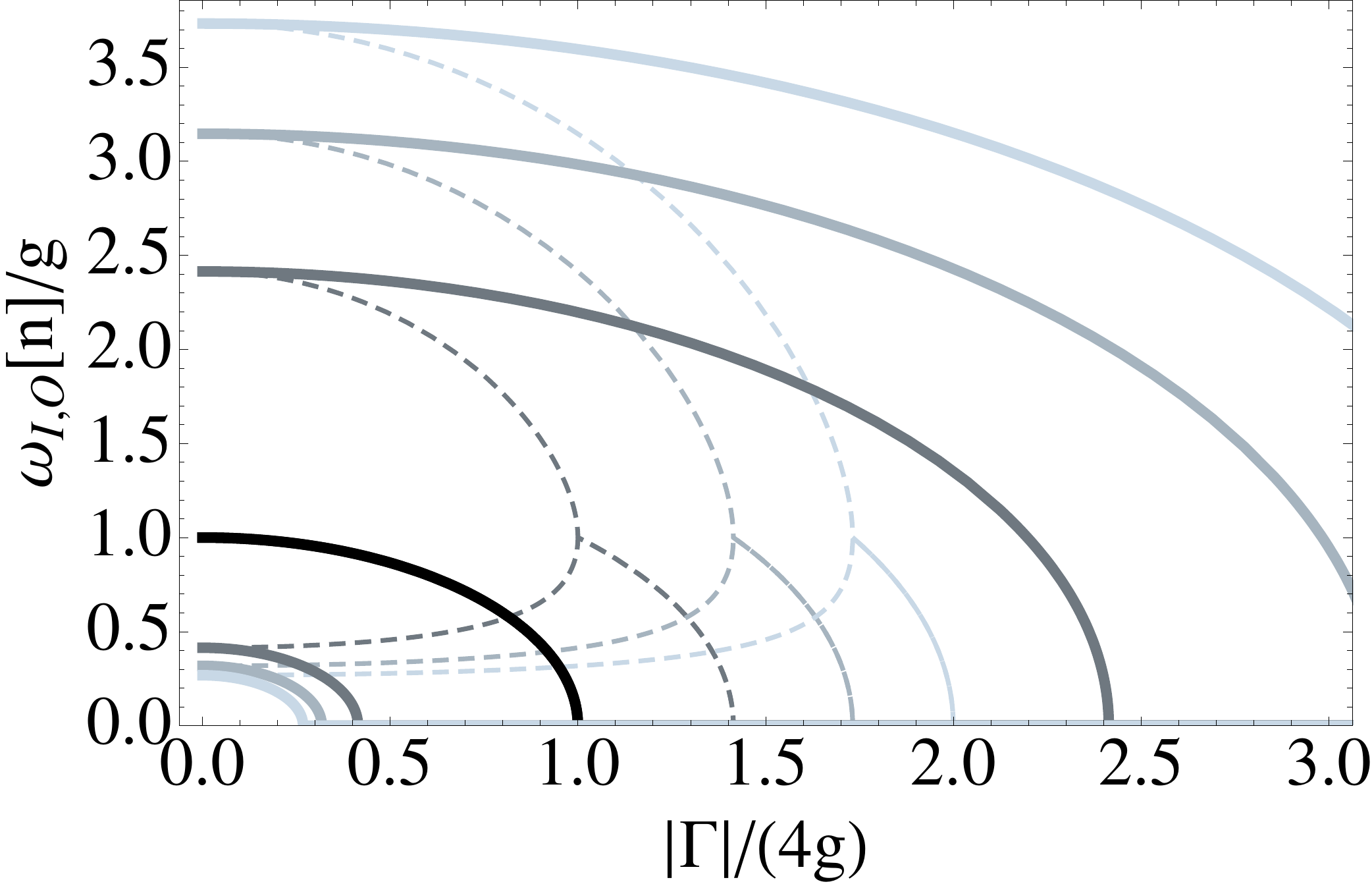}
  \caption{Transition energies (positions of the spectral peaks) as a
    function of decoherence $|\Gamma|/(4g)$ from rungs $n=1,2,3,4$
    (from dark to light lines) for two cases: spontaneous emission
    ($\Gamma=\gamma_a-\gamma_\sigma$,
    Eq.~(\ref{eq:ThuMar25145011WET2010}), thin dashed lines) and in
    the presence of pumping ($\Gamma=\Gamma_\sigma-\gamma_\phi$,
    Eq.~(\ref{eq:SunJul4155121CEST2010}) thick lines). The line
    starting at $\omega=g$ corresponds to the linear regime or
    first-rung-to-vacuum transition ($\Re(R_0)$, thick black line) and
    is common to both cases. The curves starting above (resp. below) 1
    correspond to the \emph{outer} (\emph{inner}) transitions. }
  \label{fig:ThuMar25163543WET2010}
\end{figure}

Let us recall the expression of the Jaynes-Cummings quadruplets
positions in the spontaneous regime, in order to appreciate better the
features brought by the incoherent pump~\cite{delvalle09a}:
\begin{multline}
  \label{eq:ThuMar25145011WET2010}
  \Omega_{\mathrm{O},\mathrm{I}}[n]=\Re\Big[\sqrt{g^2(n+1)-\left(\frac{\gamma_a-\gamma_\sigma}{4}\right)^2}\\
  \pm
  \sqrt{g^2n-\left(\frac{\gamma_a-\gamma_\sigma}{4}\right)^2}\Big]\,.
\end{multline}
These branches are plotted as thin dashed lines in
Fig.~\ref{fig:ThuMar25163543WET2010} (only the positive frequency
ones) as a function of their decoherence rate
$\Gamma=\gamma_a-\gamma_\sigma$. They can be exactly mapped with
transitions between rungs in the dissipative Jaynes-Cummings ladder,
exhibiting three regimes depending on their splittings: (a) at low
decoherence, both rungs---initial and final---are split (in the
strong coupling regime) and we have four peaks (two shown in the
figure). (b) When the lower rung enters weak coupling and recovers the
bare modes, only two transitions (one in the figure) can be seen from
the upper, still split, rung. (c) At large decoherence, none of the
rungs are split and all transitions are at the origin.

In the same Fig.~\ref{fig:ThuMar25163543WET2010}, we have plotted with
thick lines the positions in the presence of pumping,
Eq.~(\ref{eq:SunJul4155121CEST2010}), as a function of the
corresponding decoherence rate, $\Gamma=P_\sigma+\gamma_\sigma$.  The
positions, paired around the centre (still only the positive half is
shown), also depend on the rung.  In the linear regime, $n=0$, these
are simply the two Rabi transitions $\pm \Re(R_0)$, with the same
structure as two coupled harmonic oscillators or the spontaneous
regime that we have just described. However, for $n>0$ outer and inner
peaks have an interesting nontrivial behaviour. The inner peaks close
at low $\Gamma$, before the Rabi transitions do, forming, in the
lasing or classical regime ($n_a\gg 1$), the central peak of the
Mollow triplet. The outer peaks close much after the corresponding
Rabi peaks do, grouping to form the two side bands of the Mollow
triplet with
$\omega_\mathrm{O}[n]\approx\Re\sqrt{4ng^2-\Big(\Gamma_\sigma/4\Big)^2}$
(see Fig.~\ref{fig:FriApr1105117CEST2011}).

In the presence of pumping, it is no longer possible to map exactly
the position of the peaks composing the spectra with transitions from
the Jaynes-Cummings ladder. The effect of the incoherent pump, before
being so strong to close the dressed state splitting of the
Jaynes-Cummings rungs, is to make it homogeneous throughout the
ladder. Then, the inner peaks do not close because the rungs enter the
weak coupling regime but rather because the transitions coincide with
the cavity frequency. In any case, the Jaynes-Cummings structure is
very much distorted and the transitions between its rungs very much
mixed, in such a way that it is no longer possible to reconstruct the
ladder. Furthermore, in our derivation of
Eq.~(\ref{eq:ThuMar25104706WET2010}) we have neglected the interplay
between the cavity and emitter dynamics (by assuming them to have very
different time scales) and, therefore,
Fig.~\ref{fig:FriApr1105117CEST2011} shows the isolated effect of the
incoherent pump in contrast with the complex features we observed in
Fig.~\ref{fig:FriApr1111520CEST2011}.



\subsection{Elastic scattering $E^c$}
\label{sec:33}

It is possible to give the general expression (with detuning,
dephasing and decay) for the elastic scattering contribution to the
spectra, for both emitter and cavity emission:

\begin{subequations}
  \label{eq:TueMar29160848CEST2011}
  \begin{align}
    &\Re(E^\sigma)=\frac{4g}{\Gamma_\sigma+\gamma_\phi} \sum_{n=0}^{\infty}\\
    &\frac{\gamma_\sigma(\kappa_a+2\Gamma_\sigma) \sqrt{1+n} \mathrm{q}_\mathrm{i}[n+1]+P_\sigma(\kappa_a-2\Gamma_\sigma) \sqrt{n}\mathrm{q}_\mathrm{i}[n]}{\kappa_a^2 \Big[ 1+ (\frac{2\Delta}{\Gamma_\sigma+\gamma_\phi})^2\Big]+ 4\Gamma_\sigma^2+4\Gamma_\sigma \kappa_a(2n+1)}\,, \nonumber\\
    &\Re(E^a)=\sum_{n=0}^{\infty}\frac{2}{\kappa_a^2 \Big[ 1+ (\frac{2\Delta}{\Gamma_\sigma+\gamma_\phi})^2\Big]+ 4\Gamma_\sigma^2+4\Gamma_\sigma \kappa_a(2n+1)}\nonumber\\
    &\times\Big\{\gamma_\sigma (\kappa_a(4n+1)+2\Gamma_\sigma)(1+n)\mathrm{T}[n+1]\nonumber\\
    &+P_\sigma (\kappa_a(4n+3)+2\Gamma_\sigma)n \mathrm{T}[n]
    \Big\}\,.
  \end{align}
\end{subequations}
Note that $\Re(E^a)>0$ in all cases, while $\Re(E^\sigma)$ can be
negative, as we pointed out in the discussion of
Fig.~\ref{fig:MonMay16180915CEST2011}. The
expressions~(\ref{eq:TueMar29160848CEST2011}) support our qualitative
discussions. In the cavity emission, the scattering peak is simply the
direct emission of photons through the cavity mode and therefore, is
always positive.  For the emitter, however, these coherent cavity
photons must ``convert'' into material excitations before being
emitted, which implies the possibility of interferences that can be
positive or negative. This change of sign is even more clear when
$\gamma_\sigma=0$ where $E^\sigma<0$ if $P_\sigma<\sqrt{2}g$. At low
pumping, this carves a ``hole'' in the spectra pinned at the cavity
frequency. For $P_\sigma>\sqrt{2}g$, the delta peak is positive.  In
the case of the cavity, most of the emission comes from it, and not
the de-excitation of the dressed states, that is, from the photons
that undergo an efficient interaction with the emitter.

%

When the two-level system is not sufficiently populated/inverted (at
low $P_\sigma$), the cavity has a smaller elastic scattering
contribution. The emitter sees an interference hole being carved in
its spectrum, reminiscent of a Fano resonance. The emitter represents
an efficient pumping for the cavity in a linear way. As we already
mentioned, the cavity is sucking the cavity photons (at $\omega_a$)
out of the emitter. Although at very low pumping ($P_\sigma \approx
\gamma_a$) these approximations are not valid (a delta function weighted
negatively means a negative spectrum) and do not provide a
quantitative agreement with the numerical results, they are
interesting to understand the qualitative features that are obtained
numerically.

When the two-level system is saturated in the self-quenching regime
(at high $P_\sigma$), it is mainly in its excited state. The cavity
spectrum tends towards the bare cavity emission (thermal regime). In
the exact solution, this is a Lorentzian with the cavity linewidth
$\gamma_a$. In our approximated solution, the cavity bare emission is
a delta function and the two-level system emission loses completely
the elastic scattering component because there is weak interaction
with the cavity which provides it.

In the lasing region, $1<P_\sigma<20g$, the triplet appears in both
channels of emission: cavity and emitter. The emitter grows some small
positive elastic scattering component on top of the triplet, as a
collateral effect from the strong interaction with the cavity
(proportional to $\mathrm{q}_\mathrm{i}[n]$).

\subsection{Semiclassical approximation}
\label{sec:34}

\begin{figure}[t]
  \centering
  \includegraphics[width=\linewidth]{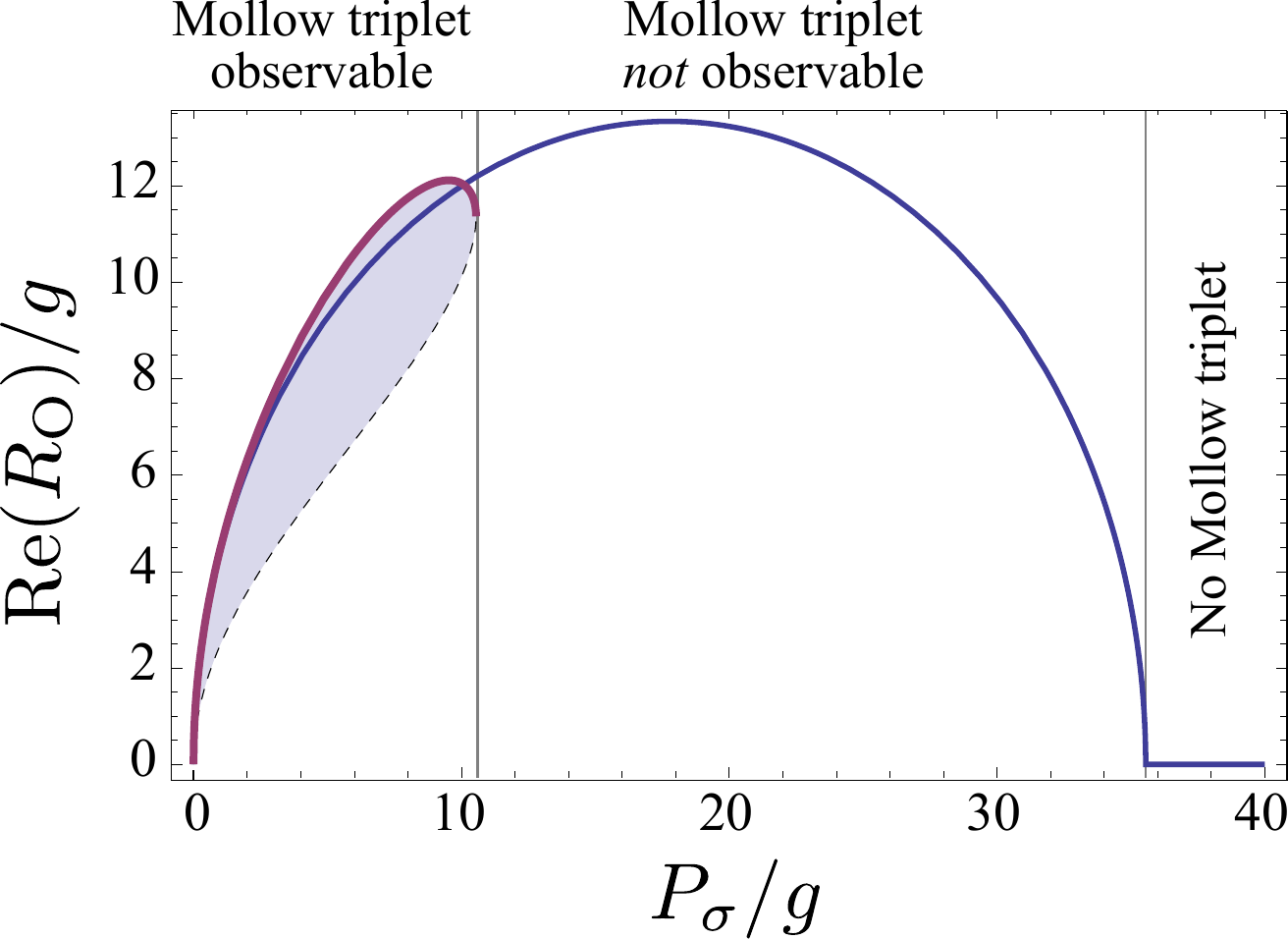}
  \caption{(Color online) Comparison between the position of the side
    peaks ($\Re(R_\mathrm{O})$, in solid blue) and those observed in
    the emitter spectrum ($\omega_\mathrm{obs}$, in solid purple,
    above the filling), as a function of pumping. In dashed is shown
    the neck of the side peak so that the filled area delimits
    approximately its half-width.}
  \label{fig:FriJul2121842MSD2010}
\end{figure}

In the lasing regime, where the $n$ and $n+1$ rungs that are close to
$n_a$ (having $n_a\gg 1$) have similar splittings and the cavity field
becomes Poissonian (coherent), simple expressions---in regard to the
complexity of the underlying machinery to obtain them---arise for the
spectra of emission~\cite{delvalle10d}. The variable $n$ becomes
continuous as compared to the large intensity $n_a$ and, given that
the distribution is Poissonian (peaked at the mean value), we can
consider only the case $n=n_a$ in
Eq.~(\ref{eq:WedMar24190127WET2010}). The integral over the
distribution $\mathrm{T}[n]$ simplifies to 1. In this regime, the
inner peaks have collapsed into the centre while the outer remain
splitted. Substituting $n_a$ (and $n_\sigma$ for the normalization)
from Eq.~(\ref{eq:WedMar24190127WET2010}), we obtain the final
expression for the emitter spectrum:
\begin{widetext}
 \begin{multline}
   \label{eq:MonJul5185141CEST2010}
   S_\sigma(\omega)=\Big(\frac{2P_\sigma}{\Gamma_\sigma+\gamma_\phi+\kappa_\sigma}-\frac{\Gamma_\sigma}{\kappa_\sigma}\Big)
   \delta(\omega)+\frac{1}{2\pi}\frac{\frac{\Gamma_\sigma+\gamma_\phi}{2}}{\big(\frac{\Gamma_\sigma+\gamma_\phi}{2}\big)^2+\omega^2}\\
   +\frac{1}{(\Gamma_\sigma+\gamma_\phi+\kappa_\sigma)[(-2P_\sigma+\Gamma_\sigma)^2\kappa_\sigma^2+ ( (3\Gamma_\sigma+\gamma_\phi)^2+4(\Gamma_\sigma-2P_\sigma)\kappa_\sigma  )\omega^2+4\omega^4 ]}\Big\{-4P_\sigma^2
   \kappa_\sigma(3\Gamma_\sigma+\gamma_\phi+\kappa_\sigma)\\+2P_\sigma
   \Gamma_\sigma [3\Gamma_\sigma^2+4
   \Gamma_\sigma(\gamma_\phi+2\kappa_\sigma)+(\gamma_\phi+\kappa_\sigma)(\gamma_\phi+3\kappa_\sigma)]
   + 4P_\sigma \kappa_\sigma
   \omega^2-(\Gamma_\sigma+\gamma_\phi+\kappa_\sigma)[\Gamma_\sigma^2(3\Gamma_\sigma+\gamma_\phi+2\kappa_\sigma)+(\Gamma_\sigma-\gamma_\phi)\omega^2]\Big\}\,.
 \end{multline}
\end{widetext}
Similarly to the coherent excitation case, it is composed of an
elastic scattering term (delta peak), a central peak (a Lorentzian peak
with FWHM $\Gamma_\sigma+\gamma_\phi$) and two side bands. When
splitted, these have a FWHM $(3\Gamma_\sigma+\gamma_\phi)/2$ and
positions given by $\omega_\mathrm{O}[n_a]\approx \Re(R_\mathrm{O})$ where
\begin{equation}
  \label{eq:FriApr1174554CEST2011}
  R_\mathrm{O}=\sqrt{\frac{(2P_\sigma-\Gamma_\sigma)\kappa_\sigma}{2}-\big(\frac{\Gamma_\sigma+\gamma_\phi}{4} \big)^2}\,.
\end{equation}
This is the Mollow splitting in the case of incoherent excitation,
analogous to $R_\mathrm{L}$ in Eq.~(\ref{eq:FriApr1183145CEST2011}).
It is plotted in Fig.~\ref{fig:FriJul2121842MSD2010}. Contrary to the
laser excitation, this splitting can now close due to the pumping
intensity $P_\sigma$. In Fig.~\ref{fig:FriJul2121842MSD2010}, we show
the domains where the Mollow triplet is clearly resolved. This
naturally requires that the system is able to enter the regime of
lasing in strong-coupling, which starts at figures of about
$\gamma_a/g\approx0.1$. This is the case shown in the Figure, where we
compare the Mollow splitting, Eq.~(\ref{eq:FriApr1183145CEST2011}),
with the observed splitting, represented as the shaded area which is
delimited by the maximum (upper boundary, solid) and the neck (lower
boundary, dashed) of a side peak. When the Mollow splitting is
maximum, decoherence has however broadened so much the satellite peaks
that no triplet is observable anymore.

\begin{figure}[tb]
  \centering
  \includegraphics[width=.85\linewidth]{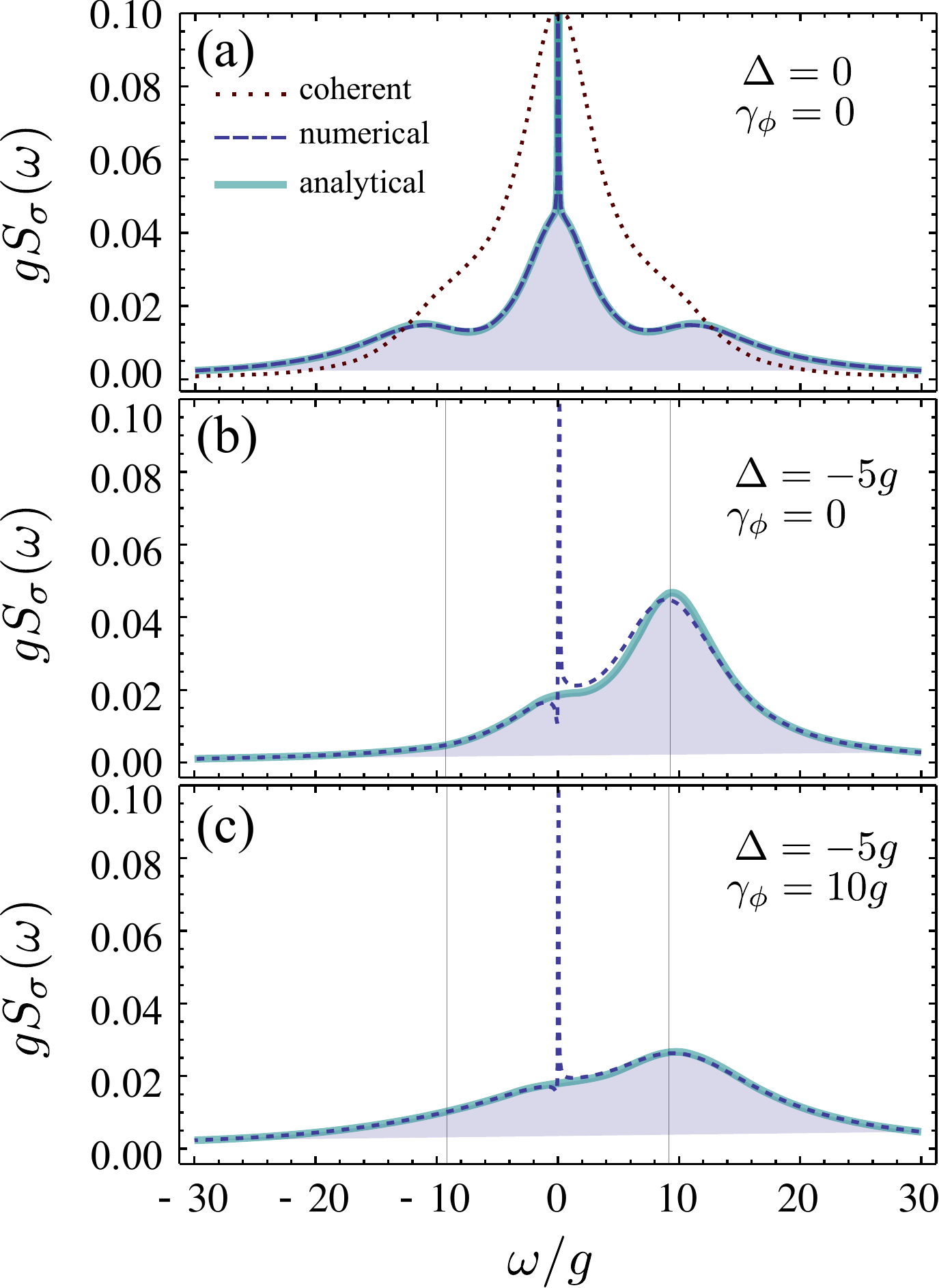}
  \caption{(Colour online) Mollow triplet in the direct PL spectra
    from the emitter, $S_\sigma(\omega)$ for $\gamma_a=0.1g$, $P_a=0$,
    $\gamma_\sigma=0.00334$ and $P_\sigma=7g$ ($n_a\approx 11.6$ and
    $n_\sigma\approx 0.53$), for various configurations of detuning
    and dephasing. The exact numerical results are plotted with a
    dark-blue dashed line, overlapping almost exactly our analytical
    formula, in light-blue solid. The coherent scattering peak is
    featured in the analytical solution in~(a) while only the
    incoherent spectrum is shown in (b) and~(c). The dotted line in
    (a) represents the Mollow triplet obtained under coherent
    excitation, with a laser intensity equivalent to the cavity
    occupation in the incoherent case, that is, for
    $\Omega_\mathrm{L}=\sqrt{n_a}g$ and also an equivalent emitter
    broadening, $\gamma_\sigma\rightarrow \gamma_\sigma+P_\sigma$. The
    two types of Mollow lineshapes are clearly different even though
    their peak positions and broadening are equal. Also, as compared
    to the coherent excitation case, the Mollow triplet under
    incoherent pumping is a resonant structure that becomes strongly
    asymmetric with detuning.}
  \label{fig:WedMay4122917CEST2011}
\end{figure}

\begin{figure}
  \centering
  \includegraphics[width=\linewidth]{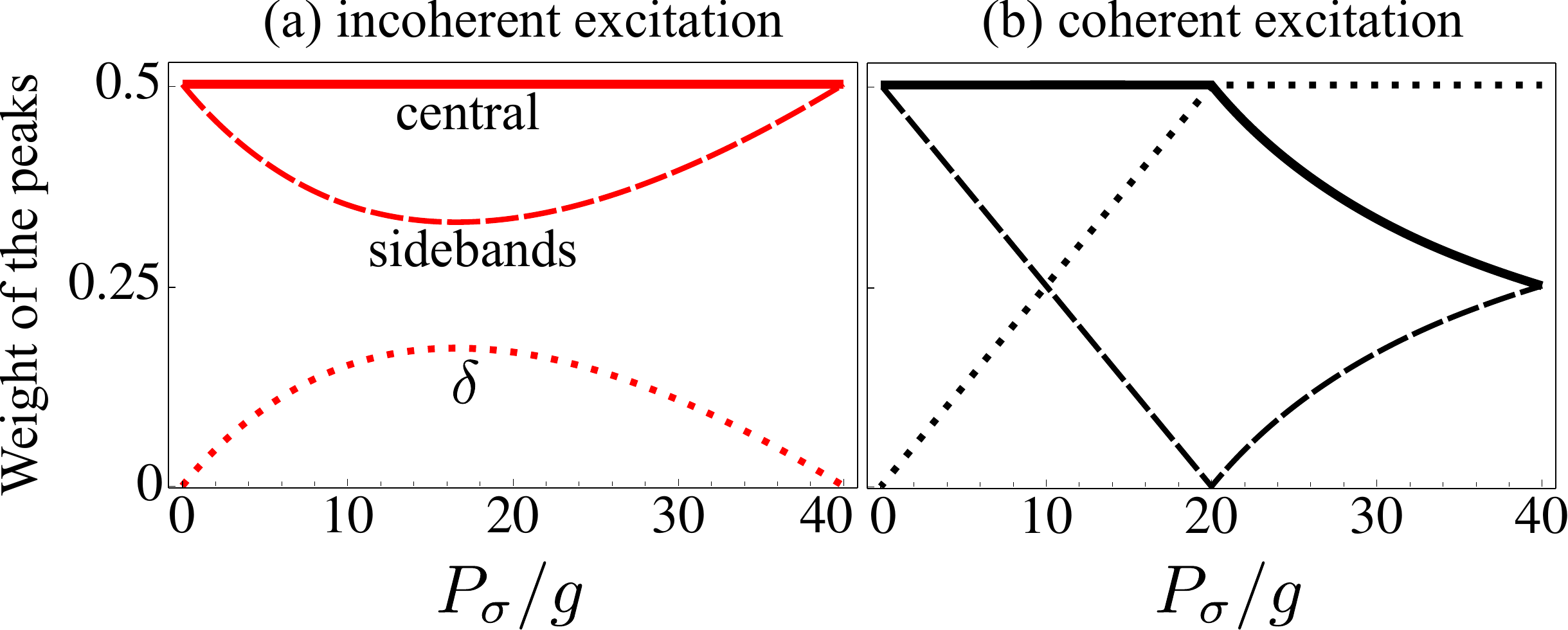}
  \caption{Contributions to the Mollow triplet under (a) incoherent
    and (b) coherent excitation for the three types of peaks: the
    central Lorentzian peak (constant at $1/2$, solid line), the side
    bands (dashed) and the elastic scattering peak (dotted). The
    parameters are chosen so that the two Mollow triplets can be
    compared on equal grounds, as explained in the text, for
    $\gamma_a=0.1g$, $P_a=0$, $\gamma_\sigma=\gamma_\phi=0$ and
    $\Delta=0$.}
  \label{fig:FriApr1120554CEST2011}
\end{figure}

Applying the same procedure, for the cavity emission we have
$S_a(\omega)=\delta(\omega)$, that is, a purely elastic spectrum with
negligible linewidth. A more accurate approximation to the FWHM of the
elastic peaks in this regime, which reproduces the typical lasing
cavity line narrowing, is given by the broadening~$\gamma_\mathrm{L}$:
\begin{equation}
  \label{eq:TueMay24012603CEST2011}
  \gamma_\mathrm{L}=2g^2\gamma_a/P_\sigma^2
\end{equation}
as derived by Poddubny \emph{et al.}~\cite{poddubny10a}.

In Fig.~\ref{fig:WedMay4122917CEST2011}, we show the excellent
agreement between the analytical approximation (in dashed lines) and
the exact numerical computation (solid lines). In
Fig.~\ref{fig:WedMay4122917CEST2011}(a), the case of resonance---the
one of most interest---is where the Mollow triplet is best
observed. Its analytical expression is given by
Eq.~(\ref{eq:MonJul5185141CEST2010}), where we also include the
scattering peak, (both as a result of the numerical procedure and
approximated by Eq.~(\ref{eq:TueMay24012603CEST2011})) seen as a very
sharp line, which we have truncated in the plot, as it extends more
than one hundred times higher than is shown. In
Fig.~\ref{fig:WedMay4122917CEST2011}(a), we also provide further
evidence that the Mollow triplet formed under incoherent pumping is of
a different nature than that formed under coherent
excitation~\cite{delvalle10d}, by superimposing the coherent
excitation Mollow triplet (dotted line). We take
$\Omega_\mathrm{L}\rightarrow\sqrt{n_a}g$ and
$\gamma_\sigma\rightarrow\gamma_\sigma+P_\sigma$ to compare both
expressions on equal grounds (both lineshapes remain dissimilar even
when parameters are left completely free). This substitution makes
both types of triplet share the same position and broadening for their
three peaks. However their relative weight is different, as shown in
Fig.~\ref{fig:FriApr1120554CEST2011}. These strong qualitative
departures result in the striking differences between the final
spectra, although the peaks have identical characteristic if taken in
isolation.

\begin{figure}[t]
  \centering
  \includegraphics[width=\linewidth]{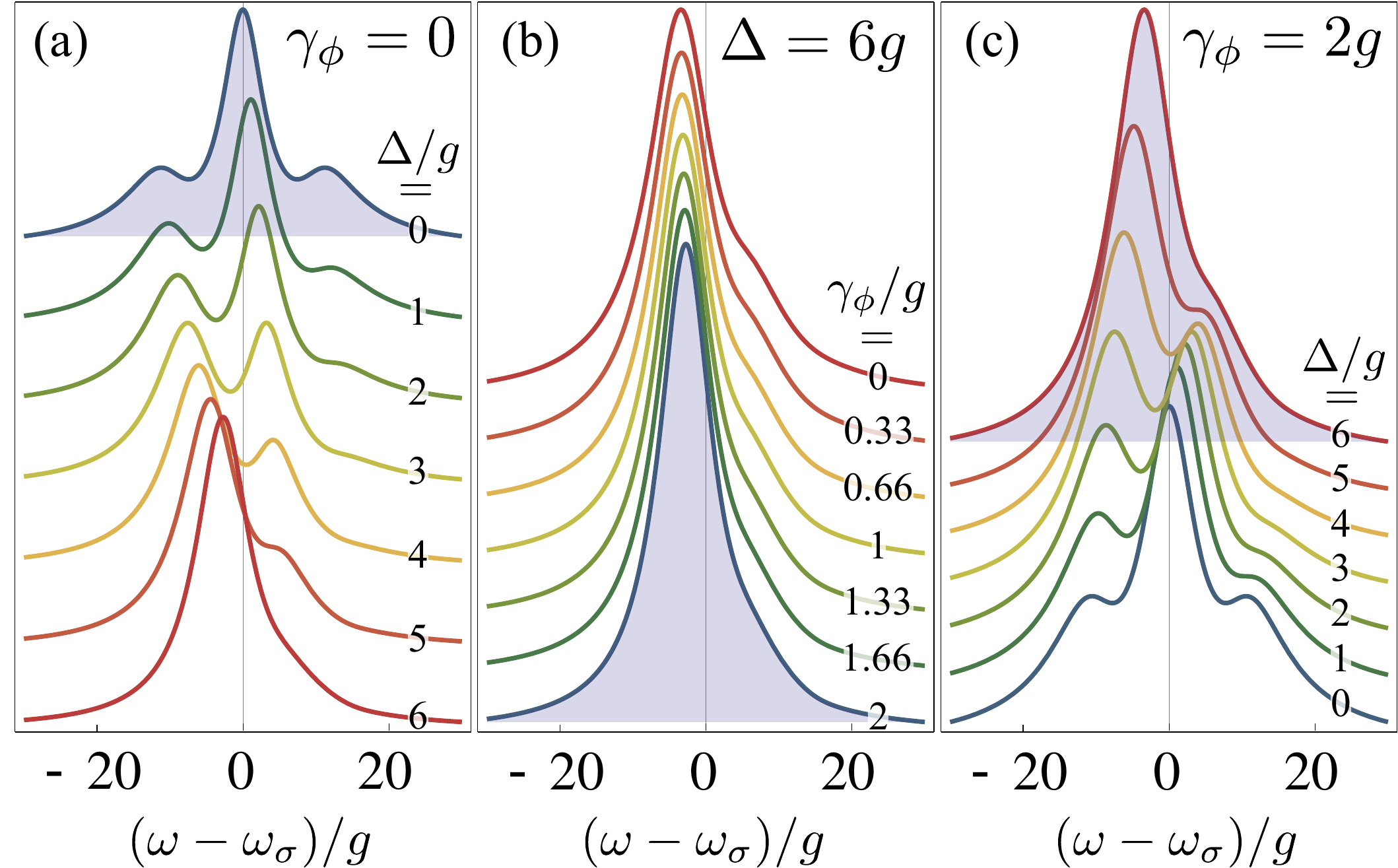}
  \caption{Counterpart of Fig.~\ref{fig:WedMay4170658CEST2011} for the
    Mollow triplet under incoherent excitation. In contrast to the
    coherent excitation case, detuning alone breaks the Mollow
    triplet. Parameters are $\gamma_a/g=0.1$, $\gamma_\sigma/g=0.01$,
    $P_a=0$ and~$P_\sigma/g=7$.}
  \label{fig:ThuMay19144815CEST2011}
\end{figure}

When the system is not at resonance, in sharp contrast with the
conventional Mollow triplet that retained its qualitative features
(cf.~Fig.~\ref{fig:WedMay4170658CEST2011}), the Mollow triplet under
incoherent pumping becomes strongly asymmetric, as it recovers the
scenario of an anticrossing of two modes. The out-of resonance case is
studied in details in Fig.~\ref{fig:ThuMay19144815CEST2011}. Whereas
both detuning and dephasing were needed to break the symmetry of the
conventional Mollow triplet, the one formed under incoherent pumping
is lost by detuning alone, pumping playing already the role of
dephasing. Dephasing has otherwise the expected effect of smearing out
and broadening the spectral features.  Analytical results can also be
given for the non-resonant case when~$\Delta\neq0$, but, as for the
conventional Mollow, they are too long to be reasonably written down.
We plot it on top of the numerical solution in
Figs.~\ref{fig:WedMay4122917CEST2011}(b) and~(c) where one can see the
semiclassical approximation is excellent there as well.  


\section{Summary and Outlooks}
\label{sec:4}

We investigated the steady states of the Jaynes--Cummings Hamiltonian
established under the interplay of decay and incoherent pumping (of
the emitter), including pure dephasing and detuning for wider
generality. This is the simplest and most fundamental realization of a
fully quantized system, realized with atoms, quantum dots or
superconducting qubits in a cavity. We identify five regimes through
semi-analytical and approximated solutions, all confronted to exact
numerical solutions. These are: $i$ the linear quantum regime, $ii$
the nonlinear quantum regime, $iii$ the nonlinear classical regime, or
lasing regime, $iv$ the self-quenching regime and $v$ the thermal
regime. We provided closed-form analytical expressions that account
for most of theses regimes and a simple numerical procedure (solving a
set of two coupled nonlinear equations) that afford an excellent
description over the entire range of excitation and all the five
regimes that we have outlined. This also allows a transparent reading
of the physics involved, namely, the first regime involves the lowest
rung of the Jaynes--Cummings ladder only and corresponds to
spontaneous emission. A linear-model (two coupled oscillators) and a
truncated Jaynes--Cummings model offer two complementary views of this
regime. The quantum regime is the one where the system starts to climb
the ladder, requiring a full record of all the quantum correlators
involved. This is therefore the most complicated regime from the point
of view of the amount of information required to describe it, since no
good approximation can synthesize the dynamics of a few quanta. In
very good systems, this manifests spectrally in a complex structure of
peaks at anharmonic frequencies. As pumping is further increased,
lasing ensues which brings back the system to a simple level of
description in a semiclassical approximation. A single narrowing line
in the cavity or a variation of the Mollow triplet for the emitter
describe accurately the system. In the fully quantized picture, lasing
appears as a condensate of dressed states, with a complex pattern that
however does not manifest in the observable, showing a breakdown of
the quantized picture in favour of a classical description.

The Jaynes--Cummings model that arose as a challenge for full-field
quantization~\cite{stroud04a} remains to this day a proficient source
of theoretical and experimental investigations into the quantum
realm. The realization in the laboratory of the nontrivial quantum
physics that it covers will shed light on foremost issues such as the
quantum to classical crossover, emergence of coherence, lasing and
quantum nonlinearities.  A solid understanding of the various regimes
it realizes may also lead to useful devices and applications, from
single-photon sources to low-threshold lasers.

\section*{Acknowledgements}

We thank Paul Gartner for comments and discussions.  We acknowledge
support from the Newton International Fellowship scheme, Emmy Noether
project HA 5593/1-1 funded by the German Research Foundation (DFG) and
EU FP-7 Marie Curie Initiative `SQOD'.

\appendix

\section{Quantum regression formula for coherent excitation}
\label{ap:qrf}

Two-time correlators of the type $\mean{\ud{\sigma}(0)\sigma(\tau)}$
can be computed by means of the quantum regression
formula~\cite{delvalle09a}. Once we find the set of
operators~$C_{\{m,n\}}={\ud{\sigma}}^m\sigma^n$ (with~$m$, $n
\in\{0,1\}$) and the regression matrix~$M_{\substack{mn\\m'n'}}$ that
satisfy~$\Tr{[C_{\{m,n\}}\mathcal{L}O]}=\sum_{\{m'n'\}}M_{\substack{mn\\m'n'}}\Tr{[C_{\{m',n'\}}O]}$
for a general operator $O$, then the equations of motion for the
two-time correlators ($\tau\geq 0$) in the steady state ($t=0$) read:
\begin{equation}
  \label{eq:TueMay5174356GMT2009}
  \partial_\tau\mean{O(0)C_{\{m,n\}}(\tau)}=\sum_{\{m'n'\}} M_{\substack{mn\\m'n'}}\mean{O(0)C_{\{m',n'\}}(\tau)}\,.
\end{equation}
The corresponding regression matrix~$M$ is given, in the case of a
coherent and classical excitation of the emitter, as explained in
section~\ref{sec:1}, by:
\begin{subequations}
  \label{eq:TueDec23114907CET2008}
  \begin{align}
    &M_{\substack{mn\\mn}}=i\Delta_L(m-n)-\frac{\gamma_\sigma}2(m+n)-\frac{\gamma_\phi}2(m-n)^2\,,\\
    &M_{\substack{mn\\1-m,n}}=i\Omega_\mathrm{L}[m+2n(1-m)]\,,\\
    &M_{\substack{mn\\m,1-n}}=-i\Omega_\mathrm{L}[n+2m(1-n)]\,.
  \end{align}
\end{subequations}
and zero everywhere else. We concentrate on computing~$g_1^{(1)}$,
which corresponds to setting~$O=\ud{\sigma}$ and
having~$\{m,n\}=\{0,1\}$ in Eq.~(\ref{eq:TueMay5174356GMT2009}). We
obtain the following differential equation:
\begin{equation}
  \label{eq:TueDec23115950CET2008}
  \partial_\tau\mathbf{v}_L(\tau)=-\mathbf{M}_L\mathbf{v}_L(\tau)+\mathbf{A}_L\mean{\ud{\sigma}}\,,
\end{equation}
where
\begin{equation}
  \label{eq:TueDec23120004CET2008}
  \mathbf{v}_L(\tau)=
  \begin{pmatrix}
    \langle\ud{\sigma}(0)\sigma(\tau)\rangle\\
    \langle\ud{\sigma}(0)\ud{\sigma}(\tau)\rangle\\
    \langle\ud{\sigma}(0)\ud{\sigma}\sigma(\tau)\rangle
\end{pmatrix}
\,,\quad
\mathbf{A}_L=i\Omega_\mathrm{L}
  \begin{pmatrix}
    -1\\
    1\\
    0
\end{pmatrix}
\end{equation}
and
\begin{equation}
  \label{eq:TueDec23120838CET2008}
  \mathbf{M}_L=
    \begin{pmatrix}
      -i\Delta_L+\frac{\gamma_\sigma+\gamma_\phi}{2} & 0 &-2i\Omega_\mathrm{L}\\
      0 &i\Delta_L+\frac{\gamma_\sigma+\gamma_\phi}{2} &2i\Omega_\mathrm{L}\\
      -i\Omega_\mathrm{L}&i\Omega_\mathrm{L}&\gamma_\sigma
\end{pmatrix}\,.
\end{equation}
The solution is
\begin{equation}
  \label{eq:TueNov30133208CET2010}
  \mathbf{v}_L(\tau)=e^{-\mathbf{M}_L\tau}[\mathbf{v}_L(0)-\mathbf{M}_L^{-1}\mathbf{A}_L\mean{\ud{\sigma}}]+\mathbf{M}_L^{-1}\mathbf{A}_L\mean{\ud{\sigma}}\,,
\end{equation}
in terms of the steady steady values $\mean{\ud{\sigma}}$ and
$\mathbf{v}_L(0)=[\mean{\ud{\sigma}\sigma},0,0 ]$ (initial condition
of Eq.~(\ref{eq:TueDec23120004CET2008})). We also need, therefore, to
compute the steady state of the system,
\begin{equation}
  \label{eq:TueNov30133159CET2010}
  \mathbf{u}_L=
  \begin{pmatrix}
    \langle\sigma\rangle\\
    \langle\ud{\sigma}\rangle\\
    \langle\ud{\sigma}\sigma\rangle
\end{pmatrix}\,,
\end{equation}
which can be done, again, by means of the general formula in
Eq.~(\ref{eq:TueMay5174356GMT2009}). This time we take~$O=1$ and
~$\{m,n\}=\{0,1\}$, $\{1,0\}$, $\{1,1\}$ and find the equation:
\begin{equation}
  \label{eq:TueNov30133456CET2010}
  \partial_t \mathbf{u}_L=0=-\mathbf{M}_L\mathbf{u}_L+\mathbf{A}_L\,.
\end{equation}
The solution is $\mathbf{u}_L=\mathbf{M}_L^{-1}\mathbf{A}_L$. It allows
us to simplify Eq.~(\ref{eq:TueNov30133208CET2010}) further as
\begin{equation}
  \label{eq:TueNov30133208CET2010b}
  \mathbf{v}_L(\tau)=e^{-\mathbf{M}_L\tau}[\mathbf{v}_L(0)- \mathbf{u}_L\mean{\ud{\sigma}}]+\mathbf{u}_L\mean{\ud{\sigma}}\,,
\end{equation}
The final explicit solutions for the mean values and correlators of
interest are presented in the main text.

\section{Derivation of the field correlators and density matrix}
\label{sec:ThuJul1223020MSD2010}

The equations of motion of the correlators in
Eq.~(\ref{eq:MoMar22172729WET2010}) can be derived from the master
equation~(\ref{eq:ThuOct18162449UTC2007}) by simply applying the
general formula $\mean{O}=\Tr{(O\rho)}$ or from the rules given by the
quantum regression formula~\cite{delvalle09a}:
\begin{subequations}
  \label{eq:MoMar22174241WET2010}
  \begin{align}
    &\partial_tN_a[n]=-n\Gamma_a N_a[n]+n^2P_aN_a[n-1]\\
    &+2g n N^i_{a\sigma}[n]\,,\nonumber\\
    &\partial_tN_\sigma[n]=-\big[\Gamma_a(n-1)+\Gamma_\sigma\big]N_\sigma[n]\\
    &+P_a(n-1)^2N_\sigma[n-1]+P_\sigma N_a[n-1]-2g N^i_{a\sigma}[n]\,,\nonumber\\
    &\partial_t N^i_{a\sigma}[n]=-\Big[\frac{\Gamma_a}{2}(2n-1)+\frac{\Gamma_\sigma+\gamma_\phi}{2}\Big]N^i_{a\sigma}[n]\\
    &+P_an(n-1)N^i_{a\sigma}[n-1]+\Delta N^r_{a\sigma}[n]\nonumber\\
    &+g(2N_\sigma[n+1]+nN_\sigma[n]-N_a[n])\nonumber\,,\\
    & \partial_t N^r_{a\sigma}[n]=-\Big[\frac{\Gamma_a}{2}(2n-1)+\frac{\Gamma_\sigma+\gamma_\phi}{2}\Big]N^r_{a\sigma}[n]\\
    &+P_an(n-1) N^r_{a\sigma}[n-1]-\Delta N^i_{a\sigma}[n]\,,\nonumber
  \end{align}
\end{subequations}
with $n\geq 1$ and also $N_a[0]=1$. In the steady state, with $P_a=0$,
one can further simplify these equations and write $N_{a\sigma}[n]$
and $N_\sigma[n]$ in terms of $N_a[n]$ to obtain
Eqs.~(\ref{eq:MoMar22180002WET2010}).

The master equation~(\ref{eq:ThuOct18162449UTC2007}) can also be
rewritten in terms of the density matrix elements in
Eq.~(\ref{eq:SunJul20203845CEST2008}) as:
\begin{subequations} 
  \label{eq:SunJul20204529CEST2008} 
\begin{align}
    &\partial_t\mathrm{p}_0[n]=-\big[(\gamma_a+P_a)n+P_a\big]\mathrm{p}_0[n]\\
    &+\gamma_a(n+1)\mathrm{p}_0[n+1]+P_an\mathrm{p}_0[n-1]\label{eq:SunNov23185347GMT2008}\nonumber\\
    &-P_\sigma\mathrm{p}_0[n]+\gamma_\sigma
    \mathrm{p}_1[n]-2g\sqrt{n}\mathrm{q}_\mathrm{i}[n]\,,\nonumber\\
    &\partial_t\mathrm{p}_1[n]=-\big[(\gamma_a+P_a)n+P_a\big]\mathrm{p}_1[n]\\
    &+\gamma_a(n+1)\mathrm{p}_1[n+1]+P_an\mathrm{p}_1[n-1]\nonumber\\
    &-\gamma_\sigma\mathrm{p}_1[n]+P_\sigma \mathrm{p}_0[n]+2g\sqrt{n+1}\mathrm{q}_\mathrm{i}[n+1]\,,\label{eq:SunNov23185436GMT2008}\nonumber\\
    &\partial_t\mathrm{q}_\mathrm{i}[n]=-\Big[(\gamma_a+P_a)n-\frac{\Gamma_a}{2}\Big]\mathrm{q}_\mathrm{i}[n]\\
    &+\gamma_a\sqrt{(n+1)n}\mathrm{q}_\mathrm{i}[n+1]+P_a\sqrt{(n-1)n}\mathrm{q}_\mathrm{i}[n-1]\nonumber\\
    &-\frac{\Gamma_\sigma+\gamma_\phi}{2}\mathrm{q}_\mathrm{i}[n]+g\sqrt{n}(\mathrm{p}_0[n]-\mathrm{p}_1[n-1])-\Delta \,\mathrm{q}_\mathrm{r}[n]\,,\label{eq:SunNov23185449GMT2008}\nonumber\\
    &\partial_t\mathrm{q}_\mathrm{r}[n]=-\Big[(\gamma_a+P_a)n-\frac{\Gamma_a}{2}\Big]\mathrm{q}_\mathrm{r}[n]\\
    &+\gamma_a\sqrt{(n+1)n}\mathrm{q}_\mathrm{r}[n+1]+P_a\sqrt{(n-1)n}\mathrm{q}_\mathrm{r}[n-1]\nonumber\\
    &-\frac{\Gamma_\sigma+\gamma_\phi}{2}\mathrm{q}_\mathrm{r}[n]+\Delta\,\mathrm{q}_\mathrm{i}[n]\,.  \nonumber
\end{align}
\end{subequations} 
At resonance, the real part of the coherence distribution,
$\mathrm{q}_\mathrm{r}$, gets decoupled and vanishes in the steady
state. As a result, only
Eqs.~(\ref{eq:SunNov23185347GMT2008})--(\ref{eq:SunNov23185449GMT2008})
need to be solved. When~$g$ vanishes, $\mathrm{q}_\mathrm{i}$ does not
couple the two modes anymore, and their statistics become thermal,
like in the boson case. Through the off-diagonal
elements~$\mathrm{q}_\mathrm{i}$, the photon density matrix can vary
between Poissonian, thermal (superpoissonian) and subpoissonian
distributions.

In order to solve these equations in the steady state, we neglect the
photonic dynamics (as explained in the main text) and further
substitute
$\mathrm{q}_\mathrm{r}[n]=[2\Delta/(\Gamma_\sigma+\gamma_\phi)]\mathrm{q}_\mathrm{i}[n]$
in the equation for $\mathrm{q}_\mathrm{i}$, and write everywhere
$\mathrm{p}_0[n]$ as $\mathrm{T}[n]-\mathrm{p}_1[n]$ and
$\mathrm{p}_1[n+1]$ as $\mathrm{T}[n+1]-\mathrm{p}_0[n+1]$, so that
$\mathrm{q}_\mathrm{r}[n]$, $\mathrm{p}_0[n]$, $\mathrm{p}_1[n+1]$ do
not appear explicitly in the remaining three equations. Then, the
equations read in matricial form:
\begin{equation}
  \label{eq:TueMar23171847WET2010}
\partial_t\mathbf{u}_{0}[n]=-\mathbf{M}_{0}[n]\mathbf{u}_{0}[n]+\mathbf{A}_0[n]
\end{equation}
with: 
\begin{multline} 
\label{eq:MoMar22202526WET2010}
  \mathbf{u}_{0}[n]= 
\begin{pmatrix}
    \mathrm{p}_0[n+1]\\
    \mathrm{p}_1[n]\\
    \mathrm{q}_\mathrm{i}[n+1] 
\end{pmatrix} \,, \quad  \mathbf{A}_0[n]= 
\begin{pmatrix}
    \gamma_\sigma\mathrm{T}[n+1]\\
    P_\sigma\mathrm{T}[n] \\ 0
\end{pmatrix} \,,\\
  \mathbf{M}_{0}[n]= 
\begin{pmatrix}
  \Gamma_\sigma&0&2g\sqrt{n+1}\\
  0&\Gamma_\sigma&-2g\sqrt{n+1}\\
  -g\sqrt{n+1}&g\sqrt{n+1}&\frac{\Gamma_\sigma+\gamma_\phi}{2}+\frac{2\Delta^2}{\Gamma_\sigma+\gamma_\phi}
\end{pmatrix}\,.
\end{multline} 
The solution in the steady state is
$\mathbf{u}_{0}[n]=(\mathbf{M}_{0}[n])^{-1}\mathbf{A}_0[n]$ which
gives the result of the text,
Eq.~(\ref{eq:MoMar22203849WET2010}). Note that this solution is only
exact in the case $\gamma_a=P_a=0$.

\section{Perturbative regime of interaction in the limit of weak coupling}
\label{sec:ThuMay26100821CEST2011}

Eq.~(\ref{eq:MoMar22173241WET2010}) can be solved exactly as a series
Taylor expansion on the pumping $P_\sigma$. For this, we rewrite
Eq.~(\ref{eq:MoMar22173241WET2010}) in terms of the fraction
$F[n]=N_a[n+1]/N_a[n]=n_a g^{(n+1)}/g^{(n)}$ (for $n>1$):
\begin{equation}
  \label{eq:ThuMay26194523CEST2011}
  F[n-1](F[n]+\frac{B_n}{A_n})=\frac{C_n}{A_n}
\end{equation}
where
\begin{subequations}
  \label{eq:ThuMay26194916CEST2011}
  \begin{align}
    A_n=&\frac{2\gamma_a}{\Gamma_\sigma+n\gamma_a}\,,\\
    B_n=&\frac{1}{C_\mathrm{eff}[n]}+\frac{n\gamma_a}{\Gamma_\sigma+(n-1)\gamma_a}-\frac{2P_\sigma}{\Gamma_\sigma+n\gamma_a}+1\,,\\
    C_n=&\frac{nP_\sigma}{\Gamma_\sigma+(n-1)\gamma_a}\,.
  \end{align}
\end{subequations}
All quantities can be expanded, or assumed to have a solution in the
case of $F[n]$, in power series of $P_\sigma$:
\begin{subequations}
  \label{eq:ThuMay26194916CEST2011}
  \begin{align}
    &\frac{B_n}{A_n}=\sum_{k=0}^{\infty}\alpha_{k}[n]
    P_\sigma^{k}\,,\quad \frac{C_n}{A_n}=\sum_{k=0}^{\infty}\beta_{k}[n] P_\sigma^{k}\,,\\
    &F[n]=\sum_{k=0}^{\infty}f_{k}[n] P_\sigma^{k}\,.
  \end{align}
\end{subequations}
A key feature of this expansion is that $\beta_0[n]=0$ and $f_0[n]=0$
for all $n$ (we recall the linear behaviour of $n_a$ at low
pumping). With this considerations,
Eq.~(\ref{eq:ThuMay26194523CEST2011}) now reads
\begin{equation}
  \label{eq:ThuMay26200230CEST2011}
  \sum_{q=1}^{\infty}\sum_{k=1}^{\infty}f_{q}[n-1](f_{k}[n]+\alpha_{k}[n])P_\sigma^{k+q}=\sum_{t=1}^{\infty}\beta_{t}[n]P_\sigma^{t}\,.
\end{equation}
We further change the sum index $q$, on the left hand side of the
equation, for $t=q+k$, so that we can get rid of it and of the pump
dependence:
\begin{equation}
  \label{eq:ThuMay26200645CEST2011}
  \sum_{k=1}^{t}f_{t-k}[n-1](f_{k}[n]+\alpha_{k}[n])P_\sigma^{k+q}=\beta_{t}[n]\,.
\end{equation}
The exact solution can be found exactly and recurrently as:
\begin{subequations}
  \label{eq:ThuMay26201409CEST2011}
  \begin{align}
    &f_{1}[n]=\frac{\beta_1[n+1]}{\alpha_0[n+1]}\,,\\
    &f_{2}[n]=\frac{\beta_2[n+1]-f_{1}[n](f_{1}[n+1]+\alpha_1[n+1])}{\alpha_0[n+1]}\,,\\
    &\hdots \quad \nonumber\\
    &f_{t}[n]=\frac{\beta_t[n+1]-f_{t-1}[n]\sum_{k=1}^{t-1}(f_k[n+1]+\alpha_k[n+1])}{\alpha_0[n+1]}\,.
   \end{align}
\end{subequations}
This method is only useful in practical terms when the effect of the
coupling is perturbative, that is, at very low pumping or in the weak
coupling regime, where only a few terms of the expansion are
needed. Otherwise, in order to reproduce non perturbative effects such
as the transition into lasing, the sum should be performed to all
orders of $P_\sigma$, which might not be practical numerically.

\section{Derivation of the two-time correlators}
\label{sec:SuSep26172348WEST2010}

We can obtain the elements $\rho^c_{k;l}(\tau)$ needed to compute the
two-time correlators in Eq.~(\ref{eq:SunNov16185703GMT2008}), in an
equivalent way as $\rho_{[k;l]}$, as they follow the same master
equation~\cite{elena_molmer96a}: $\partial_t
\rho_{[k;l]}=\sum_{i,j}\mathcal{M}_{\Big[\substack{k;l\\i;j}\Big]}\rho_{[i;j]}$. That
is, we can solve
\begin{equation} 
\label{eq:TueMar23151909WET2010} 
\partial_\tau \rho^c_{[k;l]}(t+\tau)=\sum_{i,j}\mathcal{M}_{\Big[\substack{k;l\\i;j}\Big]}
\rho^c_{[i;j]}(t+\tau)\,.  
\end{equation} 
The initial conditions are given in terms of the steady state density
matrix:
\begin{equation} 
  \label{eq:TueMar23161649WET2010} 
  \rho^c_{[k;l]}(0)=\mean{\ud{c}(\ket{l}\bra{k})(0)}=\sum_\alpha\rho_{[k;\alpha]}\bra{\alpha}\ud{c}\ket{l}\,.
\end{equation}

Let us be more specific by writing the formulas for the two
correlators of interest. For the emitter spectra, we have
$\bra{l_1,l_2}\sigma\ket{k_1,k_2}=\delta_{l_1,k_1}\delta_{l_2,0}\delta_{k_2,1}$,
which gives:
\begin{equation} 
  \label{eq:TueMar23152544WET2010}
  \langle\ud{\sigma}(0)\sigma(\tau)\rangle=\sum_{n=0}^{n_{max}}\rho^\sigma_{[n,1;n,0]}(\tau)\,.
\end{equation} 
For the cavity spectra, we have
$\bra{l_1,l_2}a\ket{k_1,k_2}=\sqrt{k_1}\delta_{l_2,k_2}\delta_{l_1,k_1-1}$,
which gives:
\begin{equation}
   \label{eq:TueMar23152544WET20102}
   \langle\ud{a}(0)a(\tau)\rangle=\sum_{n=0}^{n_{max}}\sum_{i=0,1}\sqrt{n}\rho^a_{[n,i;n-1,i]}(\tau)\,.
\end{equation} 
$\rho^c$ is obtained by solving the master
Eq.~(\ref{eq:TueMar23151909WET2010}) in both cases, but the initial
conditions that follow in each case from
Eq.~(\ref{eq:TueMar23161649WET2010}), are different. For the emitter
correlator, they are $\rho^\sigma_{[n,i;m,j]}(0)=
\rho_{[n,i;m,1]}\delta_{j,0}$, while for the photon,
$\rho^a_{[n,i;m,j]}(0)= \sqrt{m+1}\rho_{[n,i;m+1,j]}$.  In the same
way as when solving the steady state distributions, we write the
master equation for $\rho^c_{[n,i;m,j]}$ only for the elements that
will be different from zero during the evolution with $\tau$. We have
to include all the elements that are nonzero in the initial condition
plus those that are linked to them. One can check that the nonzero
elements are the same for the initial conditions of both the cavity
and the emitter correlators, those defined in
Eq.~(\ref{eq:TueMar23162915WET2010}). They follow the equations:
\begin{subequations}
  \label{eq:SunJul20204529CEST20082} 
  \begin{align} 
    &\partial_\tau
    S_0[n]=-\big[\frac{\gamma_a+P_a}{2}(2n-1)+P_a\big]S_0[n]\\
    &+\gamma_a\sqrt{n(n+1})S_0[n+1]+P_a\sqrt{n(n-1)}S_0[n-1]\nonumber\\
    &-P_\sigma
    S_0[n]+\gamma_\sigma S_1[n]+ig(\sqrt{n-1}V[n]-\sqrt{n}Q[n-1])\,,\nonumber\\
    &\partial_\tau
    S_1[n]=-\big[\frac{\gamma_a+P_a}{2}(2n-1)+P_a\big]S_1[n]\\
    &+\gamma_a\sqrt{n(n+1})S_1[n+1]+P_a\sqrt{n(n-1)}S_1[n-1]\nonumber\\
    &+P_\sigma S_0[n]-\gamma_\sigma S_1[n]-ig(\sqrt{n+1}V[n+1]-\sqrt{n}Q[n])\,,\nonumber\\
    &\partial_\tau Q[n]=-\Big[(\gamma_a+P_a)n+P_a\Big]Q[n]\\
    &+\gamma_a(n+1)Q[n+1]+P_anQ[n-1]\nonumber\\
    &-\Big(\frac{\Gamma_\sigma+\gamma_\phi}{2}-i\Delta\Big)Q[n]\nonumber\\
    &+ig(\sqrt{n}S_1[n]-\sqrt{n+1}S_0[n+1])\,,\nonumber\\
    &\partial_\tau V[n]=-\big[(\gamma_a+P_a)n-\gamma_a\big]V[n]\\
    &+\gamma_a\sqrt{(n-1)(n+1})V[n+1]+P_a\sqrt{n(n-2)}V[n-1]\nonumber\\
    &-\Big(\frac{\Gamma_\sigma+\gamma_\phi}{2}+i\Delta\Big)V[n]\nonumber\\
    &+ig(\sqrt{n-1}S_0[n]-\sqrt{n}S_1[n-1])\,.\nonumber
\end{align} 
\end{subequations} 
As explained in the main text, we solve these equations by neglecting
the very slow photonic dynamics. We then define the steady state and
slow varying function
\begin{equation}
  \label{eq:TueMar23172253WET2010}
  X[n]\equiv S_0[n](0)+S_1[n](0)\,,
\end{equation}
and substitute $S_0[n]=X[n]-S_1[n]$ and $S_1[n+1]=X[n+1]-S_0[n+1]$, so
that we can rewrite the equations in a matricial
form~(\ref{eq:TueMar23171847WET2010}):
\begin{equation}
  \label{eq:TueMar23172101WET2010}
  \partial_\tau\mathbf{u}_{1}[n](\tau)=-\mathbf{M}_{1}[n]\mathbf{u}_{1}[n](\tau)+\mathbf{A}_1[n]
\end{equation}
with:
\begin{multline}
  \label{eq:TueMar23172130WET2010}
  \mathbf{u}_1[n]=
  \begin{pmatrix}
    S_0[n+1]\\
    S_1[n]\\
    Q[n]\\
    V[n+1]
  \end{pmatrix}
  \,, \quad \mathbf{A}_1[n]=
  \begin{pmatrix}
    \gamma_\sigma X[n+1]\\
    P_\sigma X[n]\\
    0\\
    0
  \end{pmatrix}\,,\\
  \text{and\ }\mathbf{M}_{1}[n]=\\
  \begin{pmatrix}
    \Gamma_\sigma&0&ig\sqrt{n+1}&-ig\sqrt{n}\\
    0&\Gamma_\sigma&-ig\sqrt{n}&ig\sqrt{n+1}\\
    ig\sqrt{n+1}&-ig\sqrt{n}&\frac{\Gamma_\sigma+\gamma_\phi}{2}-i\Delta&0\\
    -ig\sqrt{n}&ig\sqrt{n+1}&0&\frac{\Gamma_\sigma+\gamma_\phi}{2}+i\Delta
  \end{pmatrix}\,.
\end{multline}

The solution in the steady state is
\begin{multline}
  \label{eq:WedMar24151034WET2010}
  \mathbf{u}_{1}[n](\tau)=e^{-\mathbf{M}_{1}[n]\tau}\Big(\mathbf{u}_{1}[n](0)-(\mathbf{M}_{1}[n])^{-1}\mathbf{A}_1[n]\Big)\\
  +(\mathbf{M}_{1}[n])^{-1}\mathbf{A}_1[n]\,.
\end{multline}

For the emitter correlator, the initial condition derived from
Eq.~(\ref{eq:TueMar23161649WET2010}) reads:
\begin{equation}
  \label{eq:TueMar23161517WET2010}
  \mathbf{u}_1[n](0)=
  \begin{pmatrix}
    \Big(\frac{2\Delta}{\Gamma_\sigma+\gamma_\phi}+i\Big) \mathrm{q}_\mathrm{i}[n+1]\\
    0\\
    \mathrm{p}_1[n]\\
    0
  \end{pmatrix}\,.
\end{equation}
which implies substituting
$X[n]=\Big(\frac{2\Delta}{\Gamma_\sigma+\gamma_\phi}+i\Big)\mathrm{q}_\mathrm{i}[n]$
in $\mathbf{A}_1[n]$. The final correlator is found by taking the
third element of the vectorial solution and summing contributions from
all rungs:
$\langle\ud{\sigma}(0)\sigma(\tau)\rangle=\sum_{n=0}^{\infty}(\mathbf{u}_1[n])_3$,
which corresponds to Eq.~(\ref{eq:WedMar30182503CEST2011}).

For the photonic correlator, the initial condition is:
\begin{equation}
  \label{eq:WedMar24152058WET2010}
  \mathbf{u}_1[n](0)=
  \begin{pmatrix}
    \sqrt{n+1}\mathrm{p}_0[n+1]\\
    \sqrt{n}\mathrm{p}_1[n]\\
    \Big(\frac{2\Delta}{\Gamma_\sigma+\gamma_\phi}-i\Big) \sqrt{n+1}\mathrm{q}_\mathrm{i}[n+1]\\
    \Big(\frac{2\Delta}{\Gamma_\sigma+\gamma_\phi}+i\Big) \sqrt{n}\mathrm{q}_\mathrm{i}[n+1]
  \end{pmatrix}\,.
\end{equation}
which implies substituting $X[n]=\sqrt{n}\mathrm{T}[n]$ in
$\mathbf{A}_1[n]$. The final correlator is found as:
$\langle\ud{a}(0)a(\tau)\rangle=\sum_{n=0}^{\infty}\Big[\sqrt{n+1}(\mathbf{u}_1[n])_1+\sqrt{n}(\mathbf{u}_1[n])_2\Big]$,
which corresponds to Eq.~(\ref{eq:WedMar30182553CEST2011}).

\subsection{Elastic scattering term}

The second line in Eq.~(\ref{eq:WedMar24151034WET2010}), 
\begin{equation}
\mathbf{B}_1[n]=(\mathbf{M}_{1}[n])^{-1}\mathbf{A}_1[n]
\label{eq:WedMar30181744CEST2011}
\end{equation}
is independent of $\tau$, due to the approximation of infinite
lifetime $\gamma_a\approx 0$. If $\gamma_a$ was of the order of
$g,\Gamma_\sigma$, we could not have assumed $X[n]$ to be
$\tau$-independent and, therefore, nor $\mathbf{B}_1[n]$. In the
regime where our approximation is valid, this term gives rise to the
first element in Eq.~(\ref{eq:ThuMar25104239WET2010}), $E^c$, which
turns into a delta function at the cavity frequency in the frequency
domain, Eq.~(\ref{eq:WedMar24190127WET2010}).

For the emitter, the intensity of this contribution is found from
summing the third element of $\mathbf{B}_1[n]$, for all rungs~$n\geq
0$: $E^\sigma=\sum_{n=0}^{\infty}(\mathbf{B}_1[n])_3$. For the cavity
case, it is found as
$E^\sigma=\sum_{n=0}^{\infty}\Big[\sqrt{n+1}(\mathbf{B}_1[n])_1+\sqrt{n}(\mathbf{B}_1[n])_2\Big]$.

\subsection{First rung}

In principle, we must solve separately the Rabi dynamics of the first
rung with the ground state, $n=0$, having a $2\times 2$ system
($S_1[0]=0$ and $V[0]=0$):
\begin{multline}
  \label{eq:WedMar24181802WET2010}
  \mathbf{u}_1[0]=
  \begin{pmatrix}
    S_0[1]\\
    Q[0]\\
  \end{pmatrix}
  \,, \quad \mathbf{A}_1[0]=
  \begin{pmatrix}
    \gamma_\sigma X[1]\\
    0
  \end{pmatrix}\,,\\
  \mathbf{M}_{1}[0]=
  \begin{pmatrix}
\Gamma_\sigma&ig\\
ig&\frac{\Gamma_\sigma+\gamma_\phi}{2}-i\Delta
  \end{pmatrix}\,.
\end{multline}

For the emitter, the initial conditions are:
\begin{equation}
  \label{eq:WedMar24182052WET2010}
  \mathbf{u}_1[0](0)=
  \begin{pmatrix}
    \Big(\frac{2\Delta}{\Gamma_\sigma+\gamma_\phi}+i\Big) \mathrm{q}_\mathrm{i}[1]\\
    \mathrm{p}_1[0]
  \end{pmatrix}\,.
\end{equation}
In the photonic case, the initial conditions are:
\begin{equation}
  \label{eq:WedMar24182409WET2010}
  \mathbf{u}_1[0](0)=
  \begin{pmatrix}
    \mathrm{p}_0[1]\\
    \Big(\frac{2\Delta}{\Gamma_\sigma+\gamma_\phi}-i\Big) \mathrm{q}_\mathrm{i}[1]
  \end{pmatrix}\,.
\end{equation}
One can check that the solution for $n=0$ is finally the same as
taking the limit $n\rightarrow 0$ in the expressions we obtained for
$n>1$. Therefore, there is no further need of separating the Rabi from
the rest of rungs in our expressions, although it will find simpler
expressions for the quantities of interest and we may point them
out. For instance, if $\gamma_\sigma= 0$, in both cases,
$\mathbf{A}_1[0]=0$ and the solution simplifies to:
$\mathbf{u}_{1}[0](\tau)=e^{-\mathbf{M}_{1}[0]\tau}\mathbf{u}_{1}[0](0)$.

\bibliography{Sci,Elena,books}

\begin{thebibliography}{64}
\expandafter\ifx\csname natexlab\endcsname\relax\def\natexlab#1{#1}\fi
\expandafter\ifx\csname bibnamefont\endcsname\relax
  \def\bibnamefont#1{#1}\fi
\expandafter\ifx\csname bibfnamefont\endcsname\relax
  \def\bibfnamefont#1{#1}\fi
\expandafter\ifx\csname citenamefont\endcsname\relax
  \def\citenamefont#1{#1}\fi
\expandafter\ifx\csname url\endcsname\relax
  \def\url#1{\texttt{#1}}\fi
\expandafter\ifx\csname urlprefix\endcsname\relax\def\urlprefix{URL }\fi
\providecommand{\bibinfo}[2]{#2}
\providecommand{\eprint}[2][]{\url{#2}}

\bibitem[{\citenamefont{Haroche and Raimond}(2006)}]{haroche_book06a}
\bibinfo{author}{\bibfnamefont{S.}~\bibnamefont{Haroche}} \bibnamefont{and}
  \bibinfo{author}{\bibfnamefont{J.-M.} \bibnamefont{Raimond}},
  \emph{\bibinfo{title}{Exploring the Quantum: Atoms, Cavities, and Photons}}
  (\bibinfo{publisher}{Oxford University Press}, \bibinfo{year}{2006}).

\bibitem[{\citenamefont{Raimond et~al.}(2001)\citenamefont{Raimond, Brune, and
  Haroche}}]{raimond01a}
\bibinfo{author}{\bibfnamefont{J.~M.} \bibnamefont{Raimond}},
  \bibinfo{author}{\bibfnamefont{M.}~\bibnamefont{Brune}}, \bibnamefont{and}
  \bibinfo{author}{\bibfnamefont{S.}~\bibnamefont{Haroche}},
  \bibinfo{journal}{Rev. Mod. Phys.} \textbf{\bibinfo{volume}{73}},
  \bibinfo{pages}{565} (\bibinfo{year}{2001}).

\bibitem[{\citenamefont{Khitrova et~al.}(2006)\citenamefont{Khitrova, Gibbs,
  Kira, Koch, and Scherer}}]{khitrova06a}
\bibinfo{author}{\bibfnamefont{G.}~\bibnamefont{Khitrova}},
  \bibinfo{author}{\bibfnamefont{H.~M.} \bibnamefont{Gibbs}},
  \bibinfo{author}{\bibfnamefont{M.}~\bibnamefont{Kira}},
  \bibinfo{author}{\bibfnamefont{S.~W.} \bibnamefont{Koch}}, \bibnamefont{and}
  \bibinfo{author}{\bibfnamefont{A.}~\bibnamefont{Scherer}},
  \bibinfo{journal}{Nat. Phys.} \textbf{\bibinfo{volume}{2}},
  \bibinfo{pages}{81} (\bibinfo{year}{2006}).

\bibitem[{\citenamefont{Makhlin et~al.}(2001)\citenamefont{Makhlin, Sch\"on,
  and Shnirman}}]{makhlin01a}
\bibinfo{author}{\bibfnamefont{Y.}~\bibnamefont{Makhlin}},
  \bibinfo{author}{\bibfnamefont{G.}~\bibnamefont{Sch\"on}}, \bibnamefont{and}
  \bibinfo{author}{\bibfnamefont{A.}~\bibnamefont{Shnirman}},
  \bibinfo{journal}{Rev. Mod. Phys.} \textbf{\bibinfo{volume}{73}},
  \bibinfo{pages}{357} (\bibinfo{year}{2001}).

\bibitem[{\citenamefont{Kippenberg}(2008)}]{kippenberg08a}
\bibinfo{author}{\bibfnamefont{T.}~\bibnamefont{Kippenberg}},
  \bibinfo{journal}{Nature} \textbf{\bibinfo{volume}{456}},
  \bibinfo{pages}{458} (\bibinfo{year}{2008}).

\bibitem[{\citenamefont{Mu and Savage}(1992)}]{mu92a}
\bibinfo{author}{\bibfnamefont{Y.}~\bibnamefont{Mu}} \bibnamefont{and}
  \bibinfo{author}{\bibfnamefont{C.~M.} \bibnamefont{Savage}},
  \bibinfo{journal}{Phys. Rev. A} \textbf{\bibinfo{volume}{46}},
  \bibinfo{pages}{5944} (\bibinfo{year}{1992}).

\bibitem[{\citenamefont{P.~R.~Rice}(1994)}]{rice94a}
\bibinfo{author}{\bibfnamefont{H.~J.~C.} \bibnamefont{P.~R.~Rice}},
  \bibinfo{journal}{Phys. Rev. A} \textbf{\bibinfo{volume}{50}},
  \bibinfo{pages}{4318} (\bibinfo{year}{1994}).

\bibitem[{\citenamefont{Ginzel et~al.}(1993)\citenamefont{Ginzel, Briegel,
  Martini, Englert, and Schenzle}}]{ginzel93a}
\bibinfo{author}{\bibfnamefont{C.}~\bibnamefont{Ginzel}},
  \bibinfo{author}{\bibfnamefont{H.-J.} \bibnamefont{Briegel}},
  \bibinfo{author}{\bibfnamefont{U.}~\bibnamefont{Martini}},
  \bibinfo{author}{\bibfnamefont{B.-G.} \bibnamefont{Englert}},
  \bibnamefont{and} \bibinfo{author}{\bibfnamefont{A.}~\bibnamefont{Schenzle}},
  \bibinfo{journal}{Phys. Rev. A} \textbf{\bibinfo{volume}{48}},
  \bibinfo{pages}{732} (\bibinfo{year}{1993}).

\bibitem[{\citenamefont{L\"offler et~al.}(1997)\citenamefont{L\"offler, Meyer,
  and Walther}}]{loffler97a}
\bibinfo{author}{\bibfnamefont{M.}~\bibnamefont{L\"offler}},
  \bibinfo{author}{\bibfnamefont{G.~M.} \bibnamefont{Meyer}}, \bibnamefont{and}
  \bibinfo{author}{\bibfnamefont{H.}~\bibnamefont{Walther}},
  \bibinfo{journal}{Phys. Rev. A} \textbf{\bibinfo{volume}{55}},
  \bibinfo{pages}{3923} (\bibinfo{year}{1997}).

\bibitem[{\citenamefont{Jones et~al.}(1999)\citenamefont{Jones, Ghose, Clemens,
  Rice, and Pedrotti}}]{jones99a}
\bibinfo{author}{\bibfnamefont{B.}~\bibnamefont{Jones}},
  \bibinfo{author}{\bibfnamefont{S.}~\bibnamefont{Ghose}},
  \bibinfo{author}{\bibfnamefont{J.~P.} \bibnamefont{Clemens}},
  \bibinfo{author}{\bibfnamefont{P.~R.} \bibnamefont{Rice}}, \bibnamefont{and}
  \bibinfo{author}{\bibfnamefont{L.~M.} \bibnamefont{Pedrotti}},
  \bibinfo{journal}{Phys. Rev. A} \textbf{\bibinfo{volume}{60}},
  \bibinfo{pages}{3267} (\bibinfo{year}{1999}).

\bibitem[{\citenamefont{Benson and Yamamoto}(1999)}]{benson99a}
\bibinfo{author}{\bibfnamefont{O.}~\bibnamefont{Benson}} \bibnamefont{and}
  \bibinfo{author}{\bibfnamefont{Y.}~\bibnamefont{Yamamoto}},
  \bibinfo{journal}{Phys. Rev. A} \textbf{\bibinfo{volume}{59}},
  \bibinfo{pages}{4756} (\bibinfo{year}{1999}).

\bibitem[{\citenamefont{Karlovich and Kilin}(2001)}]{karlovich01a}
\bibinfo{author}{\bibfnamefont{T.~B.} \bibnamefont{Karlovich}}
  \bibnamefont{and} \bibinfo{author}{\bibfnamefont{S.~Y.} \bibnamefont{Kilin}},
  \bibinfo{journal}{Opt. Spectrosc.} \textbf{\bibinfo{volume}{91}},
  \bibinfo{pages}{343} (\bibinfo{year}{2001}).

\bibitem[{\citenamefont{Kilin and Karlovich}(2002)}]{kilin02a}
\bibinfo{author}{\bibfnamefont{S.~Y.} \bibnamefont{Kilin}} \bibnamefont{and}
  \bibinfo{author}{\bibfnamefont{T.~B.} \bibnamefont{Karlovich}},
  \bibinfo{journal}{Sov. Phys. JETP} \textbf{\bibinfo{volume}{95}},
  \bibinfo{pages}{805} (\bibinfo{year}{2002}).

\bibitem[{\citenamefont{Clemens et~al.}(2004)\citenamefont{Clemens, Rice, and
  Pedrotti}}]{clemens04a}
\bibinfo{author}{\bibfnamefont{J.~P.} \bibnamefont{Clemens}},
  \bibinfo{author}{\bibfnamefont{P.~R.} \bibnamefont{Rice}}, \bibnamefont{and}
  \bibinfo{author}{\bibfnamefont{L.~M.} \bibnamefont{Pedrotti}},
  \bibinfo{journal}{J. Opt. Soc. Am. B} \textbf{\bibinfo{volume}{21}},
  \bibinfo{pages}{2025} (\bibinfo{year}{2004}).

\bibitem[{\citenamefont{Karlovich and Kilin}(2008)}]{karlovich08a}
\bibinfo{author}{\bibfnamefont{T.~B.} \bibnamefont{Karlovich}}
  \bibnamefont{and} \bibinfo{author}{\bibfnamefont{S.~Y.} \bibnamefont{Kilin}},
  \bibinfo{journal}{Laser Phys.} \textbf{\bibinfo{volume}{18}},
  \bibinfo{pages}{783} (\bibinfo{year}{2008}).

\bibitem[{\citenamefont{del Valle et~al.}(2009)\citenamefont{del Valle, Laussy,
  and Tejedor}}]{delvalle09a}
\bibinfo{author}{\bibfnamefont{E.}~\bibnamefont{del Valle}},
  \bibinfo{author}{\bibfnamefont{F.~P.} \bibnamefont{Laussy}},
  \bibnamefont{and} \bibinfo{author}{\bibfnamefont{C.}~\bibnamefont{Tejedor}},
  \bibinfo{journal}{Phys. Rev. B} \textbf{\bibinfo{volume}{79}},
  \bibinfo{pages}{235326} (\bibinfo{year}{2009}).

\bibitem[{\citenamefont{Poddubny et~al.}(2010)\citenamefont{Poddubny, Glazov,
  and Averkiev}}]{poddubny10a}
\bibinfo{author}{\bibfnamefont{A.~N.} \bibnamefont{Poddubny}},
  \bibinfo{author}{\bibfnamefont{M.~M.} \bibnamefont{Glazov}},
  \bibnamefont{and} \bibinfo{author}{\bibfnamefont{N.~S.}
  \bibnamefont{Averkiev}}, \bibinfo{journal}{Phys. Rev. B}
  \textbf{\bibinfo{volume}{82}}, \bibinfo{pages}{205330}
  (\bibinfo{year}{2010}).

\bibitem[{\citenamefont{Gonzalez-Tudela
  et~al.}(2010)\citenamefont{Gonzalez-Tudela, del Valle, Cancellieri, Tejedor,
  Sanvitto, and Laussy}}]{gonzaleztudela10b}
\bibinfo{author}{\bibfnamefont{A.}~\bibnamefont{Gonzalez-Tudela}},
  \bibinfo{author}{\bibfnamefont{E.}~\bibnamefont{del Valle}},
  \bibinfo{author}{\bibfnamefont{E.}~\bibnamefont{Cancellieri}},
  \bibinfo{author}{\bibfnamefont{C.}~\bibnamefont{Tejedor}},
  \bibinfo{author}{\bibfnamefont{D.}~\bibnamefont{Sanvitto}}, \bibnamefont{and}
  \bibinfo{author}{\bibfnamefont{F.~P.} \bibnamefont{Laussy}},
  \bibinfo{journal}{Opt. Express} \textbf{\bibinfo{volume}{18}},
  \bibinfo{pages}{7002} (\bibinfo{year}{2010}).

\bibitem[{\citenamefont{Ritter et~al.}(2010)\citenamefont{Ritter, Gartner,
  Gies, and Jahnke}}]{ritter10a}
\bibinfo{author}{\bibfnamefont{S.}~\bibnamefont{Ritter}},
  \bibinfo{author}{\bibfnamefont{P.}~\bibnamefont{Gartner}},
  \bibinfo{author}{\bibfnamefont{C.}~\bibnamefont{Gies}}, \bibnamefont{and}
  \bibinfo{author}{\bibfnamefont{F.}~\bibnamefont{Jahnke}},
  \bibinfo{journal}{Opt. Express} \textbf{\bibinfo{volume}{18}},
  \bibinfo{pages}{9909} (\bibinfo{year}{2010}).

\bibitem[{\citenamefont{Auff\`eves et~al.}(2010)\citenamefont{Auff\`eves,
  Gerace, G\'erard, Santos, Andreani, and Poizat}}]{auffeves10a}
\bibinfo{author}{\bibfnamefont{A.}~\bibnamefont{Auff\`eves}},
  \bibinfo{author}{\bibfnamefont{D.}~\bibnamefont{Gerace}},
  \bibinfo{author}{\bibfnamefont{J.-M.} \bibnamefont{G\'erard}},
  \bibinfo{author}{\bibfnamefont{M.~F.} \bibnamefont{Santos}},
  \bibinfo{author}{\bibfnamefont{L.~C.} \bibnamefont{Andreani}},
  \bibnamefont{and} \bibinfo{author}{\bibfnamefont{J.-P.}
  \bibnamefont{Poizat}}, \bibinfo{journal}{Phys. Rev. B}
  \textbf{\bibinfo{volume}{81}}, \bibinfo{pages}{245419}
  (\bibinfo{year}{2010}).

\bibitem[{\citenamefont{del Valle and Laussy}(2010)}]{delvalle10d}
\bibinfo{author}{\bibfnamefont{E.}~\bibnamefont{del Valle}} \bibnamefont{and}
  \bibinfo{author}{\bibfnamefont{F.~P.} \bibnamefont{Laussy}},
  \bibinfo{journal}{Phys. Rev. Lett.} \textbf{\bibinfo{volume}{105}},
  \bibinfo{pages}{233601} (\bibinfo{year}{2010}).

\bibitem[{\citenamefont{Gartner}(2011)}]{lsc_gartner11a}
\bibinfo{author}{\bibfnamefont{P.}~\bibnamefont{Gartner}},
  \bibinfo{journal}{arXiv:1105.2189}  (\bibinfo{year}{2011}).

\bibitem[{\citenamefont{Mollow}(1969)}]{mollow69a}
\bibinfo{author}{\bibfnamefont{B.~R.} \bibnamefont{Mollow}},
  \bibinfo{journal}{Phys. Rev.} \textbf{\bibinfo{volume}{188}},
  \bibinfo{pages}{1969} (\bibinfo{year}{1969}).

\bibitem[{\citenamefont{Boca et~al.}(2004)\citenamefont{Boca, Miller, Birnbaum,
  Boozer, McKeever, and Kimble}}]{boca04a}
\bibinfo{author}{\bibfnamefont{A.}~\bibnamefont{Boca}},
  \bibinfo{author}{\bibfnamefont{R.}~\bibnamefont{Miller}},
  \bibinfo{author}{\bibfnamefont{K.~M.} \bibnamefont{Birnbaum}},
  \bibinfo{author}{\bibfnamefont{A.~D.} \bibnamefont{Boozer}},
  \bibinfo{author}{\bibfnamefont{J.}~\bibnamefont{McKeever}}, \bibnamefont{and}
  \bibinfo{author}{\bibfnamefont{H.~J.} \bibnamefont{Kimble}},
  \bibinfo{journal}{Phys. Rev. Lett.} \textbf{\bibinfo{volume}{93}},
  \bibinfo{pages}{233603} (\bibinfo{year}{2004}).

\bibitem[{\citenamefont{Wallraff et~al.}(2004)\citenamefont{Wallraff, Schuster,
  Blais, Frunzio, Huang, Majer, Kumar, Girvin, and Schoelkopf}}]{wallraff04a}
\bibinfo{author}{\bibfnamefont{A.}~\bibnamefont{Wallraff}},
  \bibinfo{author}{\bibfnamefont{D.~I.} \bibnamefont{Schuster}},
  \bibinfo{author}{\bibfnamefont{A.}~\bibnamefont{Blais}},
  \bibinfo{author}{\bibfnamefont{L.}~\bibnamefont{Frunzio}},
  \bibinfo{author}{\bibfnamefont{R.-S.} \bibnamefont{Huang}},
  \bibinfo{author}{\bibfnamefont{J.}~\bibnamefont{Majer}},
  \bibinfo{author}{\bibfnamefont{S.}~\bibnamefont{Kumar}},
  \bibinfo{author}{\bibfnamefont{S.~M.} \bibnamefont{Girvin}},
  \bibnamefont{and} \bibinfo{author}{\bibfnamefont{R.~J.}
  \bibnamefont{Schoelkopf}}, \bibinfo{journal}{Nature}
  \textbf{\bibinfo{volume}{431}}, \bibinfo{pages}{162} (\bibinfo{year}{2004}).

\bibitem[{\citenamefont{Fink et~al.}(2008)\citenamefont{Fink, G{\"o}ppl, Baur,
  Bianchetti, Leek, Blais, and Wallraff}}]{fink08a}
\bibinfo{author}{\bibfnamefont{J.~M.} \bibnamefont{Fink}},
  \bibinfo{author}{\bibfnamefont{M.}~\bibnamefont{G{\"o}ppl}},
  \bibinfo{author}{\bibfnamefont{M.}~\bibnamefont{Baur}},
  \bibinfo{author}{\bibfnamefont{R.}~\bibnamefont{Bianchetti}},
  \bibinfo{author}{\bibfnamefont{P.~J.} \bibnamefont{Leek}},
  \bibinfo{author}{\bibfnamefont{A.}~\bibnamefont{Blais}}, \bibnamefont{and}
  \bibinfo{author}{\bibfnamefont{A.}~\bibnamefont{Wallraff}},
  \bibinfo{journal}{Nature} \textbf{\bibinfo{volume}{454}},
  \bibinfo{pages}{315} (\bibinfo{year}{2008}).

\bibitem[{\citenamefont{Reithmaier et~al.}(2004)\citenamefont{Reithmaier, Sek,
  L\"offler, Hofmann, Kuhn, Reitzenstein, Keldysh, Kulakovskii, Reinecker, and
  Forchel}}]{reithmaier04a}
\bibinfo{author}{\bibfnamefont{J.~P.} \bibnamefont{Reithmaier}},
  \bibinfo{author}{\bibfnamefont{G.}~\bibnamefont{Sek}},
  \bibinfo{author}{\bibfnamefont{A.}~\bibnamefont{L\"offler}},
  \bibinfo{author}{\bibfnamefont{C.}~\bibnamefont{Hofmann}},
  \bibinfo{author}{\bibfnamefont{S.}~\bibnamefont{Kuhn}},
  \bibinfo{author}{\bibfnamefont{S.}~\bibnamefont{Reitzenstein}},
  \bibinfo{author}{\bibfnamefont{L.~V.} \bibnamefont{Keldysh}},
  \bibinfo{author}{\bibfnamefont{V.~D.} \bibnamefont{Kulakovskii}},
  \bibinfo{author}{\bibfnamefont{T.~L.} \bibnamefont{Reinecker}},
  \bibnamefont{and} \bibinfo{author}{\bibfnamefont{A.}~\bibnamefont{Forchel}},
  \bibinfo{journal}{Nature} \textbf{\bibinfo{volume}{432}},
  \bibinfo{pages}{197} (\bibinfo{year}{2004}).

\bibitem[{\citenamefont{Yoshie et~al.}(2004)\citenamefont{Yoshie, Scherer,
  Heindrickson, Khitrova, Gibbs, Rupper, Ell, Shchekin, and Deppe}}]{yoshie04a}
\bibinfo{author}{\bibfnamefont{T.}~\bibnamefont{Yoshie}},
  \bibinfo{author}{\bibfnamefont{A.}~\bibnamefont{Scherer}},
  \bibinfo{author}{\bibfnamefont{J.}~\bibnamefont{Heindrickson}},
  \bibinfo{author}{\bibfnamefont{G.}~\bibnamefont{Khitrova}},
  \bibinfo{author}{\bibfnamefont{H.~M.} \bibnamefont{Gibbs}},
  \bibinfo{author}{\bibfnamefont{G.}~\bibnamefont{Rupper}},
  \bibinfo{author}{\bibfnamefont{C.}~\bibnamefont{Ell}},
  \bibinfo{author}{\bibfnamefont{O.~B.} \bibnamefont{Shchekin}},
  \bibnamefont{and} \bibinfo{author}{\bibfnamefont{D.~G.} \bibnamefont{Deppe}},
  \bibinfo{journal}{Nature} \textbf{\bibinfo{volume}{432}},
  \bibinfo{pages}{200} (\bibinfo{year}{2004}).

\bibitem[{\citenamefont{Peter et~al.}(2005)\citenamefont{Peter, Senellart,
  Martrou, Lema\^itre, Hours, G\'erard, and Bloch}}]{peter05a}
\bibinfo{author}{\bibfnamefont{E.}~\bibnamefont{Peter}},
  \bibinfo{author}{\bibfnamefont{P.}~\bibnamefont{Senellart}},
  \bibinfo{author}{\bibfnamefont{D.}~\bibnamefont{Martrou}},
  \bibinfo{author}{\bibfnamefont{A.}~\bibnamefont{Lema\^itre}},
  \bibinfo{author}{\bibfnamefont{J.}~\bibnamefont{Hours}},
  \bibinfo{author}{\bibfnamefont{J.~M.} \bibnamefont{G\'erard}},
  \bibnamefont{and} \bibinfo{author}{\bibfnamefont{J.}~\bibnamefont{Bloch}},
  \bibinfo{journal}{Phys. Rev. Lett.} \textbf{\bibinfo{volume}{95}},
  \bibinfo{pages}{067401} (\bibinfo{year}{2005}).

\bibitem[{\citenamefont{McKeever et~al.}(2003)\citenamefont{McKeever, Boca,
  Boozer, Buck, and Kimble}}]{mckeever03a}
\bibinfo{author}{\bibfnamefont{J.}~\bibnamefont{McKeever}},
  \bibinfo{author}{\bibfnamefont{A.}~\bibnamefont{Boca}},
  \bibinfo{author}{\bibfnamefont{A.~D.} \bibnamefont{Boozer}},
  \bibinfo{author}{\bibfnamefont{J.~R.} \bibnamefont{Buck}}, \bibnamefont{and}
  \bibinfo{author}{\bibfnamefont{H.~J.} \bibnamefont{Kimble}},
  \bibinfo{journal}{Nature} \textbf{\bibinfo{volume}{425}},
  \bibinfo{pages}{268} (\bibinfo{year}{2003}).

\bibitem[{\citenamefont{Astafiev et~al.}(2007)\citenamefont{Astafiev, Inomata,
  Niskanen, Yamamoto, Pashkin, Nakamura, and Tsai}}]{astafiev07a}
\bibinfo{author}{\bibfnamefont{O.}~\bibnamefont{Astafiev}},
  \bibinfo{author}{\bibfnamefont{K.}~\bibnamefont{Inomata}},
  \bibinfo{author}{\bibfnamefont{A.~O.} \bibnamefont{Niskanen}},
  \bibinfo{author}{\bibfnamefont{T.}~\bibnamefont{Yamamoto}},
  \bibinfo{author}{\bibfnamefont{Y.~A.} \bibnamefont{Pashkin}},
  \bibinfo{author}{\bibfnamefont{Y.}~\bibnamefont{Nakamura}}, \bibnamefont{and}
  \bibinfo{author}{\bibfnamefont{J.~S.} \bibnamefont{Tsai}},
  \bibinfo{journal}{Nature} \textbf{\bibinfo{volume}{449}},
  \bibinfo{pages}{588} (\bibinfo{year}{2007}).

\bibitem[{\citenamefont{Nomura et~al.}(2010)\citenamefont{Nomura, Kumagai,
  Iwamoto, Ota, and Arakawa}}]{nomura10a}
\bibinfo{author}{\bibfnamefont{M.}~\bibnamefont{Nomura}},
  \bibinfo{author}{\bibfnamefont{N.}~\bibnamefont{Kumagai}},
  \bibinfo{author}{\bibfnamefont{S.}~\bibnamefont{Iwamoto}},
  \bibinfo{author}{\bibfnamefont{Y.}~\bibnamefont{Ota}}, \bibnamefont{and}
  \bibinfo{author}{\bibfnamefont{Y.}~\bibnamefont{Arakawa}},
  \bibinfo{journal}{Nat. Phys.} \textbf{\bibinfo{volume}{6}},
  \bibinfo{pages}{279} (\bibinfo{year}{2010}).

\bibitem[{\citenamefont{Wu et~al.}(1975)\citenamefont{Wu, Grove, and
  Ezekiel}}]{wu75a}
\bibinfo{author}{\bibfnamefont{F.~Y.} \bibnamefont{Wu}},
  \bibinfo{author}{\bibfnamefont{R.~E.} \bibnamefont{Grove}}, \bibnamefont{and}
  \bibinfo{author}{\bibfnamefont{S.}~\bibnamefont{Ezekiel}},
  \bibinfo{journal}{Phys. Rev. Lett.} \textbf{\bibinfo{volume}{35}},
  \bibinfo{pages}{1426} (\bibinfo{year}{1975}).

\bibitem[{\citenamefont{Wrigge et~al.}(2008)\citenamefont{Wrigge, Gerhardt,
  Hwang, Zumofen, and Sandoghdar}}]{wrigge08a}
\bibinfo{author}{\bibfnamefont{G.}~\bibnamefont{Wrigge}},
  \bibinfo{author}{\bibfnamefont{I.}~\bibnamefont{Gerhardt}},
  \bibinfo{author}{\bibfnamefont{J.}~\bibnamefont{Hwang}},
  \bibinfo{author}{\bibfnamefont{G.}~\bibnamefont{Zumofen}}, \bibnamefont{and}
  \bibinfo{author}{\bibfnamefont{V.}~\bibnamefont{Sandoghdar}},
  \bibinfo{journal}{Nat. Phys.} \textbf{\bibinfo{volume}{4}},
  \bibinfo{pages}{60} (\bibinfo{year}{2008}).

\bibitem[{\citenamefont{Astafiev et~al.}(2010)\citenamefont{Astafiev, Zagoskin,
  Jr., Pashkin, Yamamoto, Inomata, Nakamura, and Tsai}}]{astafiev10a}
\bibinfo{author}{\bibfnamefont{O.}~\bibnamefont{Astafiev}},
  \bibinfo{author}{\bibfnamefont{A.~M.} \bibnamefont{Zagoskin}},
  \bibinfo{author}{\bibfnamefont{A.~A.~A.} \bibnamefont{Jr.}},
  \bibinfo{author}{\bibfnamefont{Y.~A.} \bibnamefont{Pashkin}},
  \bibinfo{author}{\bibfnamefont{T.}~\bibnamefont{Yamamoto}},
  \bibinfo{author}{\bibfnamefont{K.}~\bibnamefont{Inomata}},
  \bibinfo{author}{\bibfnamefont{Y.}~\bibnamefont{Nakamura}}, \bibnamefont{and}
  \bibinfo{author}{\bibfnamefont{J.~S.} \bibnamefont{Tsai}},
  \bibinfo{journal}{Science} \textbf{\bibinfo{volume}{327}},
  \bibinfo{pages}{840} (\bibinfo{year}{2010}).

\bibitem[{\citenamefont{Muller et~al.}(2007)\citenamefont{Muller, Flagg,
  Bianucci, Wang, Deppe, Ma, Zhang, Salamo, Xiao, and Shih}}]{muller07a}
\bibinfo{author}{\bibfnamefont{A.}~\bibnamefont{Muller}},
  \bibinfo{author}{\bibfnamefont{E.~B.} \bibnamefont{Flagg}},
  \bibinfo{author}{\bibfnamefont{P.}~\bibnamefont{Bianucci}},
  \bibinfo{author}{\bibfnamefont{X.~Y.} \bibnamefont{Wang}},
  \bibinfo{author}{\bibfnamefont{D.~G.} \bibnamefont{Deppe}},
  \bibinfo{author}{\bibfnamefont{W.}~\bibnamefont{Ma}},
  \bibinfo{author}{\bibfnamefont{J.}~\bibnamefont{Zhang}},
  \bibinfo{author}{\bibfnamefont{G.~J.} \bibnamefont{Salamo}},
  \bibinfo{author}{\bibfnamefont{M.}~\bibnamefont{Xiao}}, \bibnamefont{and}
  \bibinfo{author}{\bibfnamefont{C.~K.} \bibnamefont{Shih}},
  \bibinfo{journal}{Phys. Rev. Lett.} \textbf{\bibinfo{volume}{99}},
  \bibinfo{pages}{187402} (\bibinfo{year}{2007}).

\bibitem[{\citenamefont{Vamivakas et~al.}(2009)\citenamefont{Vamivakas, Zhao,
  Lu, and Atat{\"u}re}}]{vamivakas09a}
\bibinfo{author}{\bibfnamefont{A.~N.} \bibnamefont{Vamivakas}},
  \bibinfo{author}{\bibfnamefont{Y.}~\bibnamefont{Zhao}},
  \bibinfo{author}{\bibfnamefont{C.-Y.} \bibnamefont{Lu}}, \bibnamefont{and}
  \bibinfo{author}{\bibfnamefont{M.}~\bibnamefont{Atat{\"u}re}},
  \bibinfo{journal}{Nat. Phys.} \textbf{\bibinfo{volume}{5}},
  \bibinfo{pages}{198} (\bibinfo{year}{2009}).

\bibitem[{\citenamefont{Flagg et~al.}(2009)\citenamefont{Flagg, Muller,
  Robertson, Founta, Deppe, Xiao, Ma, Salamo, and Shih}}]{flagg09a}
\bibinfo{author}{\bibfnamefont{E.~B.} \bibnamefont{Flagg}},
  \bibinfo{author}{\bibfnamefont{A.}~\bibnamefont{Muller}},
  \bibinfo{author}{\bibfnamefont{J.~W.} \bibnamefont{Robertson}},
  \bibinfo{author}{\bibfnamefont{S.}~\bibnamefont{Founta}},
  \bibinfo{author}{\bibfnamefont{D.~G.} \bibnamefont{Deppe}},
  \bibinfo{author}{\bibfnamefont{M.}~\bibnamefont{Xiao}},
  \bibinfo{author}{\bibfnamefont{W.}~\bibnamefont{Ma}},
  \bibinfo{author}{\bibfnamefont{G.~J.} \bibnamefont{Salamo}},
  \bibnamefont{and} \bibinfo{author}{\bibfnamefont{C.~K.} \bibnamefont{Shih}},
  \bibinfo{journal}{Nat. Phys.} \textbf{\bibinfo{volume}{5}},
  \bibinfo{pages}{203} (\bibinfo{year}{2009}).

\bibitem[{\citenamefont{Ates et~al.}(2009)\citenamefont{Ates, Ulrich, Ulhaq,
  Reitzenstein, L{\"o}offler, H{\"o}fling, Forchel, and Michler}}]{ates09a}
\bibinfo{author}{\bibfnamefont{S.}~\bibnamefont{Ates}},
  \bibinfo{author}{\bibfnamefont{S.~M.} \bibnamefont{Ulrich}},
  \bibinfo{author}{\bibfnamefont{A.}~\bibnamefont{Ulhaq}},
  \bibinfo{author}{\bibfnamefont{S.}~\bibnamefont{Reitzenstein}},
  \bibinfo{author}{\bibfnamefont{A.}~\bibnamefont{L{\"o}offler}},
  \bibinfo{author}{\bibfnamefont{S.}~\bibnamefont{H{\"o}fling}},
  \bibinfo{author}{\bibfnamefont{A.}~\bibnamefont{Forchel}}, \bibnamefont{and}
  \bibinfo{author}{\bibfnamefont{P.}~\bibnamefont{Michler}},
  \bibinfo{journal}{Nat. Photon.} \textbf{\bibinfo{volume}{3}},
  \bibinfo{pages}{724} (\bibinfo{year}{2009}).

\bibitem[{\citenamefont{Jaynes and Cummings}(1963)}]{jaynes63a}
\bibinfo{author}{\bibfnamefont{E.}~\bibnamefont{Jaynes}} \bibnamefont{and}
  \bibinfo{author}{\bibfnamefont{F.}~\bibnamefont{Cummings}},
  \bibinfo{journal}{Proc. IEEE} \textbf{\bibinfo{volume}{51}},
  \bibinfo{pages}{89} (\bibinfo{year}{1963}).

\bibitem[{\citenamefont{Ulrich et~al.}(2011)\citenamefont{Ulrich, Ates,
  Reitzenstein, L\"offler, Forchel, and Michler}}]{lsc_ulrich11a}
\bibinfo{author}{\bibfnamefont{S.~M.} \bibnamefont{Ulrich}},
  \bibinfo{author}{\bibfnamefont{S.}~\bibnamefont{Ates}},
  \bibinfo{author}{\bibfnamefont{S.}~\bibnamefont{Reitzenstein}},
  \bibinfo{author}{\bibfnamefont{A.}~\bibnamefont{L\"offler}},
  \bibinfo{author}{\bibfnamefont{A.}~\bibnamefont{Forchel}}, \bibnamefont{and}
  \bibinfo{author}{\bibfnamefont{P.}~\bibnamefont{Michler}},
  \bibinfo{journal}{arXiv:1103.1594}  (\bibinfo{year}{2011}).

\bibitem[{\citenamefont{Roy and Hughes}(2011)}]{lsc_roy11a}
\bibinfo{author}{\bibfnamefont{C.}~\bibnamefont{Roy}} \bibnamefont{and}
  \bibinfo{author}{\bibfnamefont{S.}~\bibnamefont{Hughes}},
  \bibinfo{journal}{arXiv:1102.0254}  (\bibinfo{year}{2011}).

\bibitem[{\citenamefont{Loudon}(2000)}]{loudon_book00a}
\bibinfo{author}{\bibfnamefont{R.}~\bibnamefont{Loudon}},
  \emph{\bibinfo{title}{The quantum theory of light}}
  (\bibinfo{publisher}{Oxford Science Publications}, \bibinfo{year}{2000}),
  \bibinfo{edition}{3rd} ed.

\bibitem[{\citenamefont{del Valle}(2009)}]{delvalle_book09a}
\bibinfo{author}{\bibfnamefont{E.}~\bibnamefont{del Valle}},
  \emph{\bibinfo{title}{Microcavity Quantum Electrodynamics}}
  (\bibinfo{publisher}{VDM Verlag}, \bibinfo{year}{2009}).

\bibitem[{\citenamefont{Cohen-Tannoudji and Reynaud}(1977)}]{cohentannoudji77a}
\bibinfo{author}{\bibfnamefont{C.}~\bibnamefont{Cohen-Tannoudji}}
  \bibnamefont{and} \bibinfo{author}{\bibfnamefont{S.}~\bibnamefont{Reynaud}},
  \bibinfo{journal}{J. phys. B.: At. Mol. Phys.} \textbf{\bibinfo{volume}{10}},
  \bibinfo{pages}{345} (\bibinfo{year}{1977}).

\bibitem[{\citenamefont{Shore and Knight}(1993)}]{shore93a}
\bibinfo{author}{\bibfnamefont{B.~W.} \bibnamefont{Shore}} \bibnamefont{and}
  \bibinfo{author}{\bibfnamefont{P.~L.} \bibnamefont{Knight}},
  \bibinfo{journal}{J. Mod. Opt.} \textbf{\bibinfo{volume}{40}},
  \bibinfo{pages}{1195} (\bibinfo{year}{1993}).

\bibitem[{\citenamefont{Laucht et~al.}(2009)\citenamefont{Laucht, Hauke,
  Villas-B\^oas, Hofbauer, B\"ohm, Kaniber, and Finley}}]{laucht09b}
\bibinfo{author}{\bibfnamefont{A.}~\bibnamefont{Laucht}},
  \bibinfo{author}{\bibfnamefont{N.}~\bibnamefont{Hauke}},
  \bibinfo{author}{\bibfnamefont{J.~M.} \bibnamefont{Villas-B\^oas}},
  \bibinfo{author}{\bibfnamefont{F.}~\bibnamefont{Hofbauer}},
  \bibinfo{author}{\bibfnamefont{G.}~\bibnamefont{B\"ohm}},
  \bibinfo{author}{\bibfnamefont{M.}~\bibnamefont{Kaniber}}, \bibnamefont{and}
  \bibinfo{author}{\bibfnamefont{J.~J.} \bibnamefont{Finley}},
  \bibinfo{journal}{Phys. Rev. Lett.} \textbf{\bibinfo{volume}{103}},
  \bibinfo{pages}{087405} (\bibinfo{year}{2009}).

\bibitem[{\citenamefont{Laussy et~al.}(2009)\citenamefont{Laussy, del Valle,
  and Tejedor}}]{laussy09a}
\bibinfo{author}{\bibfnamefont{F.~P.} \bibnamefont{Laussy}},
  \bibinfo{author}{\bibfnamefont{E.}~\bibnamefont{del Valle}},
  \bibnamefont{and} \bibinfo{author}{\bibfnamefont{C.}~\bibnamefont{Tejedor}},
  \bibinfo{journal}{Phys. Rev. B} \textbf{\bibinfo{volume}{79}},
  \bibinfo{pages}{235325} (\bibinfo{year}{2009}).

\bibitem[{\citenamefont{Auff\`eves et~al.}(2009)\citenamefont{Auff\`eves,
  G\'erard, and Poizat}}]{auffeves09a}
\bibinfo{author}{\bibfnamefont{A.}~\bibnamefont{Auff\`eves}},
  \bibinfo{author}{\bibfnamefont{J.-M.} \bibnamefont{G\'erard}},
  \bibnamefont{and} \bibinfo{author}{\bibfnamefont{J.-P.}
  \bibnamefont{Poizat}}, \bibinfo{journal}{Phys. Rev. A}
  \textbf{\bibinfo{volume}{79}}, \bibinfo{pages}{053838}
  (\bibinfo{year}{2009}).

\bibitem[{\citenamefont{del Valle et~al.}(2011)\citenamefont{del Valle, Laussy,
  and Finley}}]{elena_delvalle11a}
\bibinfo{author}{\bibfnamefont{E.}~\bibnamefont{del Valle}},
  \bibinfo{author}{\bibfnamefont{F.}~\bibnamefont{Laussy}}, \bibnamefont{and}
  \bibinfo{author}{\bibfnamefont{J.}~\bibnamefont{Finley}},
  \bibinfo{journal}{Unpublished.}  (\bibinfo{year}{2011}).

\bibitem[{\citenamefont{Laussy et~al.}(2006)\citenamefont{Laussy, Shelykh,
  Malpuech, and Kavokin}}]{laussy06a}
\bibinfo{author}{\bibfnamefont{F.~P.} \bibnamefont{Laussy}},
  \bibinfo{author}{\bibfnamefont{I.~A.} \bibnamefont{Shelykh}},
  \bibinfo{author}{\bibfnamefont{G.}~\bibnamefont{Malpuech}}, \bibnamefont{and}
  \bibinfo{author}{\bibfnamefont{A.}~\bibnamefont{Kavokin}},
  \bibinfo{journal}{Phys. Rev. B} \textbf{\bibinfo{volume}{73}},
  \bibinfo{pages}{035315} (\bibinfo{year}{2006}).

\bibitem[{\citenamefont{Scully and Zubairy}(2002)}]{scully_book02a}
\bibinfo{author}{\bibfnamefont{M.~O.} \bibnamefont{Scully}} \bibnamefont{and}
  \bibinfo{author}{\bibfnamefont{M.~S.} \bibnamefont{Zubairy}},
  \emph{\bibinfo{title}{Quantum optics}} (\bibinfo{publisher}{Cambridge
  University Press}, \bibinfo{year}{2002}).

\bibitem[{\citenamefont{M{\o}lmer}(1996)}]{elena_molmer96a}
\bibinfo{author}{\bibfnamefont{K.}~\bibnamefont{M{\o}lmer}}
  (\bibinfo{year}{1996}),
  \urlprefix\url{http://www.phys.au.dk/quantop/kvanteoptik/qrtnote.pdf}.

\bibitem[{\citenamefont{Laussy et~al.}(2008)\citenamefont{Laussy, del Valle,
  and Tejedor}}]{laussy08a}
\bibinfo{author}{\bibfnamefont{F.~P.} \bibnamefont{Laussy}},
  \bibinfo{author}{\bibfnamefont{E.}~\bibnamefont{del Valle}},
  \bibnamefont{and} \bibinfo{author}{\bibfnamefont{C.}~\bibnamefont{Tejedor}},
  \bibinfo{journal}{Phys. Rev. Lett.} \textbf{\bibinfo{volume}{101}},
  \bibinfo{pages}{083601} (\bibinfo{year}{2008}).

\bibitem[{\citenamefont{Laussy and del Valle}(2010)}]{laussy10b}
\bibinfo{author}{\bibfnamefont{F.~P.} \bibnamefont{Laussy}} \bibnamefont{and}
  \bibinfo{author}{\bibfnamefont{E.}~\bibnamefont{del Valle}},
  \bibinfo{journal}{J. Phys.: Conf. Ser.} \textbf{\bibinfo{volume}{210}},
  \bibinfo{pages}{012018} (\bibinfo{year}{2010}).

\bibitem[{\citenamefont{del Valle}(2010)}]{delvalle10b}
\bibinfo{author}{\bibfnamefont{E.}~\bibnamefont{del Valle}},
  \bibinfo{journal}{Phys. Rev. A} \textbf{\bibinfo{volume}{81}},
  \bibinfo{pages}{053811} (\bibinfo{year}{2010}).

\bibitem[{\citenamefont{{\u Imamo\=glu} et~al.}(1996)\citenamefont{{\u
  Imamo\=glu}, Ram, Pau, and Yamamoto}}]{imamoglu96a}
\bibinfo{author}{\bibfnamefont{A.}~\bibnamefont{{\u Imamo\=glu}}},
  \bibinfo{author}{\bibfnamefont{R.~J.} \bibnamefont{Ram}},
  \bibinfo{author}{\bibfnamefont{S.}~\bibnamefont{Pau}}, \bibnamefont{and}
  \bibinfo{author}{\bibfnamefont{Y.}~\bibnamefont{Yamamoto}},
  \bibinfo{journal}{Phys. Rev. A} \textbf{\bibinfo{volume}{53}},
  \bibinfo{pages}{4250} (\bibinfo{year}{1996}).

\bibitem[{\citenamefont{{\u Imamo\=glu} and Ram}(1996)}]{imamoglu96b}
\bibinfo{author}{\bibfnamefont{A.}~\bibnamefont{{\u Imamo\=glu}}}
  \bibnamefont{and} \bibinfo{author}{\bibfnamefont{R.~J.} \bibnamefont{Ram}},
  \bibinfo{journal}{Physics Letter A} \textbf{\bibinfo{volume}{214}},
  \bibinfo{pages}{193} (\bibinfo{year}{1996}).

\bibitem[{\citenamefont{Laussy et~al.}(2004)\citenamefont{Laussy, Malpuech, and
  Kavokin}}]{laussy04a}
\bibinfo{author}{\bibfnamefont{F.~P.} \bibnamefont{Laussy}},
  \bibinfo{author}{\bibfnamefont{G.}~\bibnamefont{Malpuech}}, \bibnamefont{and}
  \bibinfo{author}{\bibfnamefont{A.}~\bibnamefont{Kavokin}},
  \bibinfo{journal}{Phys. Stat. Sol. C} \textbf{\bibinfo{volume}{1}},
  \bibinfo{pages}{1339} (\bibinfo{year}{2004}).

\bibitem[{\citenamefont{Berry}(1987)}]{berry87a}
\bibinfo{author}{\bibfnamefont{M.}~\bibnamefont{Berry}}, \bibinfo{journal}{New
  Scientist} \textbf{\bibinfo{volume}{44}}, \bibinfo{pages}{47}
  (\bibinfo{year}{1987}).

\bibitem[{\citenamefont{Gutzwiller}(1992)}]{gutzwiller92a}
\bibinfo{author}{\bibfnamefont{M.~C.} \bibnamefont{Gutzwiller}},
  \bibinfo{journal}{Sci. Am.} \textbf{\bibinfo{volume}{206}},
  \bibinfo{pages}{78} (\bibinfo{year}{1992}).

\bibitem[{\citenamefont{Haake}(2001)}]{elena_haake01a}
\bibinfo{author}{\bibfnamefont{F.}~\bibnamefont{Haake}},
  \emph{\bibinfo{title}{Quantum Signatures of Chaos}}, vol.
  \bibinfo{volume}{2nd ed.} (\bibinfo{publisher}{Springer-Verlag, New York},
  \bibinfo{year}{2001}).

\bibitem[{\citenamefont{Ohta et~al.}(2011)\citenamefont{Ohta, Ota, Nomura,
  Kumagai, Ishida, Iwamoto, and Arakawa}}]{ohta11a}
\bibinfo{author}{\bibfnamefont{R.}~\bibnamefont{Ohta}},
  \bibinfo{author}{\bibfnamefont{Y.}~\bibnamefont{Ota}},
  \bibinfo{author}{\bibfnamefont{M.}~\bibnamefont{Nomura}},
  \bibinfo{author}{\bibfnamefont{N.}~\bibnamefont{Kumagai}},
  \bibinfo{author}{\bibfnamefont{S.}~\bibnamefont{Ishida}},
  \bibinfo{author}{\bibfnamefont{S.}~\bibnamefont{Iwamoto}}, \bibnamefont{and}
  \bibinfo{author}{\bibfnamefont{Y.}~\bibnamefont{Arakawa}},
  \bibinfo{journal}{Appl. Phys. Lett.} \textbf{\bibinfo{volume}{98}},
  \bibinfo{pages}{173104} (\bibinfo{year}{2011}).

\bibitem[{\citenamefont{Stroud}(2004)}]{stroud04a}
\bibinfo{author}{\bibfnamefont{C.}~\bibnamefont{Stroud}},
  \emph{\bibinfo{title}{A jewel in the crown}} (\bibinfo{publisher}{Institute
  of Optics}, \bibinfo{year}{2004}), chap.~\bibinfo{chapter}{30}.

\end{thebibliography}

\end{document}